\documentclass{egpubl}
\usepackage{eg2025}

\ConferencePaper        
\CGFccbyncnd
\usepackage[T1]{fontenc}
\usepackage{dfadobe}  
\usepackage{cite}  %
\BibtexOrBiblatex
\electronicVersion
\PrintedOrElectronic
\ifpdf \usepackage[pdftex]{graphicx} \pdfcompresslevel=9
\else \usepackage[dvips]{graphicx} \fi
\usepackage{egweblnk} 

\usepackage{amsmath,amsfonts,bm}

\def\Secref#1{Section~\ref{#1}}

\def\eqref#1{equation~\ref{#1}}

\def\Algref#1{Algorithm~\ref{#1}}

\def\1{\bm{1}}

\DeclareMathAlphabet{\mathsfit}{\encodingdefault}{\sfdefault}{m}{sl}
\SetMathAlphabet{\mathsfit}{bold}{\encodingdefault}{\sfdefault}{bx}{n}

\newcommand{\KL}{D_{\mathrm{KL}}}

\hypersetup{
    colorlinks,
    linkcolor={red!50!black},
    citecolor={blue!50!black},
    urlcolor={blue!80!black},
    pdftitle={REED-VAE: RE-Encode Decode Training for Iterative Image Editing with Diffusion Models},
    pdfpagemode=FullScreen
}
\usepackage{amsmath}
\usepackage[linesnumbered,ruled,vlined,norelsize]{algorithm2e}
\usepackage{booktabs}
\usepackage{multirow}
\usepackage{soul}
\usepackage{algorithmic}
\usepackage{cleveref}
\usepackage[export]{adjustbox}
\usepackage{rotating}
\usepackage{wrapfig}
\usepackage{float}
\usepackage{array}
\usepackage{tabularx}
\usepackage{tikz}
\usetikzlibrary{spy, calc, positioning, arrows.meta}
\usepackage{enumitem}
\usepackage{colortbl}
\usepackage{arydshln}
\usepackage{siunitx}
\usepackage{url}
\title[REED-VAE: RE-Encode Decode Training for Iterative Image Editing with Diffusion Models]%
      {REED-VAE: RE-Encode Decode Training for Iterative Image Editing with Diffusion Models}

\author[G. Almog \& A. Shamir \& O. Fried]
{\parbox{\textwidth}{\centering Gal Almog$^{1}$
        \hspace{3mm} Ariel Shamir$^{1}$
        \hspace{3mm} Ohad Fried$^{1}$
        }
        \\
{\parbox{\textwidth}{\centering $^1$Reichman University, Israel\\
       }
}
}

\begin{document}
\newcommand{\ignorethis}[1]{}
\newcommand{\redund}[1]{#1}

\newcommand{\apriori    }     {\textit{a~priori}}
\newcommand{\aposteriori}     {\textit{a~posteriori}}
\newcommand{\perse      }     {\textit{per~se}}
\newcommand{\naive      }     {{na\"{\i}ve}}
\newcommand{\Naive      }     {{Na\"{\i}ve}}

\newcommand{\Identity   }     {\mat{I}}
\newcommand{\Zero       }     {\mathbf{0}}
\newcommand{\Reals      }     {{\textrm{I\kern-0.18em R}}}
\newcommand{\isdefined  }     {\mbox{\hspace{0.5ex}:=\hspace{0.5ex}}}
\newcommand{\texthalf   }     {\ensuremath{\textstyle\frac{1}{2}}}
\newcommand{\half       }     {\ensuremath{\frac{1}{2}}}
\newcommand{\third      }     {\ensuremath{\frac{1}{3}}}
\newcommand{\fourth     }     {\ensuremath{\frac{1}{4}}}

\newcommand{\Lone} {\ensuremath{L_1}}
\newcommand{\Ltwo} {\ensuremath{L_2}}

\newcommand{\mat        } [1] {{\text{\boldmath $\mathbit{#1}$}}}
\newcommand{\Approx     } [1] {\widetilde{#1}}
\newcommand{\change     } [1] {\mbox{{\footnotesize $\Delta$} \kern-3pt}#1}

\newcommand{\Order      } [1] {O(#1)}
\newcommand{\set        } [1] {{\lbrace #1 \rbrace}}
\newcommand{\inverse    } [1] {{#1}^{-1}}
\newcommand{\transpose  } [1] {{#1}^\mathrm{T}}
\newcommand{\invtransp  } [1] {{#1}^{-\mathrm{T}}}
\newcommand{\relu       } [1] {{\lbrack #1 \rbrack_+}}

\newcommand{\abs        } [1] {{| #1 |}}
\newcommand{\Abs        } [1] {{\left| #1 \right|}}
\newcommand{\norm       } [1] {{\| #1 \|}}
\newcommand{\Norm       } [1] {{\left\| #1 \right\|}}
\newcommand{\pnorm      } [2] {\norm{#1}_{#2}}
\newcommand{\Pnorm      } [2] {\Norm{#1}_{#2}}
\newcommand{\inner      } [2] {{\langle {#1} \, | \, {#2} \rangle}}
\newcommand{\Inner      } [2] {{\left\langle \begin{array}{@{}c|c@{}}
                               \displaystyle {#1} & \displaystyle {#2}
                               \end{array} \right\rangle}}

\newcommand{\twopartdef}[4]
{
  \left\{
  \begin{array}{ll}
    #1 & \mbox{if } #2 \\
    #3 & \mbox{if } #4
  \end{array}
  \right.
}

\newcommand{\fourpartdef}[8]
{
  \left\{
  \begin{array}{ll}
    #1 & \mbox{if } #2 \\
    #3 & \mbox{if } #4 \\
    #5 & \mbox{if } #6 \\
    #7 & \mbox{if } #8
  \end{array}
  \right.
}

\newcommand{\len}[1]{\text{len}(#1)}

\newlength{\w}
\newlength{\h}
\newlength{\x}

\definecolor{darkred}{rgb}{0.7,0.1,0.1}
\definecolor{darkgreen}{rgb}{0.1,0.6,0.1}
\definecolor{cyan}{rgb}{0.7,0.0,0.7}
\definecolor{otherblue}{rgb}{0.1,0.4,0.8}
\definecolor{maroon}{rgb}{0.76,.13,.28}
\definecolor{burntorange}{rgb}{0.81,.33,0}

\newif\ifdraft
\draftfalse

\ifdraft
  \newcommand{\todo}[1]{{\color{cyan}[\textbf{TODO:} #1]}}
  \newcommand{\gal}[1]{{\color{maroon}[\textbf{Gal:} #1]}}
  \newcommand{\arik}[1]{{\color{darkred}[\textbf{Arik:} #1]}}
  \newcommand{\ohad}[1]{{\color{magenta}[\textbf{Ohad:} #1]}}
  \newcommand{\almog}[1]{{\color{darkgreen}[\textbf{Almog:} #1]}}
  \newcommand{\galh}[1]{{\color{maroon}#1}}
  \newcommand{\arikh}[1]{{\color{darkred}#1}}
  \newcommand{\ohadh}[1]{{\color{magenta}#1}}
  \newcommand{\almogh}[1]{{\color{darkgreen}#1}}
  \newcommand{\dontlove}[1]{{\color{burntorange}#1}}
  \newcommand {\newstuff}[1]{{\color{blue}{#1}}}
  
\else
  \newcommand{\todo}[1]{}
  \newcommand{\gal}[1]{}
  \newcommand{\arik}[1]{}
  \newcommand{\ohad}[1]{}
  \newcommand{\newstuff}[1]{{\color{black}#1}}
  \newcommand{\galh}[1]{{\color{black}#1}}
  \newcommand{\arikh}[1]{{\color{black}#1}}
  \newcommand{\almogh}[1]{{\color{black}#1}}
  
\fi

\newcommand {\reqs}[1]{\colornote{red}{\tiny #1}}

\newcommand*\rot[1]{\rotatebox{90}{#1}}

\newcommand\todosilent[1]{}

\providecommand{\keywords}[1]
{
  \textbf{\textit{Keywords---}} #1
}

\newcommand{\expnumber}[2]{{#1}\mathrm{e}{#2}}

\newlength{\ww}

\newcommand{\fix}{\marginpar{FIX}}
\newcommand{\new}{\marginpar{NEW}}
\teaser{
 \includegraphics[width=0.9\linewidth]{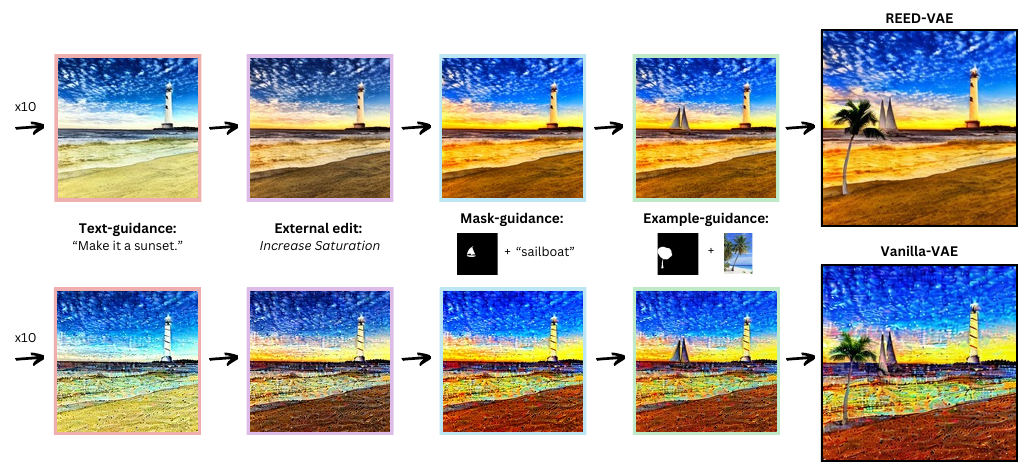}
 \centering
  \caption{REED-VAE (top) preserves image quality over multiple editing iterations, allowing users to perform multiple edit operations using a combination of frameworks and techniques. The Vanilla VAE (bottom) accumulates many artifacts and noise along the way, becoming very apparent once multiple iterative edit operations are performed. The total edit sequence consists of 14 steps, of which only the last 4 are shown here for brevity and to highlight the differences in the final picture. Four types of edit operations are performed: text-guided editing \cite{brooks2023instructpix2pix}, external editing (not diffusion-based), mask-guided editing \cite{avrahami2023blended}, and example-guided editing \cite{yang2023paint}.}
\label{fig:teaser}
}
\maketitle
\begin{abstract}
While latent diffusion models achieve impressive image editing results, their application to iterative editing of the same image is severely restricted. 
When trying to apply consecutive edit operations using current models, they accumulate artifacts and noise due to repeated transitions between pixel and latent spaces. 
Some methods have attempted to address this limitation by performing the entire edit chain within the latent space, sacrificing flexibility by supporting only a limited, predetermined set of diffusion editing operations. 
We present a \underline{re}-\underline{e}ncode \underline{d}ecode (REED) training scheme for variational autoencoders (VAEs), which promotes image quality preservation even after many iterations.
Our work enables \emph{multi-method} iterative image editing: users can perform a variety of iterative edit operations, with each operation building on the output of the previous one using both diffusion-based operations and conventional editing techniques.
We demonstrate the advantage of REED-VAE across a range of image editing scenarios, including text-based and mask-based editing frameworks.
In addition, we show how REED-VAE enhances the overall editability of images, increasing the likelihood of successful and precise edit operations.
We hope that this work will serve as a benchmark for the newly introduced task of multi-method image editing.
Our code and models will be available at: \href{https://github.com/galmog/REED-VAE}{https://github.com/galmog/REED-VAE}.
\end{abstract}

\section{Introduction}
The ability to edit high-resolution images has long been a fundamental aspect of visual content creation, enabling artists and designers to achieve desired aesthetics and convey specific messages.
Traditional image editing techniques range from basic adjustments such as color correction, cropping, and sharpening, to more advanced methods such as applying various filters and layering elements. More recently, diffusion models \cite{ho2020denoising} have led to great advancements not only in high-resolution image generation, but also in editing methods that allow controllable manipulation of existing images.
Diffusion-based editing models can receive various conditioning such as text instructions, reference images, and localization masks, and perform a wide range of tasks such as object addition/removal, object replacement, background replacement, and style or texture changes \cite{avrahami2023blended, brooks2023instructpix2pix, couairon2022diffedit, hertz2022prompt, nichol2021glide, yang2023paint}. Although many of these achieved remarkable results, they center around single-operation editing procedures.

\begin{figure*}[htpb]
    \centering
    \setlength{\tabcolsep}{-2pt}
    \renewcommand{\arraystretch}{0.5}
    \setlength{\ww}{0.153\textwidth}

    \begin{tabular}{p{0.4cm} c c c c c c}

        \rotatebox{90}{\phantom{AAA.}\scriptsize{(1) Vanilla-VAE}} &
        
        \begin{tikzpicture}[spy using outlines={}]
                \node {\includegraphics[width=\ww, frame]{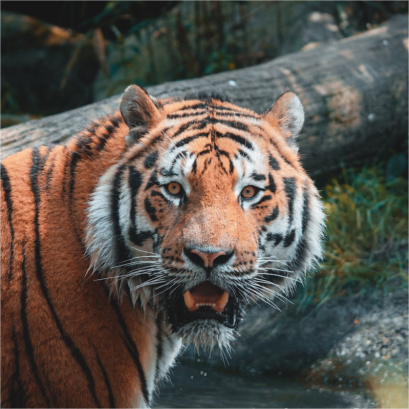}};
        \end{tikzpicture} &

        \begin{tikzpicture}[spy using outlines={}]
            \node {\includegraphics[width=\ww,frame]{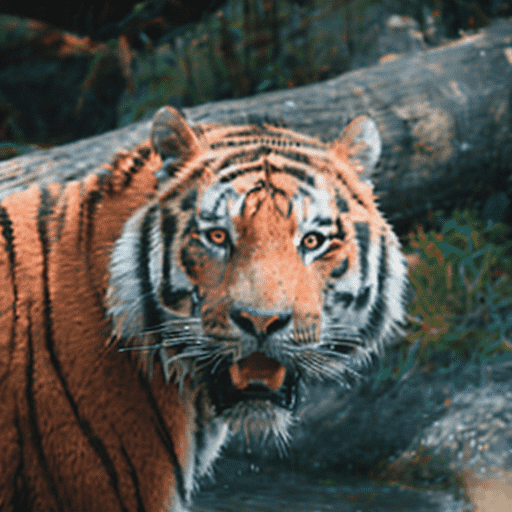}};
        \end{tikzpicture} &

        \begin{tikzpicture}[spy using outlines={}]
            \node {\includegraphics[width=\ww,frame]{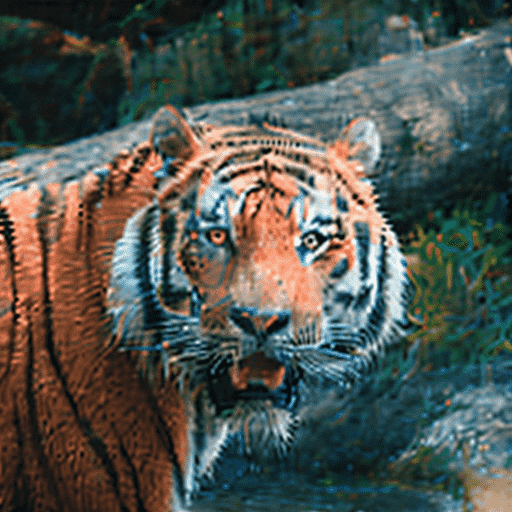}};
        \end{tikzpicture} &

        \begin{tikzpicture}[spy using outlines={}]
            \node {\includegraphics[width=\ww,frame]{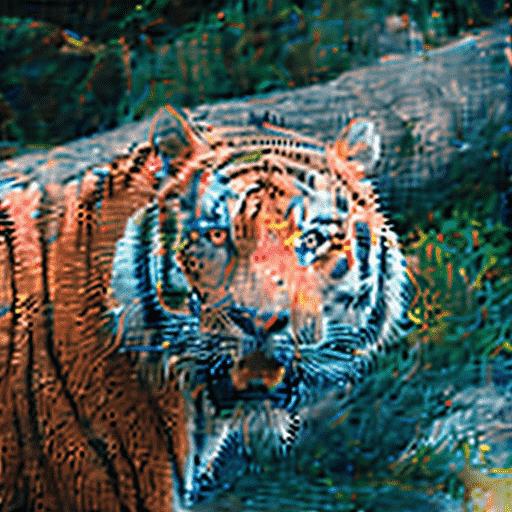}};
        \end{tikzpicture} &

        \begin{tikzpicture}[spy using outlines=]
            \node {\includegraphics[width=\ww,frame]{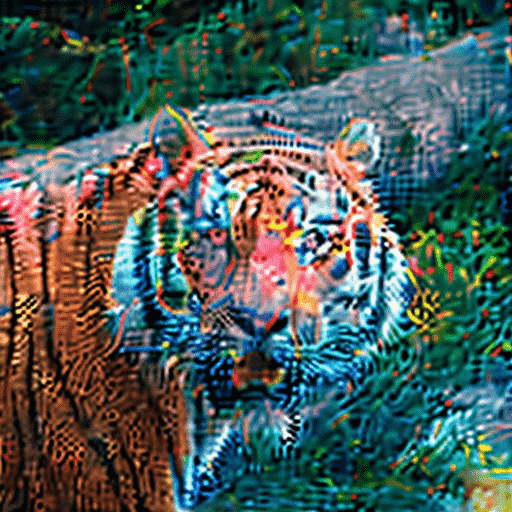}};
        \end{tikzpicture} &

        \begin{tikzpicture}[spy using outlines={circle, thick, black, magnification=2.5,size=0.8cm, connect spies}]
            \node {\includegraphics[width=\ww,frame]{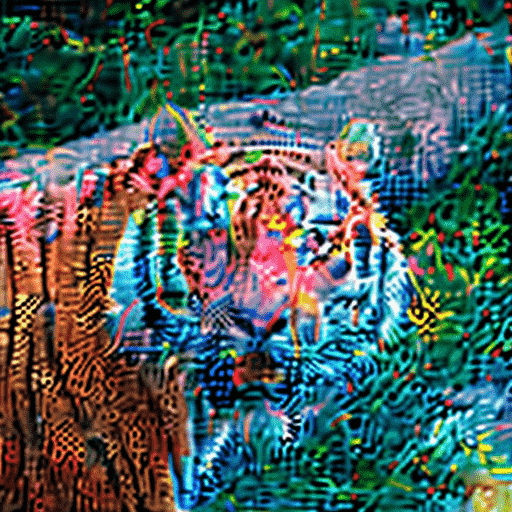}};
            \spy on (0.05,0.25) in node [left] at (1.5,-0.5);
        \end{tikzpicture}
        
        \\[-5pt]
        
        \rotatebox{90}{\phantom{AAAa}\scriptsize{(2) REED-VAE}} &
        \begin{tikzpicture}[spy using outlines={}]
            \node {\includegraphics[width=\ww,frame]{figures/iter_encode_decode/assets/tiger.png}};
        \end{tikzpicture} &

        \begin{tikzpicture}[spy using outlines={}]
            \node {\includegraphics[width=\ww,frame]{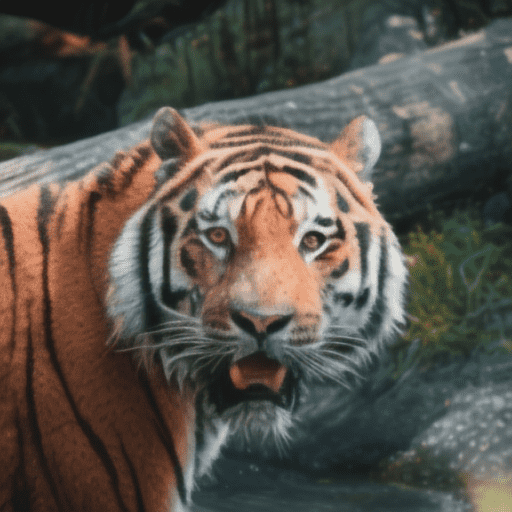}};
        \end{tikzpicture} &

        \begin{tikzpicture}[spy using outlines={}]
            \node {\includegraphics[width=\ww,frame]{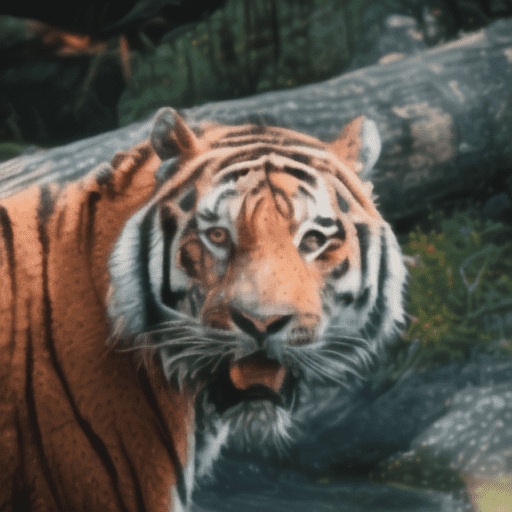}};
        \end{tikzpicture} &

        \begin{tikzpicture}[spy using outlines={}]
            \node {\includegraphics[width=\ww,frame]{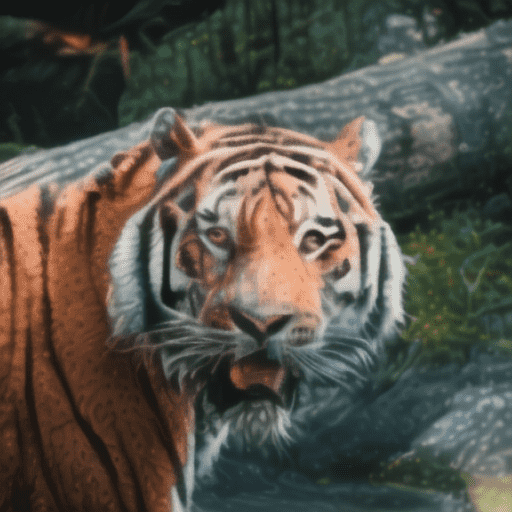}};
        \end{tikzpicture} &

        \begin{tikzpicture}[spy using outlines={}]
            \node {\includegraphics[width=\ww,frame]{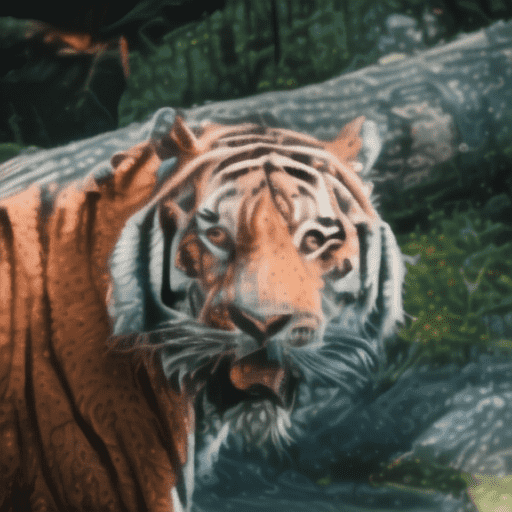}};
        \end{tikzpicture} &

        \begin{tikzpicture}[spy using outlines={circle,black,magnification=2.5,size=0.8cm, connect spies}]
            \node {\includegraphics[width=\ww,frame]{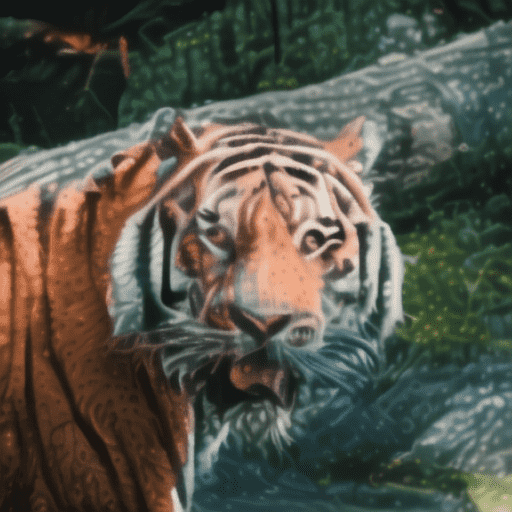}};
            \spy on (0.05,0.25) in node [left] at (1.5,-0.5);
        \end{tikzpicture}   
        
        \\
        
        &
        \scriptsize{Input } &
        \scriptsize{5 } &
        \scriptsize{10 } &
        \scriptsize{15 } &
        \scriptsize{20 } &
        \scriptsize{25 } 

        \\
        &&
        \multicolumn{5}{l}{\phantom{.}
        \begin{tikzpicture}
            \draw[->](0,0)--(13.95,0);
        \end{tikzpicture}}

        \\
        &&
        \multicolumn{5}{c}{
        \scriptsize{Num. Encode/Decode Iterations}
        }
        
    \end{tabular}
    \caption{Even without a diffusion model in the pipeline, the Vanilla-VAE (top row) accumulates artifacts and exhibits significant distortion very quickly throughout encode-decode iterations.
    The tiger's features lose their distinct shapes and edges, appearing more globular and less defined.
    The color palette is altered, with a noticeable increase in blue tones and a decrease in the richness of the orange and greens.
    Fine details such as the grass and the fur are largely lost or blurred.
    REED-VAE (bottom row), produces successive images that are robust to such artifacts and distortions.
    The tiger retains its shape, color, and surface details, demonstrating remarkably high fidelity to the original image.
    The subtle variations in orange and white hues are preserved, and fine elements remain visible.}

    \label{fig:motivation}
\end{figure*}

Ideally, users should be able to integrate the strengths of both diffusion-based models and traditional editing techniques to manipulate images, while applying and interleaving several different editing frameworks. 
In practice, there exists an inherent problem in combining diffusion-based operations with traditional methods in the same editing session. 
This is because diffusion-based models primarily work in the \emph{latent space}, while traditional methods are applied in the \emph{pixel space}.
Therefore, each time one wishes to switch between the two types of techniques, it is necessary either to encode or decode the image into the appropriate representation.
The variational autoencoder (VAE) \cite{kingma2013auto} is the most common model used for this task.
As we show in this paper, this iterative cycle of encoding and decoding destroys the quality of the image by accumulating noise and artifacts with each iteration (see Figure~\ref{fig:motivation}). 

We define \emph{multi-method} iterative image editing as the process of performing multiple (e.g. more than $5$) successive edit operations on an input image; each operation uses the previous output as its input, leveraging both diffusion-based models (latent space) and conventional editing techniques (pixel space).

Our work aims to enable  such \emph{multi-method} iterative image editing by mitigating the artifacts introduced by the VAE in the iterative autoencoding process.
We train a new VAE using a novel \underline{re}-\underline{e}ncode \underline{d}ecode (REED) training scheme.
Our training procedure utilizes an iterative training process together with dynamic incrementation and a first-step loss, that together improve image quality retention over many encode-decode iterations.
We demonstrate the impact of replacing the Vanilla-VAE with our REED-VAE through experiments using a wide range of diffusion-based image editing models.
As our REED-VAE is based on the architecture of the vanilla VAE used in Stable Diffusion \cite{rombach2022high}, it can be easily swapped into the vast majority of models.

In addition to improving multi-method iterative image editing, our REED-VAE facilitates integration between different editing paradigms; for example, between GAN-based editing methods and diffusion-based methods.
Furthermore, this improvement facilitates a seamless transition between editing with multiple different diffusion models that may have different latent spaces, for example, SD2 (4-channel) \cite{rombach2022high} and SD3 (16-channel) \cite{esser2024scaling}.
Beyond image editing, reducing the noise and artifacts accumulated through iterative VAE use has applications in other domains, such as NeRF editing methods in which it is very common to apply the VAE multiple times.
In ED-NeRF \cite{park2023ed}, it has been demonstrated that performing the NeRF editing process entirely in the latent space, thus avoiding repeated applications of the VAE, leads to significantly better results.
We will publish our code and trained REED-VAE model in hope that they can be a useful contribution to the community.

\section{Related work}
\label{related_work}
\paragraph*{Latent Diffusion Models}
Diffusion models \cite{ho2020denoising, ramesh2022hierarchical, saharia2022photorealistic} are a class of deep generative models trained to convert random noise to an image sample from a given distribution. 
These models generate images in an iterative manner, removing a small amount of noise at each time step.
To reduce the computational cost of high-resolution image generation with diffusion models, \cite{rombach2022high} introduced their latent diffusion model (LDM), which utilize a separately-trained autoencoding model that learns to map images to a latent space. 
This latent space is perceptually equivalent to the original pixel space, but significantly reduces computational complexity. 
LDMs achieve impressive results in both image generation and editing, and thus form the basis of all state-of-the-art editing models we evaluate REED on.
We refer to \cite{ho2020denoising} for more details on diffusion models and their implementation.

\paragraph*{Variational Autoencoders}
Variational autoencoders (VAEs), introduced by \cite{kingma2013auto}, are a class of probabilistic models designed to find a low-dimensional (latent) representation of data, widely used in LDMs for converting images to and from their latent representations.
Unlike traditional (deterministic) autoencoders that encode a vector $x$ into a single latent vector $z$ and decode $z$ back to the original space, VAEs encode the input image as a \emph{distribution} over the latent space. 
This regularizes the latent space and ensures that the model generates data following a specified distribution.
Common VAE architectures include the Vector Quantized Variational Autoencoder (VQ-VAE) \cite{van2017neural} and VQ-GAN \cite{esser2021taming}.
Stable Diffusion models \cite{rombach2022high} commonly use a traditional VAE regularized with either KL-divergence \cite{kingma2013auto} or vector quantization (VQ-GAN, \cite{esser2021taming}) in their diffusion models.
Since the VAE model with KL loss is the most prevalent in editing models, we use it as the baseline for REED to ensure maximal compatibility with other models. 

\paragraph*{Image Editing with Diffusion Models}
Diffusion models \cite{ho2020denoising, ramesh2022hierarchical, saharia2022photorealistic} have significantly advanced high-resolution image generation and editing methods, enabling tasks such as object addition/removal, object replacement, background changes, and style or texture changes \cite{huang2024diffusion}.
These models are typically conditioned on text instructions \cite{brooks2023instructpix2pix, parmar2023zero, Zhang2023MagicBrush, gal2022textual, zhang2024hive, pan2023effective, kawar2023imagic} or reference images \cite{yang2023paint, meng2021sdedit, ju2023direct}, sometimes with additional masks for localizing edits or inpainting \cite{avrahami2023blended, avrahami2022blended, nichol2021glide, rombach2022high, lugmayr2022repaint, cao_2023_masactrl}

Prompt-to-Prompt (P2P) \cite{hertz2022prompt} introduced attention modification as a framework for image editing by identifying that the cross-attention layers in the diffusion model link prompt tokens to the image layout.
By swapping attention masks between the source and target images, P2P allows specific elements to be edited while keeping the rest static.
However, P2P was limited to generated (synthetic) images; to enable real-image editing, \emph{inversion} techniques such as DDIM Inversion \cite{dhariwal2021diffusion, song2020denoising} are required to map real images into the latent space of pre-trained diffusion models.
Inversion is the task of finding the latent vector such that denoising it with the pre-trained diffusion model will return the original image, allowing image latents to be edited throughout the denoising process.
Originally, DDIM Inversion suffered from notable limitations in preserving high-frequency details and achieving exact image reconstrctions, which are crucial for editing workflows.
To address these limtiations, Null-Text Inversion (NTI) \cite{mokady2023null} was introduced as an improvement over DDIM Inversion.
NTI refines the inversion process by leveraging null-text guidance to achieve highly accurate reconstructions, thus enabling real-image editing with methods such as P2P.
InstructPix2Pix \cite{brooks2023instructpix2pix} introduced instructional image editing \cite{Zhang2023MagicBrush, parmar2023zero} by training a fully supervised diffusion model that can edit based on human instructions --- for example, ``swap the car with a motorcycle''.

In addition to text, other methods utilize masked regions with corresponding text or a reference image for local editing.
Blended Latent Diffusion \cite{avrahami2023blended} achieves smooth edits by blending the edited region within the mask with the background at each diffusion step.
Paint by Example (PbE) \cite{yang2023paint} performs subject-driven editing using an input mask and a reference image, utilizing self-supervised learning to generate training data.
DragDiffusion \cite{shi2023dragdiffusion} enables interactive point-based image editing that achieve accurate spatial control.

Each of the described editing models builds upon a latent diffusion model, employing a vanilla-VAE to transition in and out of the latent space for each edit operation. 
Few works address the limitation of long editing sequences imposed by the VAE \cite{joseph2024iterative, yang2023rerender}, yet their solutions are not \emph{multi-method}, disallowing a combination of edit methods that are diffusion-based and those that operate in the pixel space in the same edit session. 
In contrast, REED-VAE can be seamlessly integrated into any diffusion-based editing method, replacing the original VAE to facilitate iterative image editing that better retains image quality, while also allowing to interleave non-diffusion-based editing operations.
We demonstrate its effectiveness on different types of editing models in~\Secref{experiments}.

\paragraph*{Iterative Image Editing}
To the best of our knowledge, only one prior work has directly addressed iterative image editing.
Joseph et al.~\cite{joseph2024iterative} extend InstructPix2Pix~\cite{brooks2023instructpix2pix} to support iterative multi-granular editing by staying inside the latent space throughout the entire editing session, only decoding back into pixel space once at the end.
At each iterative step, they overcome the autoencoder-induced degradation by using the previous encoded latent and the current edit instruction as input to the diffusion model, rather than decoding it after each step.
There are several notable limitations of this approach:
\begin{enumerate}
    \item \textbf{Restriction to diffusion-based editing}: Editing sessions are restricted to diffusion-based editing methods that operate in the latent space, excluding manual image manipulations, traditional editing techniques, and editing models that operate directly in the pixel space.

    \item \textbf{Restriction to a single model:} Edit operations are limited to models that use the same latent space. This prevents applying both diffusion-based editing and GAN-based editing, for example, or even mixing diffusion models that have different latent spaces, such as SD2 \cite{rombach2022high} and SD3 \cite{esser2024scaling}

    \item \textbf{Rigid and inconvenient workflow}: Users must either predetermine all edits and their locations, limiting the exploratory nature of the creative process, or save an additional latent vector along with each output image, complicating storage and sharing.
\end{enumerate}
Our work addresses these limitations by enabling users to iteratively edit images using any combination of methods that operate in either the pixel space or latent space, without the need to predetermine all edit operations or handle extra latent vectors.

\section{Problem Definition}
We call our new problem setting \emph{multi-method iterative image editing}.
At step $i$, our goal is to apply the edit operation $e^i$ to the input image $x^i$, such that the output $x^{i+1}$ can serve as the input for the next iteration while minimizing the accumulation of noise.
We use the term \emph{multi-method} to emphasize that each $e^i$ may refer to any type of diffusion-based editing operation, but also any type of operation employed in traditional or commercial image editing tools.
Generally, we aim for each step in the editing process to be independent, so that users have the creative freedom to apply each $e^i$ at any stage during the editing session, using a different tool and without the need to pre-define each edit in advance.
We believe that this setting better represents real-life workflows and is more conducive to the artistic process of image editing.

\begin{figure}[htpb]

    \setlength{\tabcolsep}{-2pt}
    \renewcommand{\arraystretch}{0.5}
    \setlength{\ww}{0.11\textwidth}

    \begin{tabular}{c c c c}
        \scriptsize{(a) Input } &
        \scriptsize{(b) Vanilla-VAE } &
        \scriptsize{(c) Vanilla-VAE (Blurred)} &
        \scriptsize{(d) REED-VAE } 
        \\
        \begin{tikzpicture}[spy using outlines={}]
                \node {\includegraphics[width=\ww, frame]{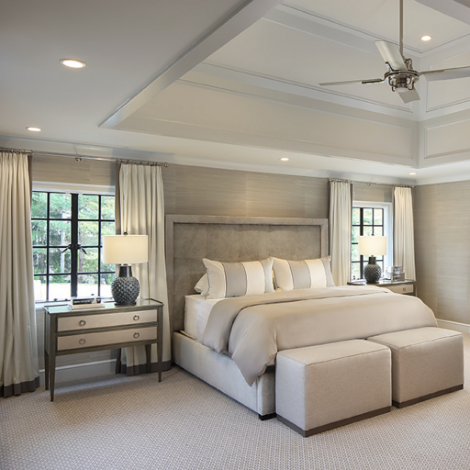}};
        \end{tikzpicture} &

        \begin{tikzpicture}[spy using outlines={}]
            \node {\includegraphics[width=\ww,frame]{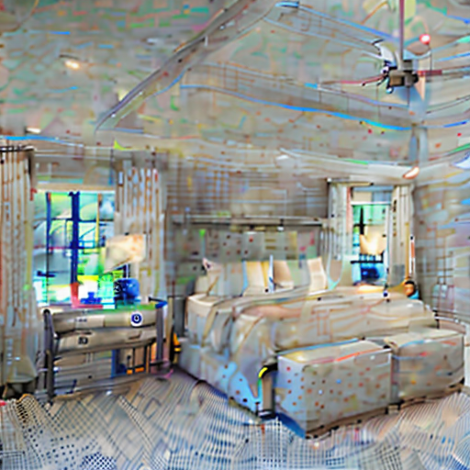}};
        \end{tikzpicture} &

        \begin{tikzpicture}[spy using outlines={}]
            \node {\includegraphics[width=\ww,frame]{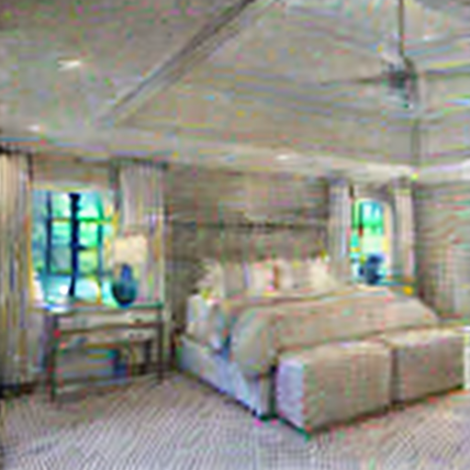}};
        \end{tikzpicture} &

        \begin{tikzpicture}[spy using outlines={}]
            \node {\includegraphics[width=\ww,frame]{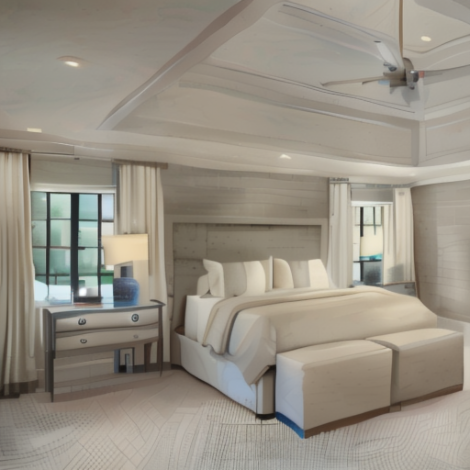}};
        \end{tikzpicture}
        
        \\[-5pt]

        \begin{tikzpicture}[spy using outlines={}]
            \node {\includegraphics[width=\ww,frame]{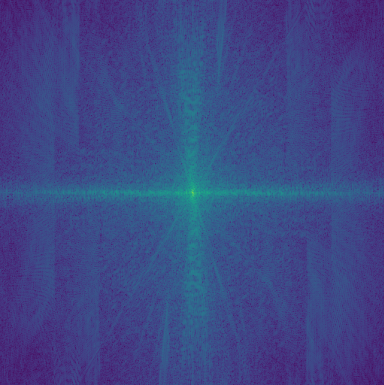}};
        \end{tikzpicture} &

        \begin{tikzpicture}[spy using outlines={}]
            \node {\includegraphics[width=\ww,frame]{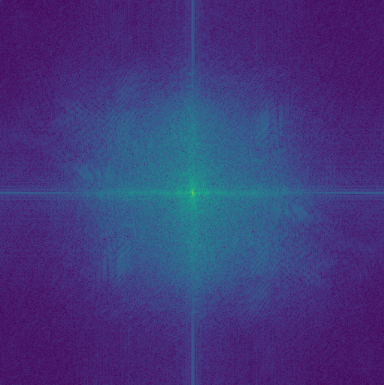}};
        \end{tikzpicture} &

        \begin{tikzpicture}[spy using outlines={}]
            \node {\includegraphics[width=\ww,frame]{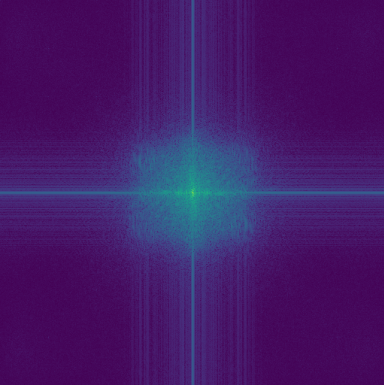}};
        \end{tikzpicture} &

        \begin{tikzpicture}[spy using outlines={}]
            \node {\includegraphics[width=\ww,frame]{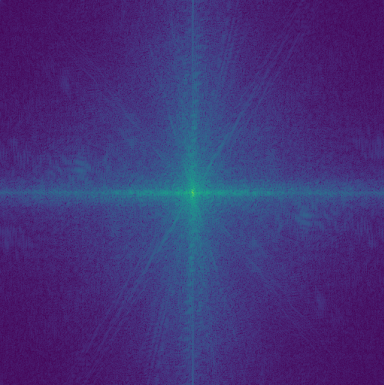}};
        \end{tikzpicture}

        \\
        
    \end{tabular}
    \caption{
    Given an input image (a) we perform 20 encode-decode iterations and present the the results in image (top) and frequency domain (bottom). Vanilla-VAE (b) exhibits significant loss of high-frequency information (evidenced by the dimming and blurring of the outer regions of the spectrum), and dominance of low-frequency features (evidenced by the enlarged central bright region). In addition, it also introduced new high-frequency features that are not seen in the input image, indicating an introduction of repetitive artifacts.
    Trying to apply smoothing after each encode-decode iteration (c) solves some of these problems at the cost of blurring the image.
    REED-VAE (d) demonstrates superior performance in preserving image fidelity across all frequency bands.
    }
    \label{fig:fourier}
\end{figure}

Na\"ively passing iterative image outputs to the diffusion model accumulates artifacts and renders the images essentially destroyed when performing editing beyond a few  operations.
As demonstrated in~\Cref{fig:motivation}, even simply encoding and decoding the same image iteratively (without editing or using the diffusion model) is enough to accumulate significant artifacts after only 5-10 iterations.
Similar to previous work \cite{joseph2024iterative, yang2023rerender}, we find that this degradation is a result of the lossy VAE used in the diffusion process. 
Figure~\ref{fig:fourier}(a-b) demonstrates this in the frequency domain: the Vanilla-VAE exhibits significant loss of high-frequency information after several encode-decode cycles, while also accumulating high-frequency noise and artifacts.
Therefore, to successfully support multi-method iterative image-editing, we focus our efforts on preventing the degradation that occurs due to the reconstruction error of the VAE in LDMs.

\section{REED: RE-Encode Decode Training}
\label{method}
We train a VAE that, when paired with a diffusion-based image editing model, can maintain image quality and editability over iterations.
VAEs consist of an encoder network that defines a posterior distribution $q(z|x)$, a prior distribution $p(z)$, and a decoder network that models $p(x|z)$.
Typically, both the posterior and prior distributions are chosen to be normal with diagonal covariance for efficient parameterization by the Gaussian reparameterization trick \cite{kingma2013auto}.
Training is regularized with a KL-divergence term between the returned posterior and a standard Gaussian distribution.

Fine-tuning VAE models on specific or niche datasets has been demonstrated to outperform training models from scratch in image generation and editing pipelines. 
Consequently, we initialize our model's weights with a pretrained VAE checkpoint from Stable Diffusion to leverage its extensive encoding and decoding capabilities.
However, to ensure wide compatibility with many image editing models, we fine-tune only the decoder of the VAE, leaving the encoder weights frozen.
This ensures that the latent embeddings, on which the diffusion model itself is highly dependent, remain consistent and aligned with the model's training distribution.
This strategy also reduces the computational resources required for training.
Additional training details are provided in the supplemental material.

\subsection{Iterative training}
As our goal is to reduce the quality degradation that occurs in the iterative encoding and decoding process, we train our model with this task in mind.
We define a parameter $k$ to indicate the number of encode-decode iterations performed on each sample in the training loop.
For each training iteration, we begin by encoding ($\mathcal{E}$) the source image $x^0$, sampling $z^0$ 
from the encoded latent distribution, and decoding ($\mathcal{D}$) $z^0$ to acquire the output image $x^1$.
We define this as one encode-decode iteration, and repeat this process for each $x^i$ for $i=0$ to $k$, where $x^k$ is the final output image as follows:
\begin{equation}
    \mathcal{D}(\mathcal{E}(x^i)) = x^{i+1}
\end{equation}
Whereas the reconstruction loss of the Vanilla VAE is computed after one iteration between $x^0$ and $x^{1}$, we perform $k$ encode-decode iterations and compute the loss, detailed in \Cref{par:loss}, on $x^k$.
We hypothesize that explicitly training the model to reconstruct images after multiple encode-decode iterations will allow it to generalize to the image editing task, i.e., will also improve the model's ability to reconstruct successive images when edits are performed.
First, iterative encoding and decoding is a simpler task than iterative editing, yet the VAE still exhibits a large amount of degradation, as evident in Figure~\ref{fig:motivation}.
Therefore, this is a good intermediate goal for our model.
Second, our experiments show that a similar artifact accumulation occurs in both image editing and simple encode-decode cycles, therefore, it is reasonable to assume that a VAE that overcomes one challenge will also improve in the other.
Our results indeed confirm that our model demonstrates improved reconstruction accuracy during both iterative encode-decode cycles \emph{and} iterative edit operations (see Section \ref{experiments}). 
We further analyze and validate the advantage of having $k>1$ in our ablation studies (\Cref{ablations}).

\subsection{Dynamic incrementation}
We initially find that a higher $k$ value imparts a greater reduction in the training loss, however; past $k=6$, the task becomes too difficult and the model does not converge.
To overcome this, we propose a dynamic loss progression that makes use of increasingly higher values of $k$ to achieve better convergence.
We begin by initializing $k=4$ and computing the training loss against $x^k$ in each training iteration.
At the end of each epoch, we compute a validation loss on a separate validation set, also against the same $x^k$ (details on the exact loss implementation in~\cref{par:loss}).
If the validation loss does not improve in 5 consecutive iterations, signifying a plateau in the training, we increment $k \gets k+1$.
Effectively, we perform a variation of curriculum learning \cite{bengio2009curriculum}, which attempts to mimic human learning by gradually increasing the complexity of data samples used when training a model.
Empirically, we find $k<4$ to be too trivial of a task, leading to no convergence (see ablations in~\Cref{ablations}).
We stop training once $k$ passes $20$.
We find that this training method achieves the best results, allowing the model to learn to reconstruct images almost perfectly after $10$ iterations with good generalisability to higher iterations.

\subsection{First-step loss} 
When using any existing model to compress an image into a latent representation, there is an inherent and unavoidable loss of information, such that even after a single encode/decode iteration the reconstruction will not be perfect.
We are interested in improving the model's \emph{iterative} performance, and not its general performance (i.e., we are satisfied if we match the original model's performance after one iteration, as long as we improve it for consecutive iterations).
For this reason, rather than computing the training loss between the last iteration output $x^k$ and the source image $x^0$, we instead compute the training loss between $x^k$ and $x^1$.
Our experiments and qualitative analysis show that the first-step loss helps the VAE learn the iterative task more easily, which we believe is due to the reduced complexity of the task (discussed further in our ablation studies, \Cref{ablations}).
Note that the validation loss and test metrics are still computed against $x^0$, as this is the true performance indicator.

\subsection{Training objective}\label{par:loss}
Putting the three components (iterative training, dynamic incrementation and first-step loss) together, our full REED training algorithm is presented in~\Algref{algo1}.
We use the same training loss as in the vanilla-VAE \cite{rombach2022high, kingma2013auto}, which is composed of an MSE reconstruction term $(\mathcal{L}_{MSE})$ along with a weighted perceptual loss ($\mathcal{L}_{LPIPS}$, \cite{zhang2018unreasonable}) term.
An additional KL-divergence term $(\KL)$ is computed between the latent vector and the standard normal distribution for regularization. 
We find that despite only training the decoder, the $\KL$ term still helps to improve the VAE's performance.
We believe this is due to the iterative nature of our training method: the output of the decoder indirectly impacts the next latent vector, therefore latent space regularization is still beneficial.
We add additional scaling parameters $\alpha$ and $\beta$ to scale the LPIPS and $\KL$ terms, respectively. 
The full objectives for training and validating the VAE are as follows:
\begin{equation}
    \resizebox{0.9\hsize}{!}{$
    \mathcal{L}_{train} = \mathcal{L}_{MSE}(x^1, x^k) + \alpha \cdot \mathcal{L}_{LPIPS}(x^1, x^k) + \beta \cdot \KL \left(z^k, \mathcal{N}(0, I) \right)
    $}
\end{equation}
\begin{equation}
    \mathcal{L}_{val} = \mathcal{L}_{MSE}(x^0, x^k) + \alpha \cdot \mathcal{L}_{LPIPS}(x^0, x^k)
\end{equation}

\newcommand{\hrulealg}[0]{\vspace{1mm} \hrule \vspace{1mm}}
\begin{algorithm}
\SetAlgoLined
\KwIn{Training data $\mathbf{X}$, number of epochs $N$, pretrained encoder $\mathcal{E}$ and decoder $\mathcal{D}$ of vanilla VAE model}
\KwOut{Trained REED-VAE decoder $\mathcal{D}$ parameters}
\hrulealg
Initialize number of encode/decode iterations $k \gets 4$\;
\For{$epoch \leftarrow 1$ \KwTo $N$}{
    \For{each image $x^0 \in \mathbf{X}$}{
        \For{$i=0,1,\dots,k-1$}{
            Encode $z^{i} \gets \mathcal{E}(x^{i})$\;
            Decode $x^{i+1} \gets \mathcal{D}(z^i)$\;
        }
        Take gradient descent step on $\nabla \mathcal{L}_{\text{train}}(x^1, x^k, z^k)$
        }
            \If{$\mathcal{L}_{\text{val}}(x^0, x^k)$ \text{reaches plateau}}{
            $k \gets k+1$\;
            \If{$k > 20$}{
            End training
            }
            }
}
\caption{Re-Encode Decode Training}
\label{algo1}
\end{algorithm}

\begin{figure*}[htpb]
  \centering
  \includegraphics[width=0.9\linewidth]{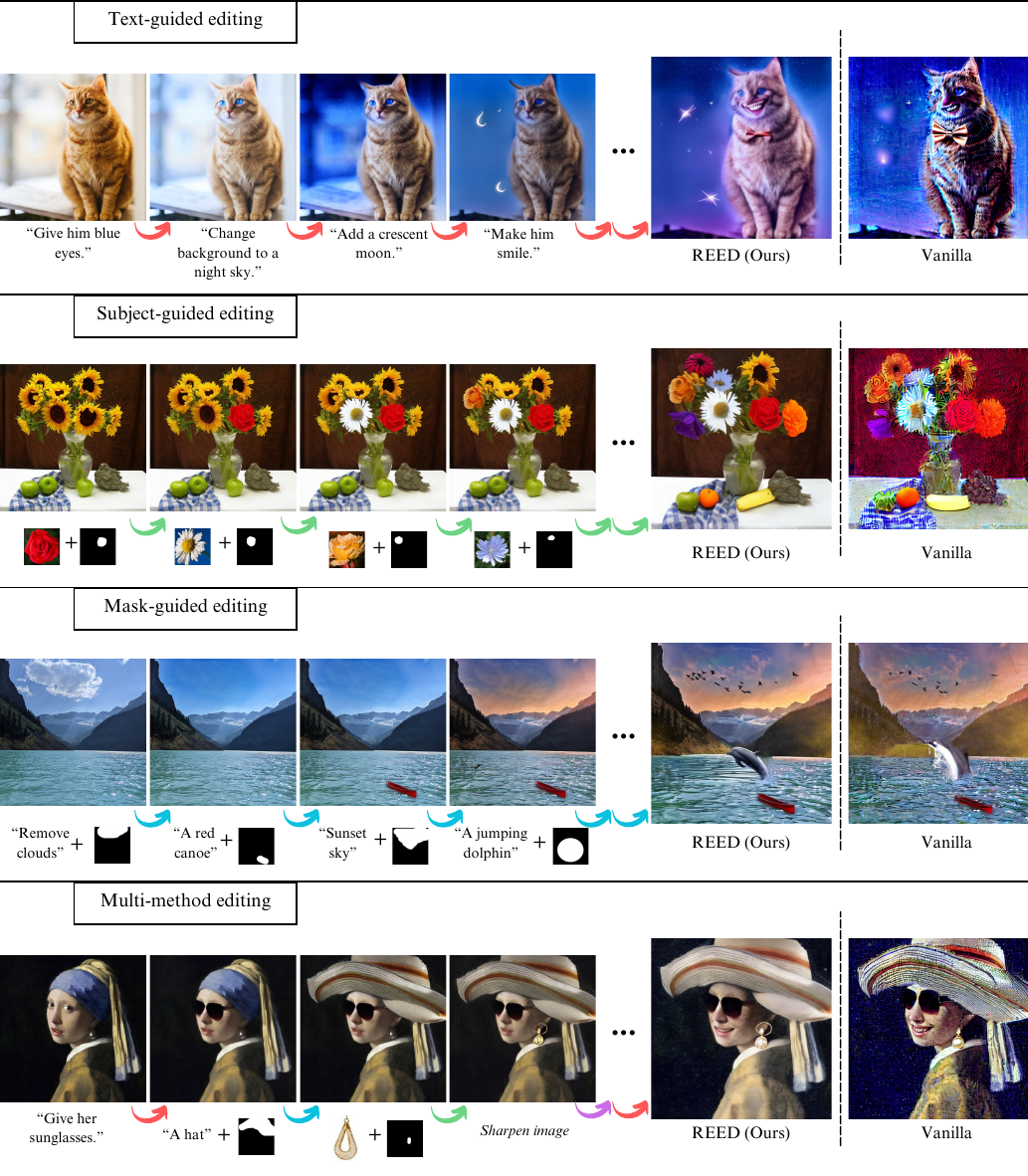}
  \caption{\label{fig:editsessions}
           Examples of types of edit sessions made possible with REED-VAE. Using the Vanilla-VAE (right), significant noise and artifacts accumulate quickly after multiple edit operations. Intermediate edit operations are omitted to highlight the final edited image. Four types of edit operations are performed: text-guided editing \cite{brooks2023instructpix2pix}, external editing (not diffusion-based), mask-guided editing \cite{avrahami2023blended}, and example-guided editing \cite{yang2023paint}.
           }
\end{figure*}

\section{Experiments}
\label{experiments}
We conduct a comprehensive evaluation of our proposed REED-VAE across various image editing models, comparing their performance with and without REED-VAE integration. 
To evaluate the effectiveness of our method, we require a dataset that contains images with ground-truth edits for many ($20+$) steps.
Unfortunately, as we are the first to perform comprehensive iterative image editing, such a dataset does not exist (to the best of our knowledge).
We address this by adapting the recently released ImagenHub dataset \cite{ku2024imagenhub} to an iterative editing process.
ImagenHub aims to standardize the evaluation process for image editing and generation models, providing a comprehensive dataset of 7 task subsets each with 100--200 images.
The dataset images are manually annotated for various image editing scenarios: single-turn, multi-turn, mask-guided, text-guided, and subject-guided image editing.
We refer to the ImagenHub paper \cite{ku2024imagenhub} for more implementation details. 
We leverage ImagenHub as a starting point for our evaluations, making adaptations (detailed below) to each evaluation task to accommodate for the iterative nature of our problem setting.
We note that these adaptations may sometimes result in edits that are not entirely sensible or visually appealing.
However, since our primary concern is quantitatively comparing the performance of REED-VAE to the Vanilla-VAE,
the realism of the editing result is less important, and this procedure provides a fair comparison.

\paragraph*{Evaluation metrics}
Our goal is to apply many iterative edit operations on a single image while maintaining the image quality as best as possible.
For evaluation, we perform a pair of ``inverse edit operations'' $\{e^1, e^2\}$ on each test image, iterating back and forth through these operations for multiple cycles.
For example, $e^1$ might be some edit operation to `change the car into a bus', then $e^2$ will be to `change the bus into a car'.
For all tasks, we use the mean squared error (MSE), LPIPS \cite{zhang2018unreasonable}, SSIM \cite{wang2004image}, and FID \cite{heusel2017gans, Seitzer2020FID} as metrics to evaluate our model's ability to preserve image quality over successive iterations, when compared to the vanilla-VAE.
These metrics are commonly used to quantify reconstruction quality of images and generation quality.
In all experiments, we observe the improvement that REED-VAE imparts at three different iterative editing stages (5, 15, and 25 iterative edit operations) using a diverse set of editing models. 
\begin{figure*}
    \centering
    \setlength{\tabcolsep}{-2pt}
    \renewcommand{\arraystretch}{0.5}
    \setlength{\ww}{0.153\textwidth}

    \begin{tabular}{p{0.4cm} c c c c c c}

        \rotatebox{90}{\phantom{AAA.}\scriptsize{(1) Vanilla NTI}} &
        
        \begin{tikzpicture}[spy using outlines={}]
                \node {\includegraphics[width=\ww, frame]{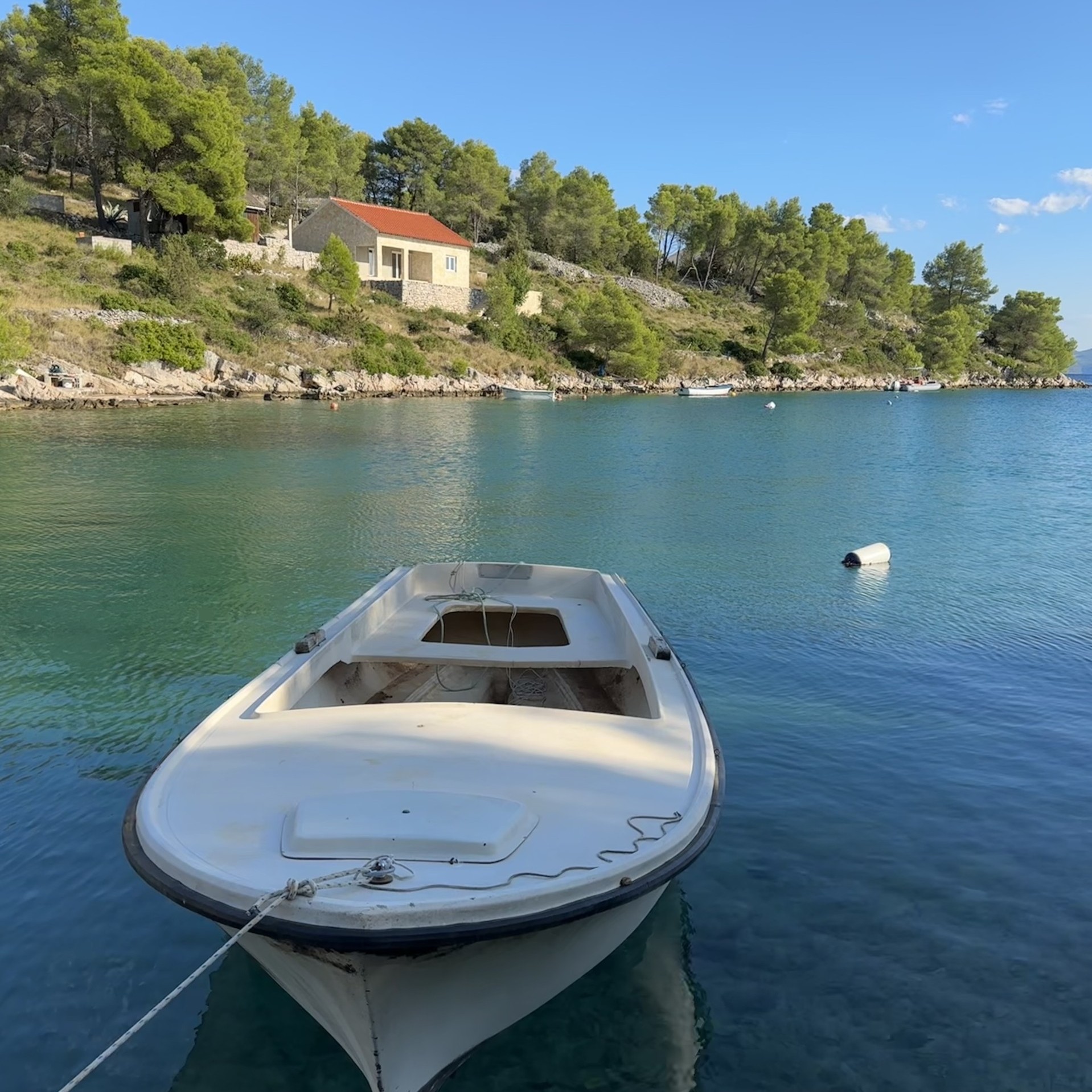}};
        \end{tikzpicture} &

        \begin{tikzpicture}[spy using outlines={}]
            \node {\includegraphics[width=\ww,frame]{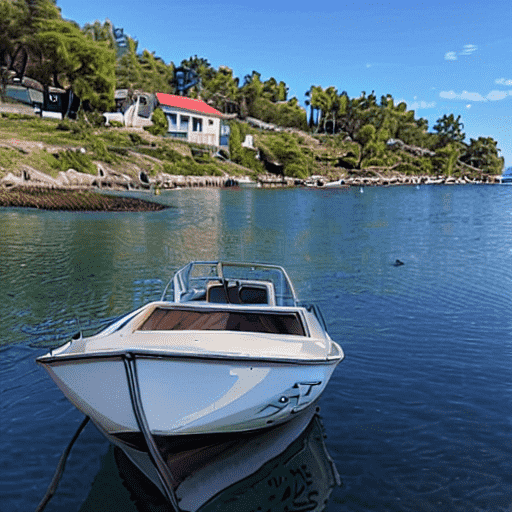}};
        \end{tikzpicture} &

        \begin{tikzpicture}[spy using outlines={}]
            \node {\includegraphics[width=\ww,frame]{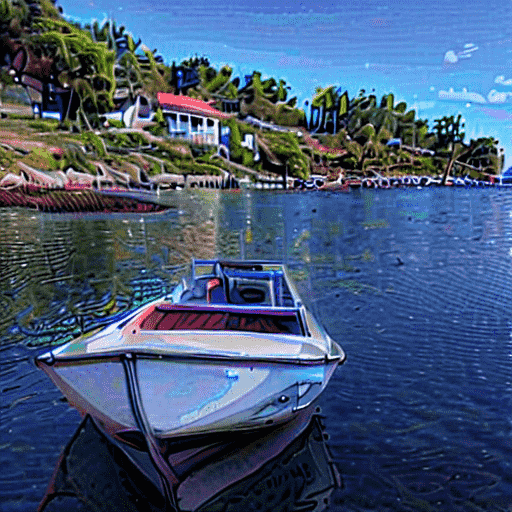}};
        \end{tikzpicture} &

        \begin{tikzpicture}[spy using outlines={}]
            \node {\includegraphics[width=\ww,frame]{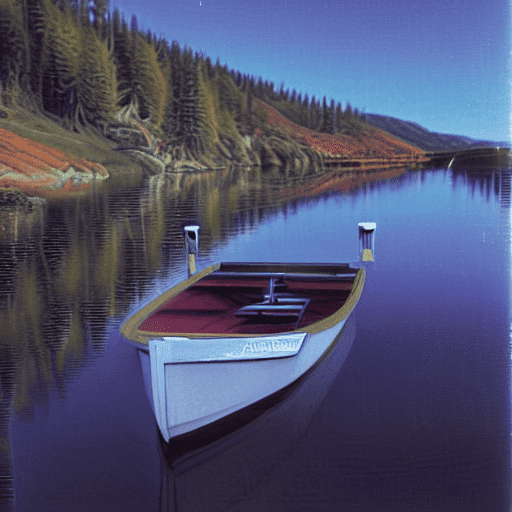}};
        \end{tikzpicture} &

        \begin{tikzpicture}[spy using outlines=]
            \node {\includegraphics[width=\ww,frame]{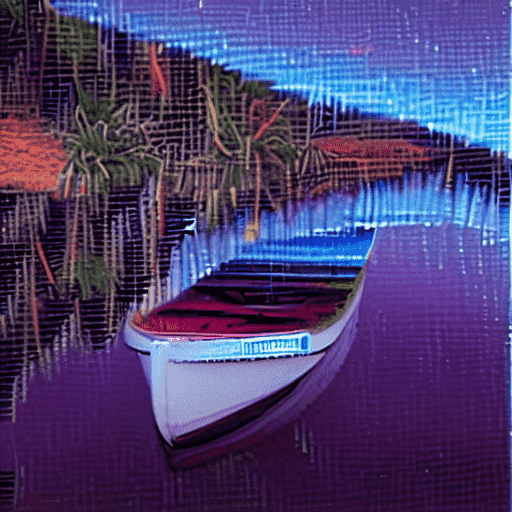}};
        \end{tikzpicture} &

        \begin{tikzpicture}[spy using outlines=]
            \node {\includegraphics[width=\ww,frame]{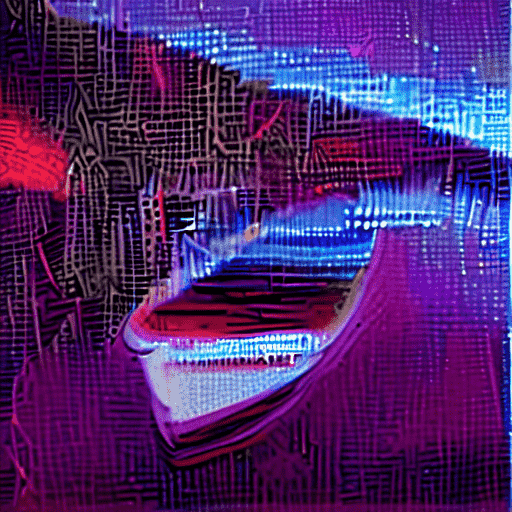}};
        \end{tikzpicture}  
        
        \\[-5pt]

        \rotatebox{90}{\phantom{A.}\scriptsize{(2) Vanilla encode/decode}} &
        
        \begin{tikzpicture}[spy using outlines={}]
                \node {\includegraphics[width=\ww, frame]{figures/iterative_ddim/assets/boat_original.jpg}};
        \end{tikzpicture} &

        \begin{tikzpicture}[spy using outlines={}]
            \node {\includegraphics[width=\ww,frame]{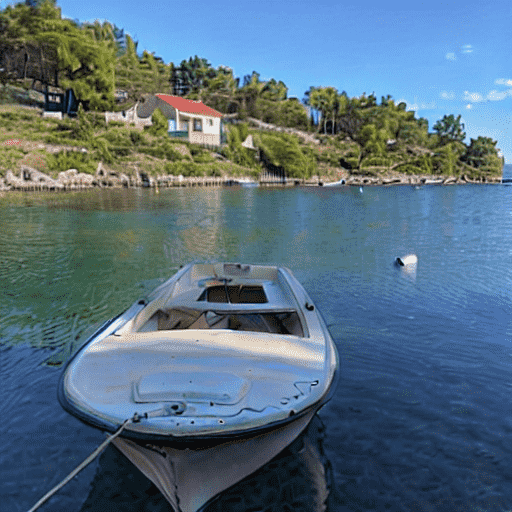}};
        \end{tikzpicture} &

        \begin{tikzpicture}[spy using outlines={}]
            \node {\includegraphics[width=\ww,frame]{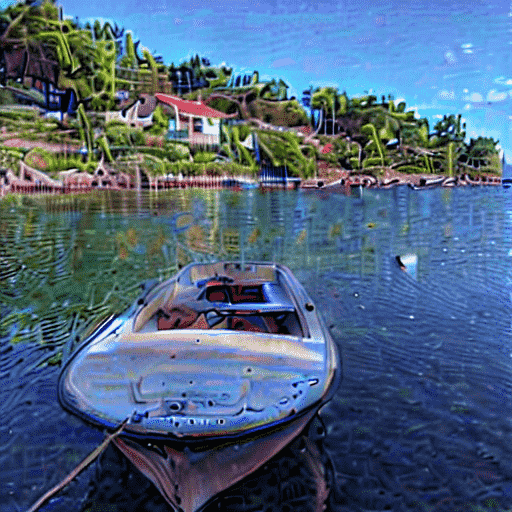}};
        \end{tikzpicture} &

        \begin{tikzpicture}[spy using outlines={}]
            \node {\includegraphics[width=\ww,frame]{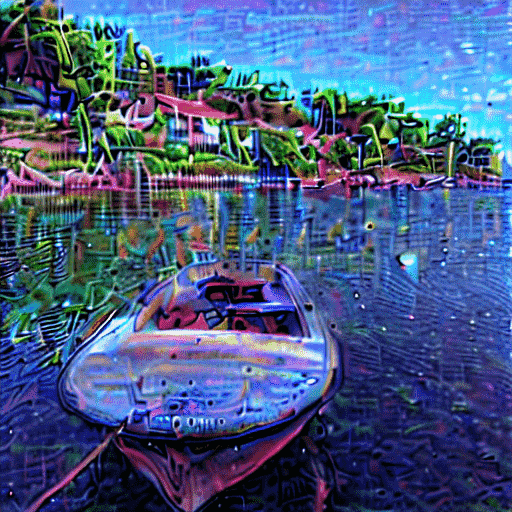}};
        \end{tikzpicture} &

        \begin{tikzpicture}[spy using outlines=]
            \node {\includegraphics[width=\ww,frame]{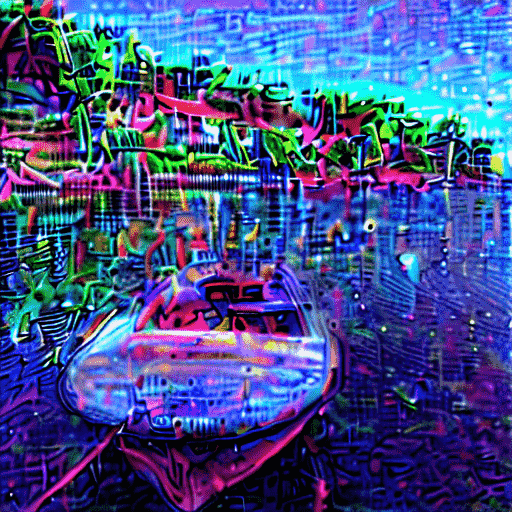}};
        \end{tikzpicture} &

        \begin{tikzpicture}[spy using outlines=]
            \node {\includegraphics[width=\ww,frame]{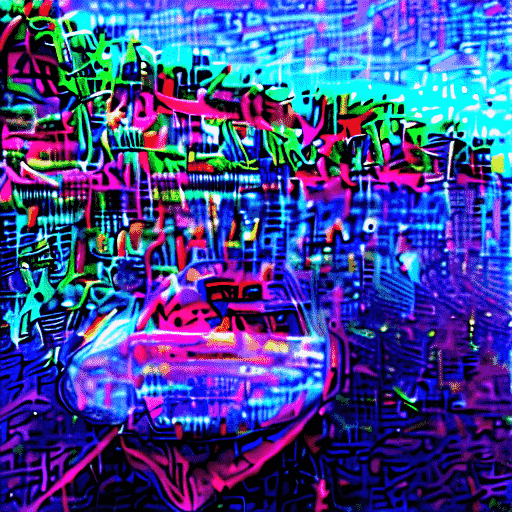}};
        \end{tikzpicture}
        
        \\[-5pt]

        \rotatebox{90}{\phantom{AAA.}\scriptsize{(3) \textbf{NTI + REED}}} &
        
        \begin{tikzpicture}[spy using outlines={}]
                \node {\includegraphics[width=\ww, frame]{figures/iterative_ddim/assets/boat_original.jpg}};
        \end{tikzpicture} &

        \begin{tikzpicture}[spy using outlines={}]
            \node {\includegraphics[width=\ww,frame]{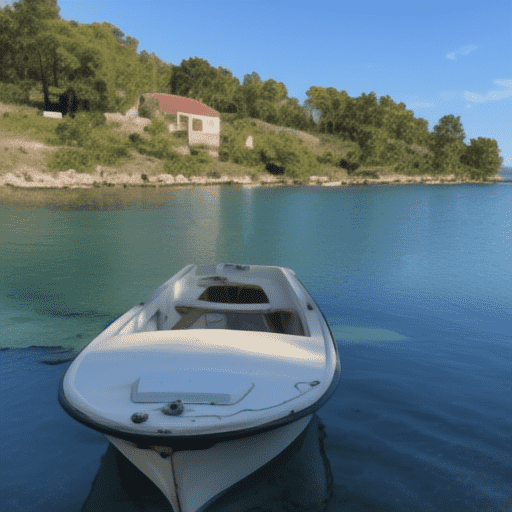}};
        \end{tikzpicture} &

        \begin{tikzpicture}[spy using outlines={}]
            \node {\includegraphics[width=\ww,frame]{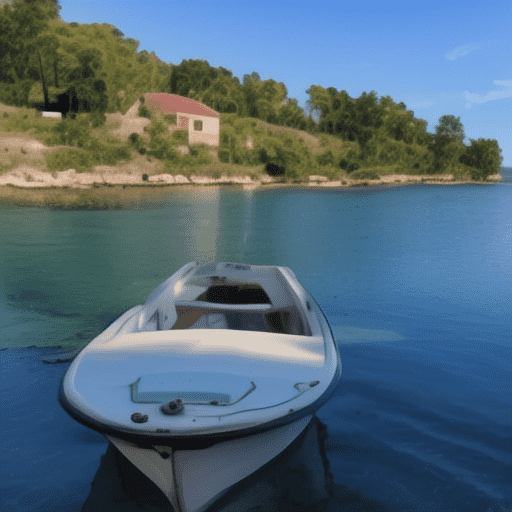}};
        \end{tikzpicture} &

        \begin{tikzpicture}[spy using outlines={}]
            \node {\includegraphics[width=\ww,frame]{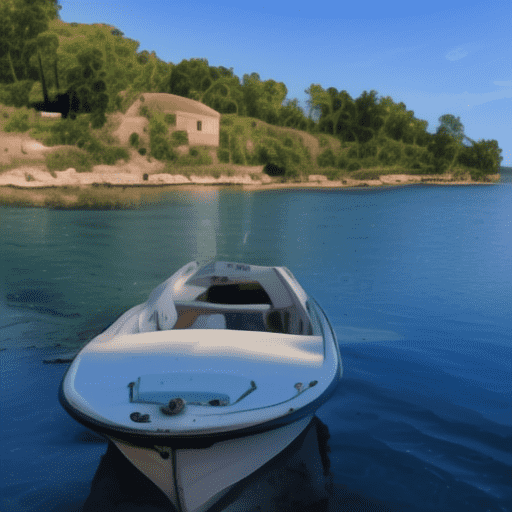}};
        \end{tikzpicture} &

        \begin{tikzpicture}[spy using outlines=]
            \node {\includegraphics[width=\ww,frame]{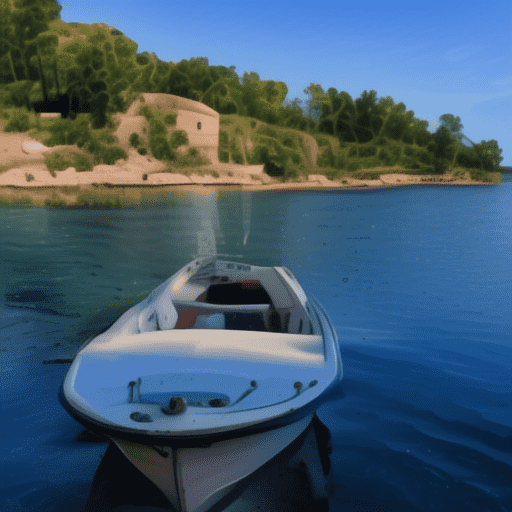}};
        \end{tikzpicture} &

        \begin{tikzpicture}[spy using outlines=]
            \node {\includegraphics[width=\ww,frame]{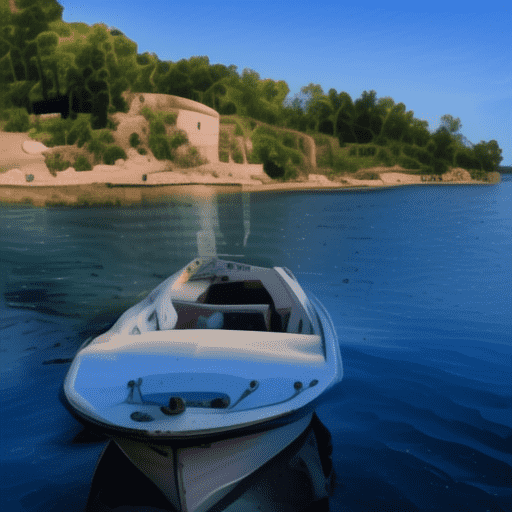}};
        \end{tikzpicture}
        
        \\
        
        &
        \scriptsize{Input } &
        \scriptsize{5 } &
        \scriptsize{10 } &
        \scriptsize{15 } &
        \scriptsize{20 } &
        \scriptsize{25 } 

        \\
        &&
        \multicolumn{5}{l}{\phantom{.}
        \begin{tikzpicture}
            \draw[->](0,0)--(13.95,0);
        \end{tikzpicture}}

        \\
        &&
        \multicolumn{5}{c}{
        \scriptsize{Num. Encode/Decode Iterations}
        }
        
    \end{tabular}
    \caption{Row 1: Null-Text Inversion (NTI) is used to iteratively invert the image and regenerate it from the inverted latent. Row 2: the Vanilla-VAE is used to iteratively encode and decode the image. Row 3: NTI is used with REED-VAE to iteratively invert the image and regenerate it from the inverted latent. Vanilla NTI loses fidelity to the original image and is not resilient to iterative degradation. Full sequences for Vanilla NTI and NTI + REED are available in the Supplementary Material.}
    \label{fig:iterative_ddim}
\end{figure*}

\paragraph*{Notation summary}
Unless stated otherwise, we use the following notation for all following discussions. 
Each sample in our evaluation set has a source image $\mathbf{x}_{s} \in \mathbb{R}^{H \times W \times 3}$ and a target image $\mathbf{x}_{t} \in \mathbb{R}^{H \times W \times 3}$.
The source caption $\mathbf{C}_{s}$ globally describes $\mathbf{x}_{s}$, the target caption $\mathbf{C}_{t}$ globally describes $\mathbf{x}_{t}$, and the local target caption $\mathbf{C}_{t}^{local}$ describes the local object to be edited.
The sample is also annotated with a human instruction $\mathbf{I}_{s}$ for transitioning from $\mathbf{x}_{s}$ to $\mathbf{x}_{t}$.
Some images are additionally annotated with a binary mask $\mathbf{m} \in \{0, 1\}^{H \times W}$ representing the pixels to be edited with a value of 1.
\begin{table*}[htpb]
\centering
    \caption{Comparisons on image editing models. The addition of REED consistently improves the performance of various image editing models across multiple quality metrics and through different iterative editing stages (5,15, and 25 iterations).MSE, PSNR, and LPIPS are computed with each image sample normalized to the [0,1] range prior to evaluation, ensuring consistency in comparison. Other metrics are computed as per their standard definitions. Metrics are computed based on experiments done on the ImagenHub dataset, consisting of 179 images.}
    \begin{adjustbox}{max width=\textwidth}
    \sisetup{detect-all=true,round-mode=places,round-precision=2}
    \begin{tabular}{lSSSSSSSSSSSSSSS}
        \toprule
        \multirow{2}{*}{Method} & \multicolumn{3}{c}{MSE $\downarrow$} & \multicolumn{3}{c}{LPIPS $\downarrow$} & \multicolumn{3}{c}{SSIM $\uparrow$} & \multicolumn{3}{c}{FID $\downarrow$} & \multicolumn{3}{c}{PSNR $\uparrow$} \\
        \cmidrule(lr){2-4} \cmidrule(lr){5-7} \cmidrule(lr){8-10} \cmidrule(lr){11-13} \cmidrule(lr){14-16}
         & {5} & {15} & {25} & {5} & {15} & {25} & {5} & {15} & {25} & {5} & {15} & {25} & {5} & {15} & {25}\\
        \midrule
        IP2P \cite{brooks2023instructpix2pix} & 0.024 & 0.11 & 0.15 & 0.33 & 0.69 & 0.76 & 0.60 & 0.23 & 0.18 & 105.75 & 246.98 & 271.72 & 17.78 & 10.15 & 8.59 \\
        {\hspace{1em}+ REED} & \bfseries 0.017 & \bfseries 0.06 & \bfseries 0.09 & \bfseries 0.18 & \bfseries 0.45 & \bfseries 0.58 & \bfseries 0.80 & \bfseries 0.53 & \bfseries 0.41 & \bfseries 62.84 & \bfseries 138.90 & \bfseries 187.97 & \bfseries 19.81 & \bfseries 13.36 & \bfseries 11.36 \\

        \midrule

        MagicBrush \cite{Zhang2023MagicBrush} & 0.02 & 0.08 & 0.14 & 0.31 & 0.71 & 0.80 & 0.65 & 0.21 & 0.13 & 103.55 & 266.99 & 295.81 & 18.84 & 11.35 & 8.75 \\
        {\hspace{1em}+ REED} & \bfseries 0.01 & \bfseries 0.03 & \bfseries 0.05 & \bfseries 0.19 & \bfseries 0.51 & \bfseries 0.69 & \bfseries 0.81 & \bfseries 0.60 & \bfseries 0.45 & \bfseries 74.70 & \bfseries 174.69 & \bfseries 223.75 & \bfseries 21.53 & \bfseries 16.66 & \bfseries 14.09 \\
        \midrule
        DiffEdit \cite{couairon2022diffedit} & \bfseries 0.03 & 0.06 & 0.08 & 0.34 & 0.62 & 0.73 & 0.65 & 0.36 & 0.21 & 160.73 & 246.33 & 301.91 & 15.99 & 12.59 & 11.21 \\
        {\hspace{1em}+ REED} & \bfseries 0.03 & \bfseries \bfseries 0.07 & 0.08 & \bfseries 0.30 & \bfseries 0.55 & \bfseries 0.68 & \bfseries 0.69 & \bfseries 0.48 & \bfseries 0.40 & \bfseries 160.28 & \bfseries 226.91 & \bfseries 246.52 & \bfseries 16.24 & \bfseries 12.91 & \bfseries 11.44 \\
        
        \midrule
        
        PbE \cite{yang2023paint} & 0.02 & 0.04 & 0.07 & 0.26 & 0.60 & 0.71 & 0.66 & 0.33 & 0.22 & 83.49 & 209.29 & 253.57 & 18.55 & 13.88 & 11.74 \\
        {\hspace{1em}+ REED} & \bfseries 0.02 & \bfseries 0.03 & \bfseries 0.04 & \bfseries 0.20 & \bfseries 0.44 & \bfseries 0.59 & \bfseries 0.77 & \bfseries 0.61 & \bfseries 0.54 & \bfseries 74.24 & \bfseries 141.09 & \bfseries 178.29 & \bfseries 19.62 & \bfseries 16.19 & \bfseries 14.43 \\
        \midrule
        
        SD Inpainting \cite{rombach2022high} & 0.01 & 0.06 & 0.11 & 0.29 & 0.69 & 0.78 & 0.67 & 0.22 & 0.14 & 95.09 & 255.78 & 283.36 & 20.46 & 12.31 & 9.72 \\
        {\hspace{1em}+ REED} & \bfseries 0.01 & \bfseries 0.03 & \bfseries 0.05 & \bfseries 0.17 & \bfseries 0.47 & \bfseries 0.65 & \bfseries 0.80 & \bfseries 0.57 & \bfseries 0.41 & \bfseries 72.73 & \bfseries 166.06 & \bfseries 210.42 & \bfseries 23.14 & \bfseries 16.66 & \bfseries 13.61 \\
        \bottomrule
        
    \end{tabular}
    \end{adjustbox}
    \label{table:exp_results}
\end{table*}

\subsection{Ablations}
\label{ablations}
First, we validate the contribution of our main components by measuring their performance in the iterative encode-decode task (for now, without editing).
We compute metrics for $5,15$ and $25$ iterative encode-decode cycles and provide the results in Table \ref{table:ablations}.
Our full REED-VAE model, including all of our main components, is able to best maintain the image quality over many iterative encode-decode operations.
Please see the supplementary material for a visual example of the improvement provided by each component.
Both the quantitative and qualitative results show a clear improvement from simply introducing iterative training (IT).
Here, the iterative training is static, i.e., $k$ does not change during training.
When trained with only $(k=2)$, the model performs marginally better than the Vanilla VAE model --- increasing to $(k=5)$, the results are more significant.
The first-step loss (FSL) further improves the model performance.
It is worth noting that when evaluated at only 5 iterations, the IT + FSL $(k=5)$ model exhibits marginally better performance in MSE, FID and LPIPS; this makes sense as the model is optimized specifically for this task (5 encode/decode iterations).
However, when evaluated after 15 or 25 encode/decode iterations, it is clear in \Cref{table:ablations} that the dynamic incrementation (DI) component yields a significant improvement in all metrics.
\newcommand{\bftab}{\fontseries{b}\selectfont}
\sisetup{detect-all=true}
\begin{table*}[htpb]
\centering
    \caption{Ablation on individual components. Static Iterative training (IT) improves metrics more significantly when $k$ is increased from $2$ to $5$. First-step loss (FSL) further improves the model performance. The final component, dynamic incrementation (DI), imparts a greater  improvements as the number of editing iterations is increased. Overall, the full REED-VAE model most effectively mitigates quality degradation and maintains image fidelity and realism even after numerous edits. MSE, PSNR, and LPIPS are computed with each image sample normalized to the [0,1] range prior to evaluation, ensuring consistency in comparison. Other metrics are computed as per their standard definitions. Metrics are computed based on experiments done on the ImagenHub dataset, consisting of 179 images.}

    \begin{adjustbox}{max width=\textwidth}
    \sisetup{detect-all=true,round-mode=places,round-precision=2}
    \begin{tabular}{l
        S[round-precision=4]
        S[round-precision=4]
        S[round-precision=3]
        SSSSSSSSSSSSSSS}

        \toprule
        \multirow{2}{*}{Model} & \multicolumn{3}{c}{MSE $\downarrow$} & \multicolumn{3}{c}{LPIPS $\downarrow$} & \multicolumn{3}{c}{SSIM $\uparrow$} & \multicolumn{3}{c}{FID $\downarrow$} & \multicolumn{3}{c}{PSNR $\uparrow$}\\
        \cmidrule(lr){2-4} \cmidrule(lr){5-7} \cmidrule(lr){8-10} \cmidrule(lr){11-13} \cmidrule(lr){14-16}
         & {5} & {15} & {25} & {5} & {15} & {25} & {5} & {15} & {25} & {5} & {15} & {25} & {5} & {15} & {25} \\
        \midrule

        Vanilla-VAE & 0.0031 & 0.0126 & 0.0337 & 0.1757 & 0.5384 & 0.7007 & 0.7654 & 0.4867 & 0.2677 & 2.5401 & 47.6765 & 137.0407 & 26.0885 & 19.3184 & 14.8407 \\
        \midrule

        IT $(k=2)$  & 0.0024 & 0.0069 & 0.0141 & 0.1633 & 0.3507 & 0.4698 & 0.7805 & 0.6187 & 0.4694 & 5.5679 & 17.6085 & 38.4442 & 27.1475 & 21.8887 & 18.6563 \\
        \midrule

        IT $(k=5)$ & 0.0023 & 0.0059 & 0.0112 & 0.1150 & 0.2418 & 0.3385 & 0.7976 & 0.6726 & 0.5531 & 4.9127 & 7.6003 & 13.9402 & 27.2779 & 22.7086 & 19.7744 \\
        \midrule

        IT + FSL $(k=5)$ & 0.0021 & 0.0064 & 0.0120 & 0.1311 & 0.2332 & 0.3507 &  0.7908 & \bfseries 0.6883 & 0.5881 & 3.8004 & 7.3692 & 11.0774 & 27.0206 & 22.3944 & 19.4942\\
        \midrule

        IT + FSL + DI $(k=5)$ & \bfseries 0.0019 & \bfseries \bfseries 0.0055 & \bfseries 0.0103 & \bfseries 0.1059 & \bfseries 0.2141 & \bfseries 0.2754 & \bfseries 0.8077 & \bfseries 0.69 & \bfseries 0.62 & \bfseries 1.7788 & \bfseries 3.4029 & \bfseries 7.8124 & \bfseries 28.8581 & \bfseries 22.7876 & \bfseries 20.1629 \\
        \bottomrule
        
    \end{tabular}
    \end{adjustbox}
    \label{table:ablations}
\end{table*}

\subsection{Text-guided image editing}
To evaluate REED-VAE on text-guided image editing, we consider InstructPix2Pix \cite{brooks2023instructpix2pix}, DiffEdit \cite{couairon2022diffedit}, and MagicBrush \cite{Zhang2023MagicBrush}.
DiffEdit takes an input of $\{\mathbf{x}_s,\mathbf{C}_s, \mathbf{C}_t\}$; we treat this as an iterative task by repeatedly editing $\mathbf{x}_s$ in alternating directions of either $\mathbf{C}_t$ or $\mathbf{C}_s$.
This is straight-forward as both $\mathbf{C}_s$ and $\mathbf{C}_t$ are provided in the ImagenHub dataset.
On the other hand, InstructPix2Pix and MagicBrush take an input of the source image and an instruction prompt $\{\mathbf{x}_s, \mathbf{I}_s\}$.
Here, in order to evaluate REED-VAE, we manually add ``reverse prompts'' to perform each given edit in the opposite direction.
For some examples of reverse edit prompts, please see the supplementary material.
We will make our full, revised dataset available with our code.
Across all evaluated text-editing models, the integration of REED-VAE demonstrates consistent improvements in image quality and editing stability over multiple iterations, as shown in~\Cref{table:exp_results}.
These improvements are particularly pronounced in perceptual quality metrics (LPIPS and FID) and become more significant as the number of editing iterations increases.

\subsection{Mask-guided image editing}
We evaluate our method on the mask-guided image editing model Stable Diffusion (SD) Inpaint \cite{rombach2022high}. 
For this evaluation, we adopt an iterative editing procedure similar to our approach for text-guided editing.
Specifically, we repeatedly edit images back and forth between different target states across multiple iterations.
Given $\{\mathbf{x}_s, \mathbf{C}_{t}^{local}, \mathbf{m}\}$, the SD Inpaint model generates an output image that aims to depict the target object $\mathbf{C}_{t}^{local}$ within the masked region of $\mathbf{x}_s$.
To simplify the evaluation, we repeat the same inpainting task across all iterations, inpainting the same object $\mathbf{C}_{t}^{local}$ into the masked region at each step.
Our results show that the SD inpaint model achieves clear improvements across all metrics, with particularly strong enhancements in PSNR and FID scores as the number of iterations increases (\Cref{table:exp_results}).

\subsection{Exemplar-driven image editing}
We use REED-VAE with Paint by Example (PbE) \cite{yang2023paint} to evaluate our method on the task of subject-driven image editing.
PbE takes as input $\{\mathbf{x}_s, \mathbf{C}_t, \mathbf{m}\}$, where $\mathbf{x}_r^1$ is an additional reference image, representing the object to be depicted in the masked region of $\mathbf{x}_s$.
To transform this into an iterative task, we apply the initial $\mathbf{m}$ to $\mathbf{x}_s$ and generate a tight bounding box around the mask.
We use this generated bounding box to create $\mathbf{x}_r^2$, a new reference image containing the original object from $\mathbf{x}_s$ that we replaced in the previous edit operation.
Now, we can alternate iteratively in a similar fashion: at each iterative step we provide $\{\mathbf{x}_s$, $\mathbf{m}\}$, and one of either $\mathbf{x}_r^1$ or $\mathbf{x}_r^2$.
As demonstrated in \Cref{table:exp_results}, enhancing PbE with REED-VAE improved all metrics during this iterative editing procedure.
Most notably, at 15 and 25 iterations the LPIPS metric was reduced by 50\% or more when REED was used instead of the Vanilla VAE, suggesting a substantial improvement in perceptual similarity to the target images.

\subsection{Comparison with Inversion-Based Methods}
Inversion-based methods, such as DDIM-inversion \cite{song2020denoising, dhariwal2021diffusion} and Null-Text Inversion (NTI) \cite{mokady2023null}, are widely used in diffusion-based editing methods.
Inversion attempts to find the initial noise vector that will produce the input image when fed into the diffusion model along with the original image prompt.
Doing this accurately is crucial for editing real images with methods that function by manipulating the latent vectors throughout the denoising process, such as Prompt-to-Prompt \cite{hertz2022prompt} and DiffEdit \cite{couairon2022diffedit}.
By regenerating the latent through the denoising process, inversion methods provide an alternative pathway to the image latent, as opposed to directly using the VAE encoder before applying edits (as done in other methods \cite{brooks2023instructpix2pix, Zhang2023MagicBrush, yang2023paint, rombach2022high}.
This raises the question of whether inversion-based editing methods can inherently mitigate the degradation observed in iterative editing tasks.

To explore this, we perform iterative inversion using NTI and compare its performance to NTI combined with REED-VAE, as well as to iterative encoding/decoding with the Vanilla-VAE (all without applying edits).
For iterative NTI, we first invert the image using NTI with a source prompt and then regenerate the image using the diffusion model with the same source prompt (thus, no edits are performed, the process regenerates the input image).
The results, shown in \Cref{fig:iterative_ddim}, reveal that NTI in fact accumulates considerable noise and artifacts over iterations.
As well, the initial reconstructions with NTI are not perfect, while NTI$+$REED achieves a much more faithful reconstruction at iteration 5.
At about iteration 10, Vanilla NTI significantly loses fidelity to the original image, resulting in heavy distortions.
We refer to the supplementary material for figures demonstrating the full iterative process which provides more insights into this phenomenon.
It is important to note that NTI (as well as regular DDIM-inversion) relies on the VAE to encode the image as the starting point for inversion.
The image is then generated by the diffusion model using the inverted latent as the starting point instead of random noise.
The VAE decoder is then used as usual to bring the generated image back to pixel space.
This highlights that the VAE remains integral to the editing process even when inversion is employed.
Furthermore, inversion methods themselves introduce unique noise and challenges that contribute to iterative degradation, explaining why they are not inherently more resilient than the Vanilla-VAE in such scenarios.

To further explore the degradation patterns associated with DDIM-inversion, we also replicate the iterative text-guided image editing task described previously, using NTI \cite{mokady2023null} and P2P editing \cite{hertz2022prompt}.
As detailed in the Supplementary Material, DDIM-inversion-based editing methods also suffer from iterative degradation, similar to methods that bypass inversion entirely.
While inversion methods are valuable components of many diffusion-based editing pipelines, they do not inherently solve the challenges posed by multi-method iterative image editing, as described in this paper.
Conversely, REED-VAE directly mitigates these issues, enabling high-fidelity editing across both pixel and latent spaces without requiring rigid workflows or sacrificing flexibility.
This ensures compatibility with a broader range of editing techniques, supporting creative and iterative editing scenarios.

\section{Conclusion}
We introduce a novel problem setting of \emph{comprehensive iterative image editing} where a user can perform iterative edit operations on a real image, each time using the previous output to perform another edit operation using the same or a different model or conventional editing techniques.
We solve the problem of accumulating artifacts with our REED-VAE, which implements a novel iterative training algorithm to enhance the VAE's ability to reconstruct images faithfully over many iterations.
We demonstrate the ease of using REED-VAE with any Stable Diffusion-based editing model in place of the vanilla VAE, and its marked effectiveness in improving image quality retention over iterations.
We make REED-VAE publicly available, and hope that it will serve as a valuable contribution to the community.

\paragraph*{Limitations and future work}
There are, nevertheless, limitations to our method in its current form.
As with any VAE, the reconstruction remains imperfect, and after a very large number of iterations (30+),  the REED-VAE will also begin to deteriorate. 
In the future, we would like to explore how leveraging REED-VAE to generate synthetic training data for diffusion models can help alleviate the model collapse problem, which is prevalent in iterative generative processes \cite{yoon2024model}.
By using REED-VAE to better align the diffusion latent space, we anticipate improved stability and performance in downstream editing tasks.
As well, applying the REED training algorithm to improve the VAEs of newer latent diffusion models (e.g., SDXL \cite{podell2023sdxl}, SD3 \cite{esser2024scaling}, and Flux \cite{flux2023}, as discussed in the Supplementary Material) will provide additional insights into the robustness and generalizability of our approach.
These models employ advanced architectures and improved, 16-channel latent spaces, which may further benefit from REED-VAE's iterative stability.

\paragraph*{Ethical considerations}
We acknowledge that all diffusion-based image editing techniques inevitably raise ethical concerns, and may reflect biases inherent in the training data used for the underlying model. 
It follows that the method presented in this paper, which allows more of such edits to be performed on a single image, can amplify these concerns. When implementing and using such models and techniques, it is crucial to establish proper safeguards, particularly concerning permissible prompts and guidance, to prevent malicious use and ensure compliance with legal standards. In preliminary tests, we observe that our model maintains watermarks aiming to detect synthetic images \cite{wen2024tree}. 
We are also actively researching methods for detecting synthetic images and videos \cite{sinitsa2024deep, knafo2022fakeout, 9151013}

\paragraph*{Acknowledgments}
We thank Almog Friedlander for the thoughtful suggestions and discussions that helped improve our research.
This work was supported by the Israel Science Foundation (Ggrant No. 1574/21) and by the Joint NSFC-ISF Research Grant (no. 3077/23).
\clearpage

{
    \small
    \bibliographystyle{eg-alpha-doi}
    \bibliography{references}
}
\clearpage
\appendix
\twocolumn[
    \centering
    \Large
    \textbf{REED-VAE: RE-Encode Decode Training for Iterative Image Editing with Diffusion Models} \\
    Supplementary Material \\
    \vspace{1.0em}
]
\setcounter{table}{0}
\renewcommand{\thetable}{A\arabic{table}}
\section{Implementation details}
\subsection{Training}
All experiments conducted are based on the released \verb|v2.1| of Stable Diffusion \cite{rombach2022high} along with the default VAE using a single NVIDIA A100 GPU card.
We finetune the VAE using the same Diffusers \cite{von-platen-etal-2022-diffusers} implementation named \verb|"AutoencoderKL"|.
In accordance with the training setting used for both the original VAE and the original diffusion model, we finetune our REED-VAE on a subset of the LAION-5B dataset \cite{schuhmann2022laion}.
During training, we preprocess the image resolution to $512 \times 512$ and train for 35 epochs, which took approximately 1 day.
Scaling parameters $\alpha$ and $\beta$ were used to scale the LPIPS and $\KL$ terms of the training loss, respectively. 
To train the final model, we set $\alpha=0.01$ and $\beta=1$.

\subsection{Backpropagation strategy}
Initially, we backpropagated through all iterations, calculating gradients for each intermediate step.
Although this approach achieved maximal learning at each iteration, it was very memory-intensive and limited the maximum number of iterations ($k$) to approximately 7 on our A100 GPU, even with gradient checkpointing.
To address this, we optimized memory usage by computing gradients only for the final iteration, significantly reducing computational overhead and enabling training with larger $k$ values.
This adjustment leverages the dynamic incrementation in our loss design, which encourages the model to learn progressively from intermediate iterations without requiring full-gradient computation at each step.

\subsection{Experiments and comparisons}
When calculating metrics, we try to isolate errors and noise that occur due to the iterative autoencoding process from those that occur due to imperfect performance by the editing model.
To do this, we compute metrics between the given target image (one of iterations 5,15, or 25) and $x^1$ --- \emph{not} to the source image.
This guarantees that (1) we always compare images from aligned edit operations (i.e., edits in the same direction, such as changing the bus into a car and not the other way around) and (2) metrics are more dependent on the model's ability to maintain image quality over iterations than the model's general editing capabilities.
In other words, if a model performs a non-sensical edit operation from the given inputs, as long as it is consistent (which it should be if it does not degrade images), then this alone should not harm its performance in our experiments.

\subsection{Metric calculations}
For all reported metrics, MSE, PSNR, and LPIPS are computed with each image sample normalized to the [0,1] range prior to evaluation, ensuring consistency in comparison. 
Other metrics are computed as per their standard definitions.
FID is calculated using the pytorch-fid implementation \cite{Seitzer2020FID}.

\subsection{Iterative prompt list used for iterative text-guided image editing}
We provide some examples of iterative prompts used in the iterative text-guided image editing task in Table. \ref{tab_appendix_1}. 
The full enhanced dataset will be made public along with our code.
\begin{table}[htbp]
\centering
\small
\caption{Example of edit prompts and corresponding reverse edit prompts used iteratively to evaluate InstructPix2Pix \cite{brooks2023instructpix2pix} and DiffEdit \cite{couairon2022diffedit}}

\begin{tabular}{c p{3.5cm} p{3.5cm}}
  \toprule
  & \textbf{Prompt} & \textbf{Reverse Prompt} \\
  \midrule
  1 & Change the frisbee into a ball & Change the ball into a frisbee \\
  2 & Put a lion in the place of the donkey & Put a donkey in the place of the lion \\
  3 & Add a pedestrian & Remove the pedestrian \\
  4 & Make it a black sheep & Make it a white sheep \\
  5 & Replace the coffee with beer & Replace the beer with a coffee \\
    \bottomrule
    \end{tabular}

\label{tab_appendix_1}

\end{table}

\subsection{Varying metric scales across editing methods}
In our main editing experiments, the scale of improvements observed with REED may vary due to editing methods differing in conditioning and inputs (text/mask/image), scope (local/global edits), and VAE use. 
For instance, DiffEdit \cite{couairon2022diffedit} automatically generates latent masks from text prompts, introducing ambiguity regarding the edit location especially as noise increases in higher iterations. 
Such ambiguity can result in edits being applied to different regions of the image, potentially inflating computed metrics - this will be noticeable even when REED is used, as metrics are evaluated against the first edit iteration.
In contrast, PbE \cite{yang2023paint} employs predefined masks, ensuring that edits remain localized and consistent regardless of accumulated noise. 
This provides a more controlled editing scenario, reducing variability in computed metrics.

\section{Additional Experiments}

\sisetup{detect-all=true}
\begin{table*}[htpb]
\centering
    \caption{Comparison of performance metrics for several state-of-the-art latent diffusion models and Stable Diffusion 2.1 with/without our REED-VAE. The metrics are calculated on the ImagenHub dataset \cite{ku2024imagenhub} (179 images) and are reported for various iteration steps (5,15,25) on an iterative encode/decode task (without editing). Despite the more advanced latent spaces in models such as SD3 and Flux (with 16 channels), these newer models still exhibit the problem of iterative degradation.}

    \begin{adjustbox}{max width=\textwidth}
    \sisetup{detect-all=true,round-mode=figures,round-precision=2}
    \begin{tabular}{lSSSSSSSSS
        S[round-precision=3]
        S[round-precision=4]
        S[round-precision=4]
        S[round-precision=3]
        S[round-precision=4]
        S[round-precision=4]}
        \toprule
        \multirow{2}{*}{Model} & \multicolumn{3}{c}{MSE $\downarrow$} & \multicolumn{3}{c}{LPIPS $\downarrow$} & \multicolumn{3}{c}{SSIM $\uparrow$} & \multicolumn{3}{c}{FID $\downarrow$} & \multicolumn{3}{c}{PSNR $\uparrow$} \\
        \cmidrule(lr){2-4} \cmidrule(lr){5-7} \cmidrule(lr){8-10} \cmidrule(lr){11-13} \cmidrule(lr){14-16}
         & {5} & {15} & {25} & {5} & {15} & {25} & {5} & {15} & {25} & {5} & {15} & {25} & {5} & {15} & {25}\\
        \midrule
        SD3 \cite{esser2024scaling} & 0.0025 & 0.0131 & 0.031 & 0.11 & 0.55 & 0.76 & 0.83 & 0.51 & 0.26 & 2.53 & 24.67 & 68.79 & 26.72 & 19.07 & 15.19 \\
        \midrule
        SDXL \cite{podell2023sdxl} & 0.0026 & 0.0067 & 0.0126 & 0.1895 & 0.4465 & 0.6067 & 0.7805 & 0.6359 & 0.5171 & 7.3321 & 18.2699 & 29.7116 & 26.9252 & 22.1197 & 19.2041 \\
        \midrule
        Flux.1 \cite{flux2023} & 0.0014 & 0.0075 & 0.0172 & \bfseries 0.0639 & 0.2716 & 0.5553 & \bfseries 0.8961 & 0.7431 & 0.5319 & \bfseries 1.0200 & 9.7245 & 28.6170 & 28.8850 & 21.5055 & 17.8769 \\
        \midrule
        Vanilla SD 2.1 \cite{rombach2022high} & 0.0031 & 0.013 & 0.034 & 0.19 & 0.55 & 0.71 & 0.76 & 0.49 & 0.26 & 2.3536 & 45.9486 & 133.8590 & 25.9875 & 19.3048 & 14.8338 \\
        {\hspace{1em}+ REED} & \bfseries 0.0011 & \bfseries 0.0042 & \bfseries 0.0086 & 0.0752 & \bfseries 0.1808 & \bfseries 0.2544 & 0.8852 & \bfseries 0.7617 & \bfseries 0.6754 & 1.0811 & \bfseries 2.7903 & \bfseries 3.8730 & \bfseries 30.4624 & \bfseries 24.1040 & \bfseries 20.9326 \\ 

        \bottomrule
        
    \end{tabular}
    \end{adjustbox}
    \label{table:new_ldms}
\end{table*}

\subsection{Comparison with newer latent diffusion models}
\begin{figure*}[htpb]
    \centering
    \setlength{\tabcolsep}{-2pt}
    \renewcommand{\arraystretch}{0.5}
    \setlength{\ww}{0.18\textwidth}

    \begin{tabular}{p{0.4cm} c c c c c}
        \rotatebox{90}{\phantom{WWWWWW.}\scriptsize{Input} } &
        \begin{tikzpicture}[spy using outlines={circle, thick, red, magnification=2.5,size=0.9cm, connect spies}]
            \node {\includegraphics[width=\ww,frame]{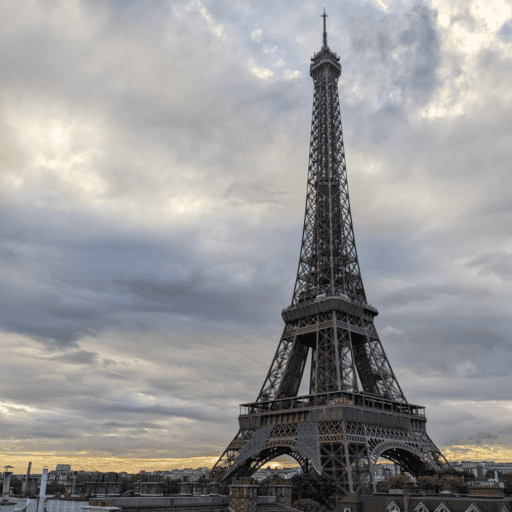}};
            \spy on (0.45,1.3) in node [left] at (1.6,0.1);
        \end{tikzpicture} &&&&
        \\[-2pt]
        \rotatebox{90}{\phantom{AAA.}\scriptsize{(1) Vanilla-VAE SD2 }} &

        \begin{tikzpicture}[spy using outlines={}]
            \node {\includegraphics[width=\ww,frame]{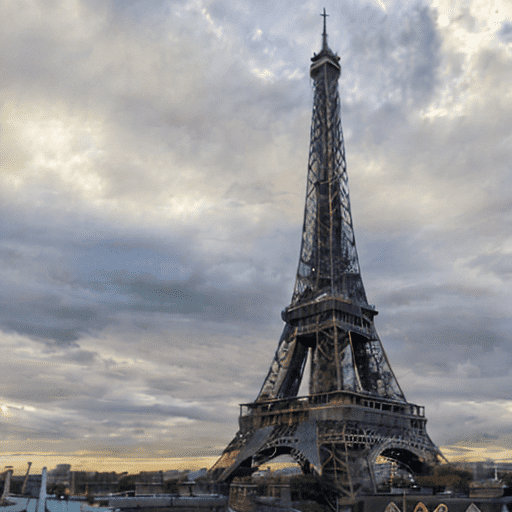}};
        \end{tikzpicture} &

        \begin{tikzpicture}[spy using outlines={}]
            \node {\includegraphics[width=\ww,frame]{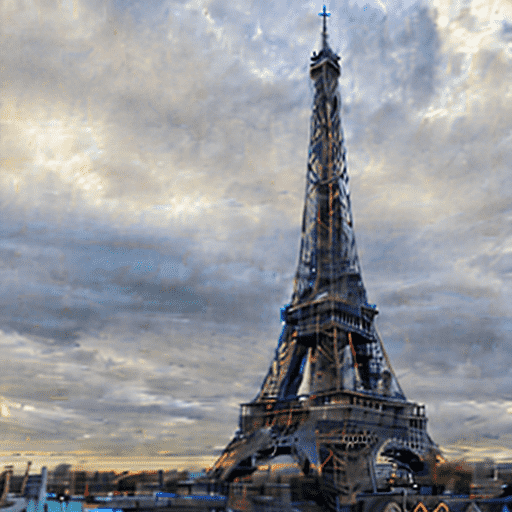}};
        \end{tikzpicture} &

        \begin{tikzpicture}[spy using outlines={}]
            \node {\includegraphics[width=\ww,frame]{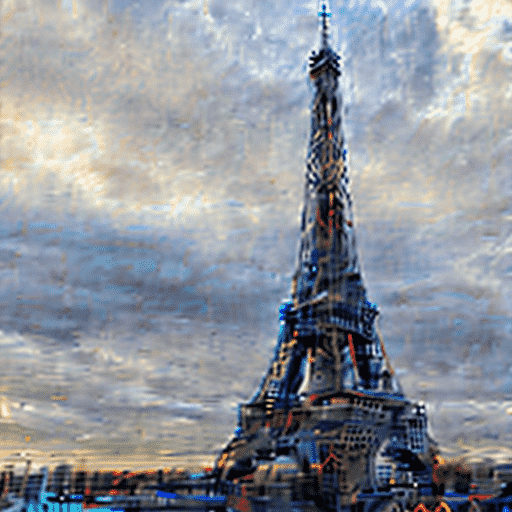}};
        \end{tikzpicture} &

        \begin{tikzpicture}[spy using outlines=]
            \node {\includegraphics[width=\ww,frame]{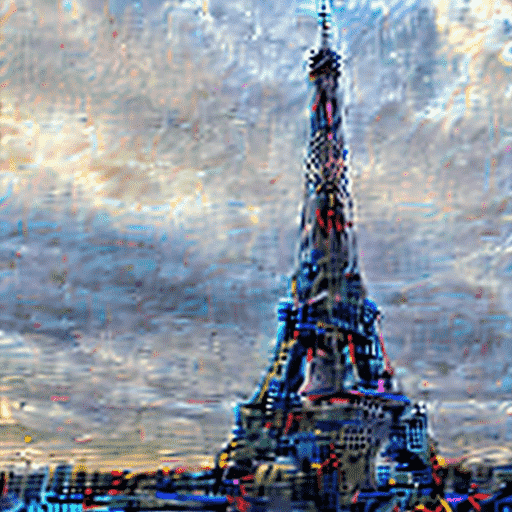}};
        \end{tikzpicture} &

        \begin{tikzpicture}[spy using outlines={circle, thick, red, magnification=2.5,size=0.9cm, connect spies}]
            \node {\includegraphics[width=\ww,frame]{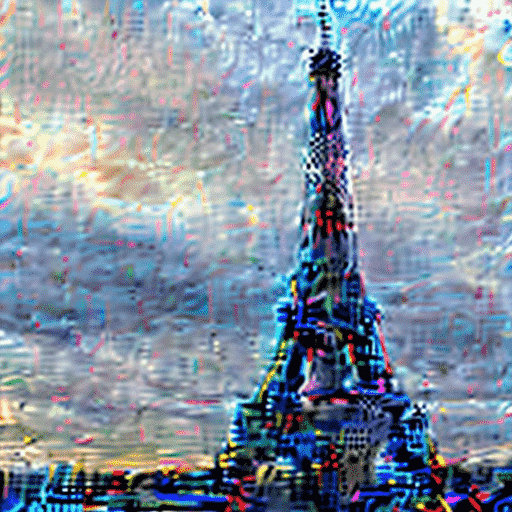}};
            \spy on (0.45,1.3) in node [left] at (1.6,0.1);
        \end{tikzpicture}
        
        \\[-5pt]

        \rotatebox{90}{\phantom{AAA.}\scriptsize{(2) Vanilla-VAE SD3 }} &

        \begin{tikzpicture}[spy using outlines={}]
            \node {\includegraphics[width=\ww,frame]{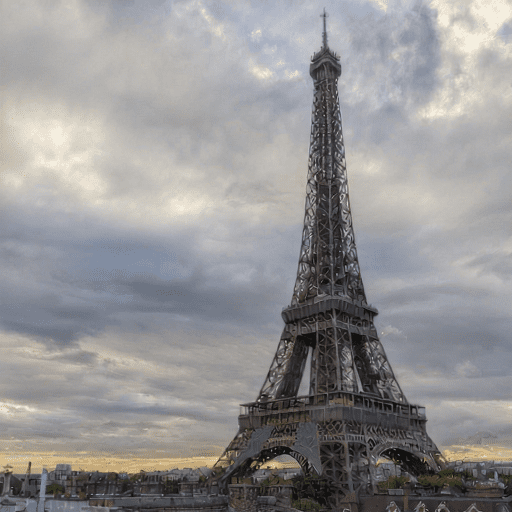}};
        \end{tikzpicture} &

        \begin{tikzpicture}[spy using outlines={}]
            \node {\includegraphics[width=\ww,frame]{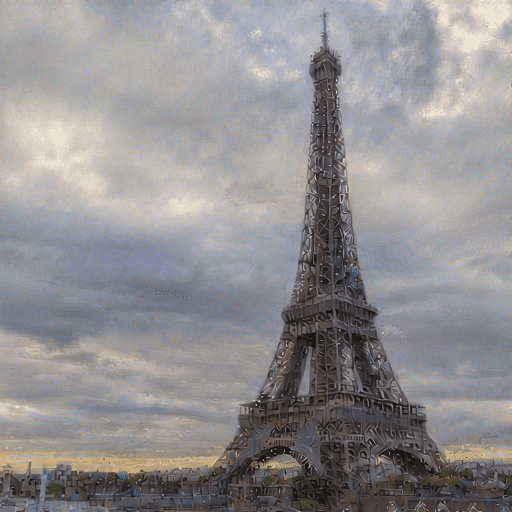}};
        \end{tikzpicture} &

        \begin{tikzpicture}[spy using outlines={}]
            \node {\includegraphics[width=\ww,frame]{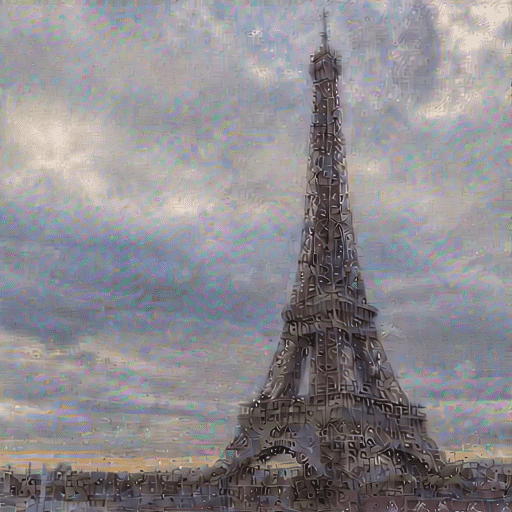}};
        \end{tikzpicture} &

        \begin{tikzpicture}[spy using outlines=]
            \node {\includegraphics[width=\ww,frame]{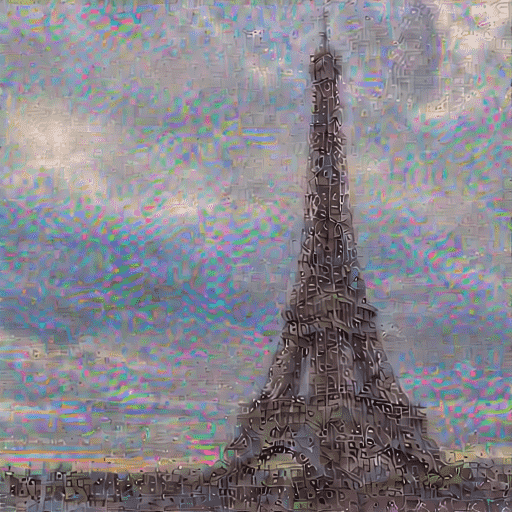}};
        \end{tikzpicture} &

        \begin{tikzpicture}[spy using outlines={circle, thick, red, magnification=2.5,size=0.9cm, connect spies}]
            \node {\includegraphics[width=\ww,frame]{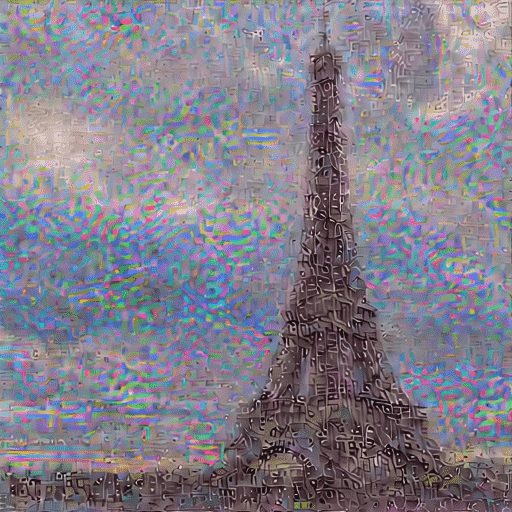}};
            \spy on (0.45,1.3) in node [left] at (1.6,0.1);
        \end{tikzpicture}
        
        \\[-5pt]

        \rotatebox{90}{\phantom{AAA.}\scriptsize{(3) Vanilla-VAE SDXL }} &

        \begin{tikzpicture}[spy using outlines={}]
            \node {\includegraphics[width=\ww,frame]{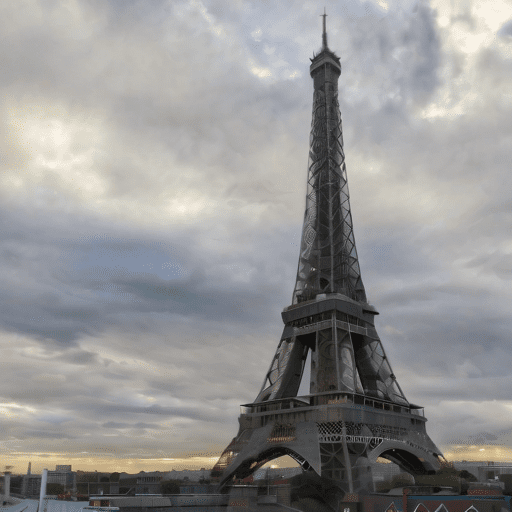}};
        \end{tikzpicture} &

        \begin{tikzpicture}[spy using outlines={}]
            \node {\includegraphics[width=\ww,frame]{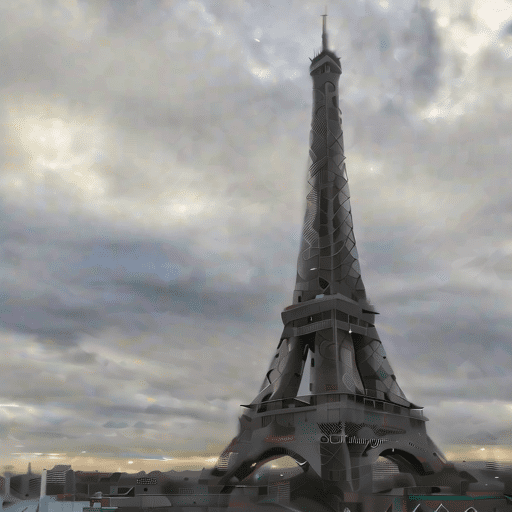}};
        \end{tikzpicture} &

        \begin{tikzpicture}[spy using outlines={}]
            \node {\includegraphics[width=\ww,frame]{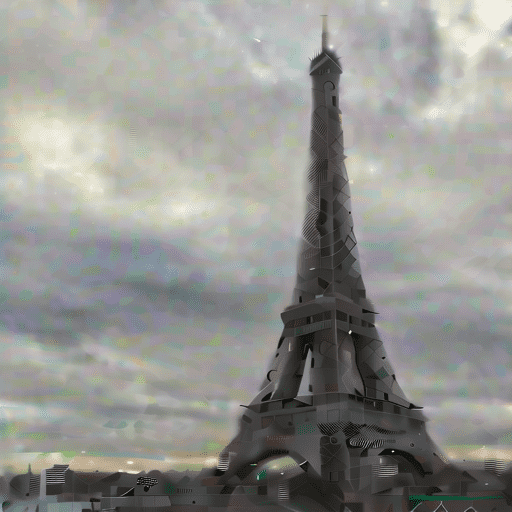}};
        \end{tikzpicture} &

        \begin{tikzpicture}[spy using outlines=]
            \node {\includegraphics[width=\ww,frame]{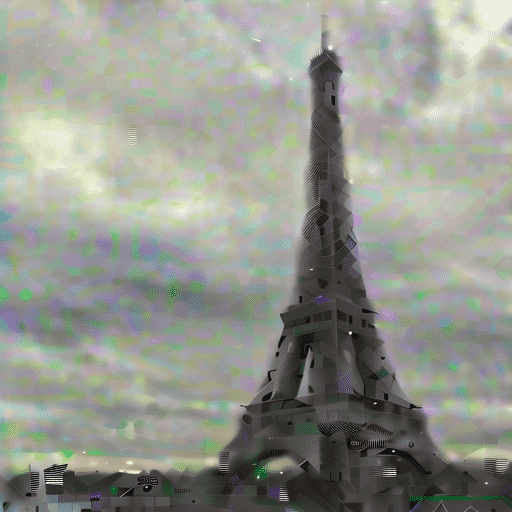}};
        \end{tikzpicture} &

        \begin{tikzpicture}[spy using outlines={circle, thick, red, magnification=2.5,size=0.9cm, connect spies}]
            \node {\includegraphics[width=\ww,frame]{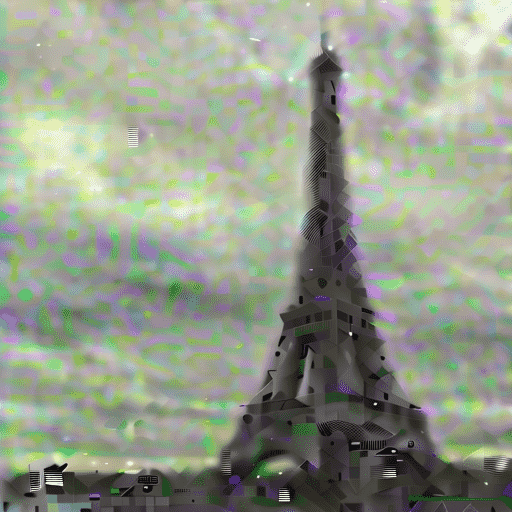}};
            \spy on (0.45,1.3) in node [left] at (1.6,0.1);
        \end{tikzpicture}

        \\[-5pt]

        \rotatebox{90}{\phantom{AAA.}\scriptsize{(4) Vanilla-VAE Flux }} &

        \begin{tikzpicture}[spy using outlines={}]
            \node {\includegraphics[width=\ww,frame]{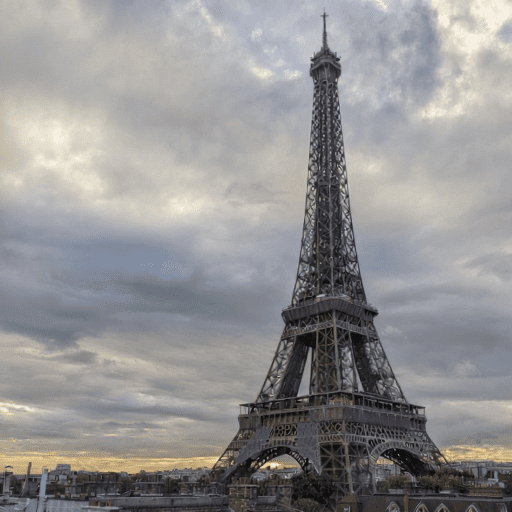}};
        \end{tikzpicture} &

        \begin{tikzpicture}[spy using outlines={}]
            \node {\includegraphics[width=\ww,frame]{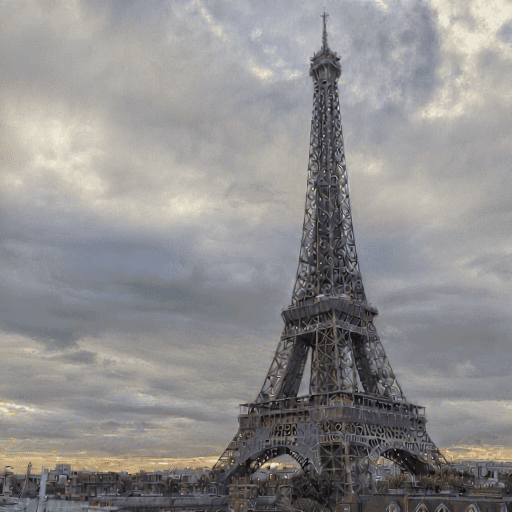}};
        \end{tikzpicture} &

        \begin{tikzpicture}[spy using outlines={}]
            \node {\includegraphics[width=\ww,frame]{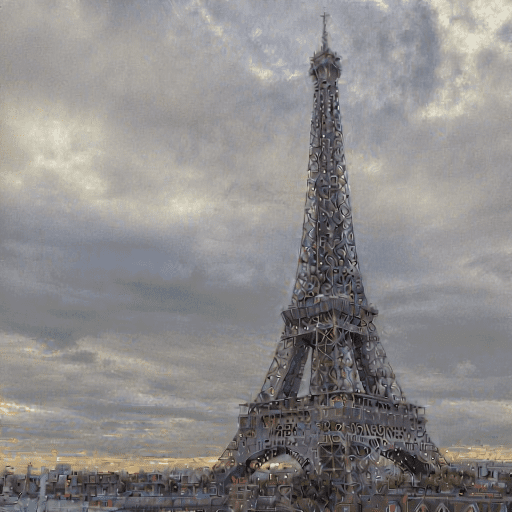}};
        \end{tikzpicture} &

        \begin{tikzpicture}[spy using outlines=]
            \node {\includegraphics[width=\ww,frame]{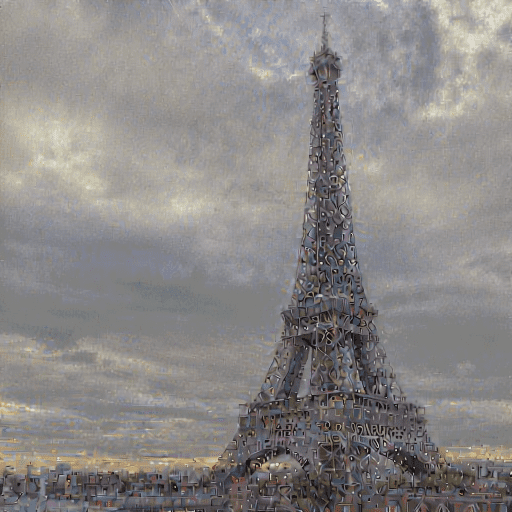}};
        \end{tikzpicture} &

        \begin{tikzpicture}[spy using outlines={circle, thick, red, magnification=2.5,size=0.9cm, connect spies}]
            \node {\includegraphics[width=\ww,frame]{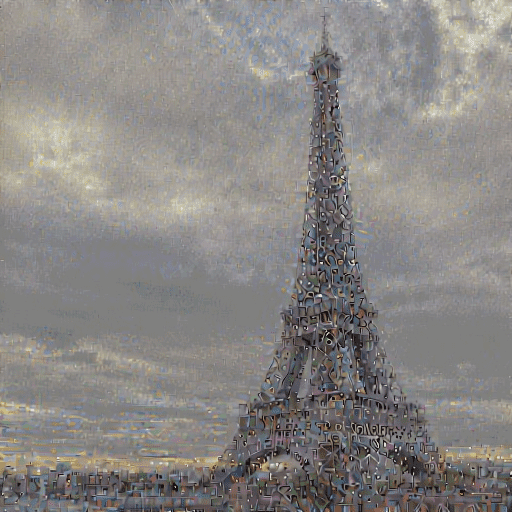}};
            \spy on (0.45,1.3) in node [left] at (1.6,0.1);
        \end{tikzpicture}
        
        \\[-5pt]

        \rotatebox{90}{\phantom{AAA.}\scriptsize{\textbf{(5) REED-VAE SD2 }}} &

        \begin{tikzpicture}[spy using outlines={}]
            \node {\includegraphics[width=\ww,frame]{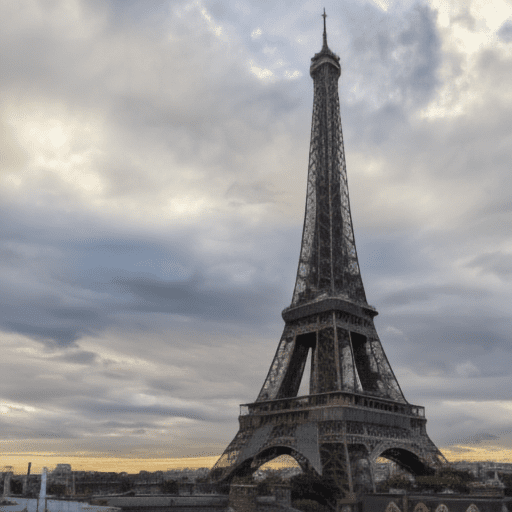}};
        \end{tikzpicture} &

        \begin{tikzpicture}[spy using outlines={}]
            \node {\includegraphics[width=\ww,frame]{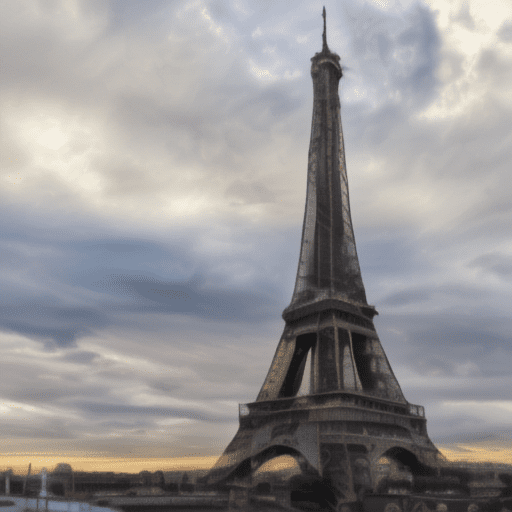}};
        \end{tikzpicture} &

        \begin{tikzpicture}[spy using outlines={}]
            \node {\includegraphics[width=\ww,frame]{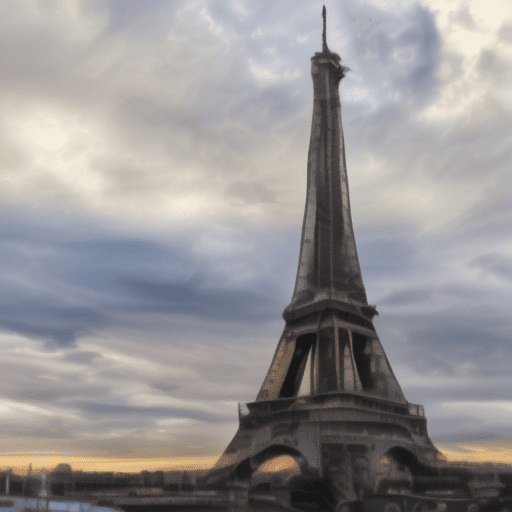}};
        \end{tikzpicture} &

        \begin{tikzpicture}[spy using outlines={}]
            \node {\includegraphics[width=\ww,frame]{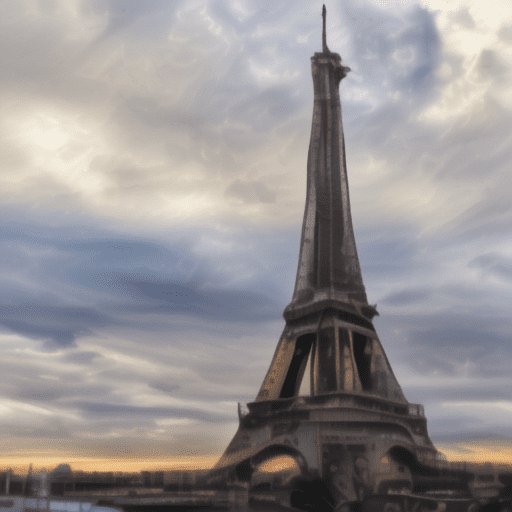}};
        \end{tikzpicture} &

        \begin{tikzpicture}[spy using outlines={circle, thick, red, magnification=2.5,size=0.9cm, connect spies}]
            \node {\includegraphics[width=\ww,frame]{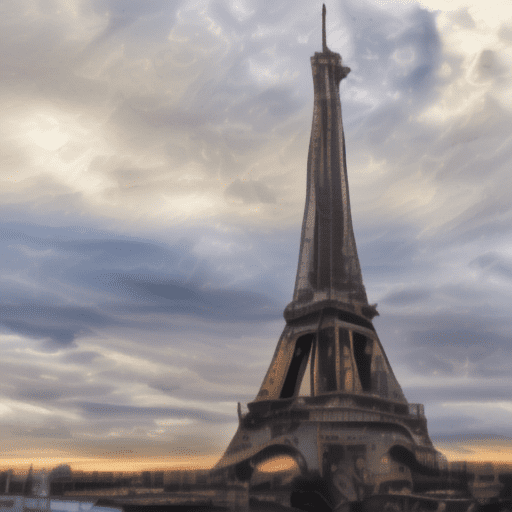}};
            \spy on (0.45,1.3) in node [left] at (1.6,0.1);
        \end{tikzpicture}
        
        \\
        
        &
        \scriptsize{5 } &
        \scriptsize{10 } &
        \scriptsize{15 } &
        \scriptsize{20 } &
        \scriptsize{25 } 

        \\
        &
        \multicolumn{5}{l}{\phantom{.}
        \begin{tikzpicture}
            \draw[->](0,0)--(16.4,0);
        \end{tikzpicture}}

        \\
        &
        \multicolumn{5}{c}{
        \scriptsize{Num. Encode/Decode Iterations}
        }
        
    \end{tabular}
    \caption{Comparison on iterative encode/decode task with more recent latent diffusion models, reported at 5,10,15,20,25 iterations. REED-VAE is able to outperform even the newest models with 16-channel latent spaces, suggesting training these new model's VAEs with the REED algorithm can improve them even further.}

    \label{fig:new_models_eiffel}
\end{figure*}

\begin{figure*}[htpb]
    \centering
    \setlength{\tabcolsep}{-2pt}
    \renewcommand{\arraystretch}{0.5}
    \setlength{\ww}{0.18\textwidth}

    \begin{tabular}{p{0.4cm} c c c c c}
        \rotatebox{90}{\phantom{WWWWWW.}\scriptsize{Input} } &
        \begin{tikzpicture}[spy using outlines={circle, thick, red, magnification=2.5,size=0.9cm, connect spies}]
            \node {\includegraphics[width=\ww,frame]{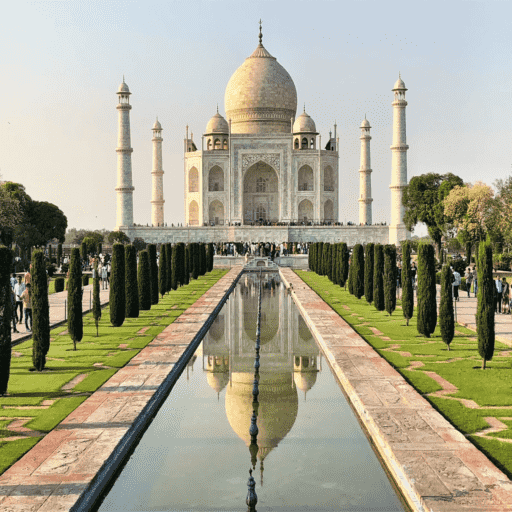}};
            \spy on (0.05,1.35) in node [left] at (1.6,0.2);
        \end{tikzpicture} &&&&
        \\[-2pt]
        \rotatebox{90}{\phantom{AAA.}\scriptsize{(1) Vanilla-VAE SD2 }} &

        \begin{tikzpicture}[spy using outlines={}]
            \node {\includegraphics[width=\ww,frame]{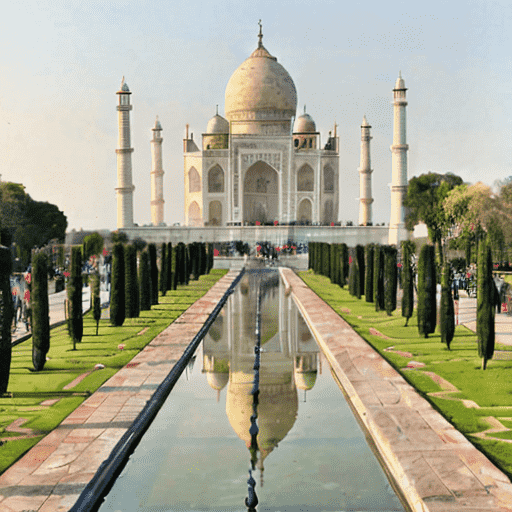}};
        \end{tikzpicture} &

        \begin{tikzpicture}[spy using outlines={}]
            \node {\includegraphics[width=\ww,frame]{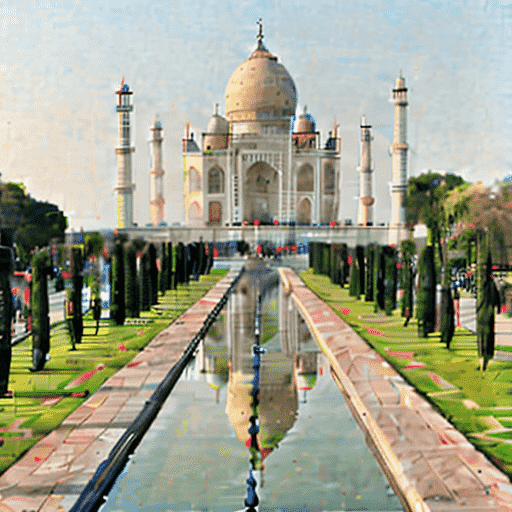}};
        \end{tikzpicture} &

        \begin{tikzpicture}[spy using outlines={}]
            \node {\includegraphics[width=\ww,frame]{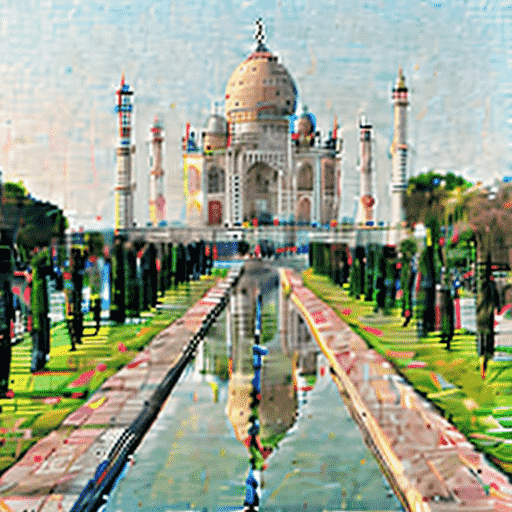}};
        \end{tikzpicture} &

        \begin{tikzpicture}[spy using outlines=]
            \node {\includegraphics[width=\ww,frame]{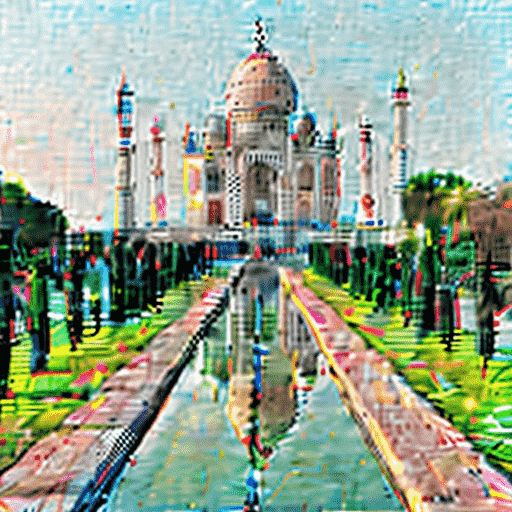}};
        \end{tikzpicture} &

        \begin{tikzpicture}[spy using outlines={circle, thick, red, magnification=2.5,size=0.9cm, connect spies}]
            \node {\includegraphics[width=\ww,frame]{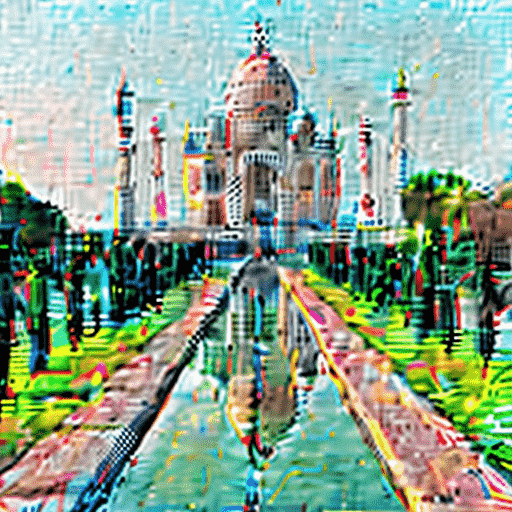}};
            \spy on (0.05,1.35) in node [left] at (1.6,0.2);
        \end{tikzpicture}
        
        \\[-5pt]

        \rotatebox{90}{\phantom{AAA.}\scriptsize{(2) Vanilla-VAE SD3 }} &

        \begin{tikzpicture}[spy using outlines={}]
            \node {\includegraphics[width=\ww,frame]{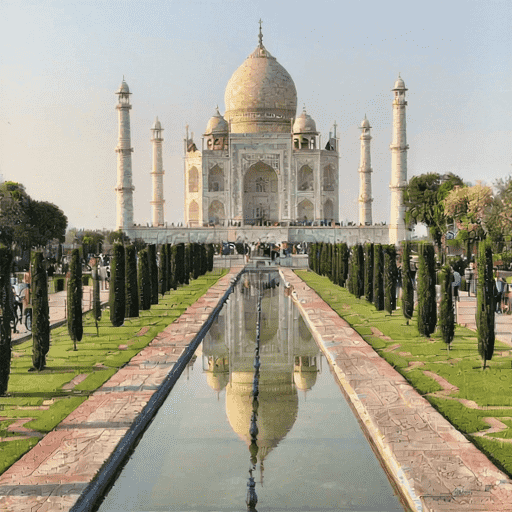}};
        \end{tikzpicture} &

        \begin{tikzpicture}[spy using outlines={}]
            \node {\includegraphics[width=\ww,frame]{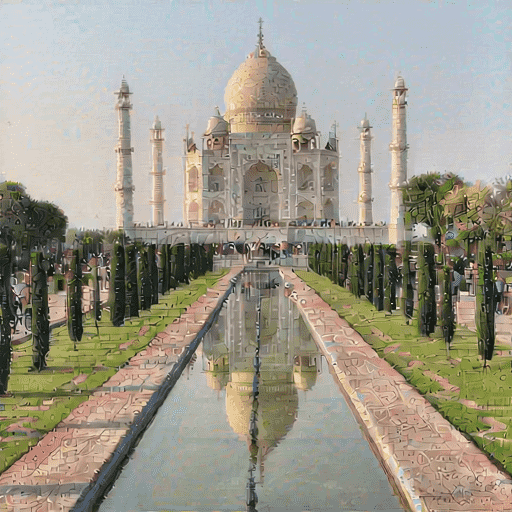}};
        \end{tikzpicture} &

        \begin{tikzpicture}[spy using outlines={}]
            \node {\includegraphics[width=\ww,frame]{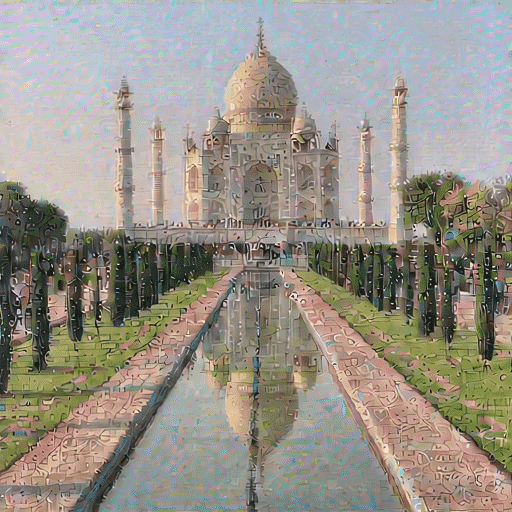}};
        \end{tikzpicture} &

        \begin{tikzpicture}[spy using outlines=]
            \node {\includegraphics[width=\ww,frame]{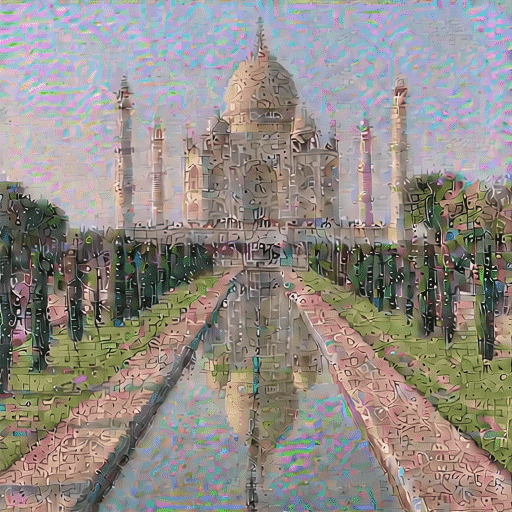}};
        \end{tikzpicture} &

        \begin{tikzpicture}[spy using outlines={circle, thick, red, magnification=2.5,size=0.9cm, connect spies}]
            \node {\includegraphics[width=\ww,frame]{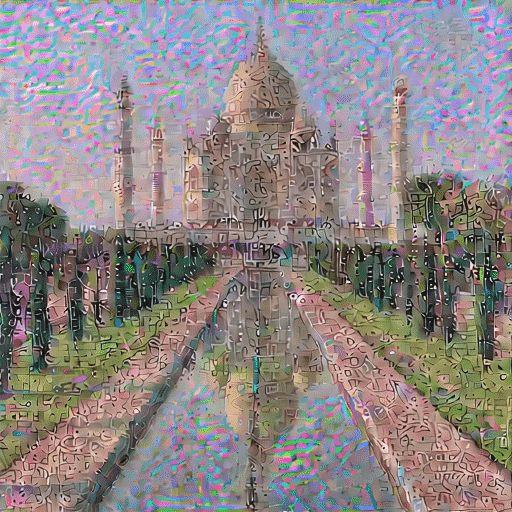}};
            \spy on (0.05,1.35) in node [left] at (1.6,0.2);
        \end{tikzpicture}
        
        \\[-5pt]

        \rotatebox{90}{\phantom{AAA.}\scriptsize{(3) Vanilla-VAE SDXL }} &

        \begin{tikzpicture}[spy using outlines={}]
            \node {\includegraphics[width=\ww,frame]{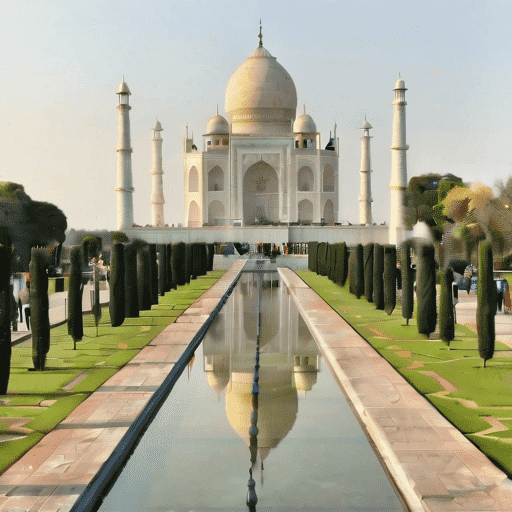}};
        \end{tikzpicture} &

        \begin{tikzpicture}[spy using outlines={}]
            \node {\includegraphics[width=\ww,frame]{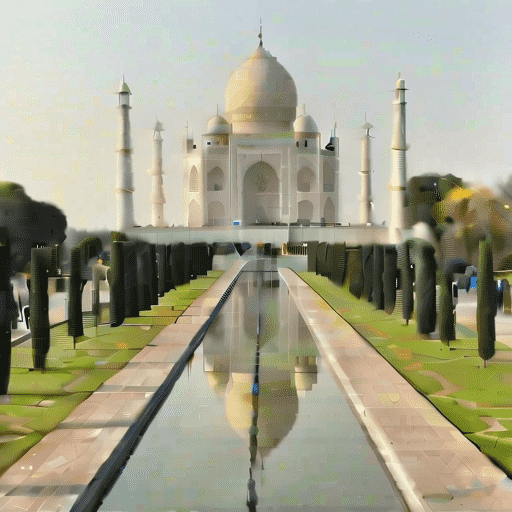}};
        \end{tikzpicture} &

        \begin{tikzpicture}[spy using outlines={}]
            \node {\includegraphics[width=\ww,frame]{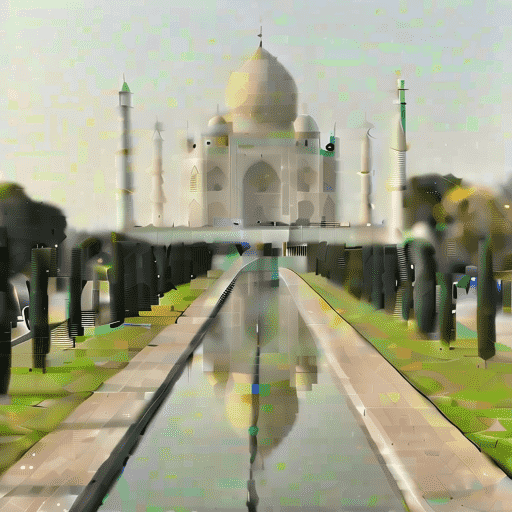}};
        \end{tikzpicture} &

        \begin{tikzpicture}[spy using outlines=]
            \node {\includegraphics[width=\ww,frame]{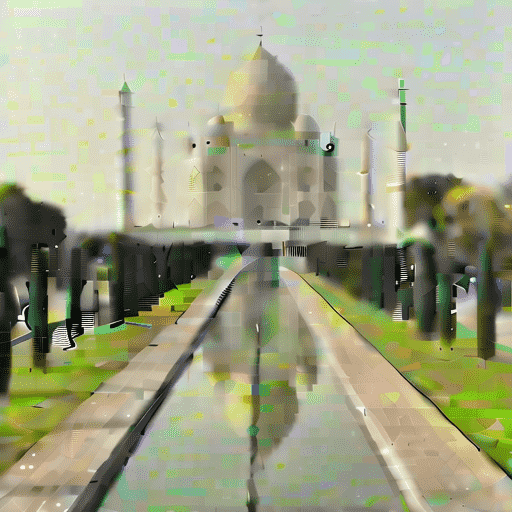}};
        \end{tikzpicture} &

        \begin{tikzpicture}[spy using outlines={circle, thick, red, magnification=2.5,size=0.9cm, connect spies}]
            \node {\includegraphics[width=\ww,frame]{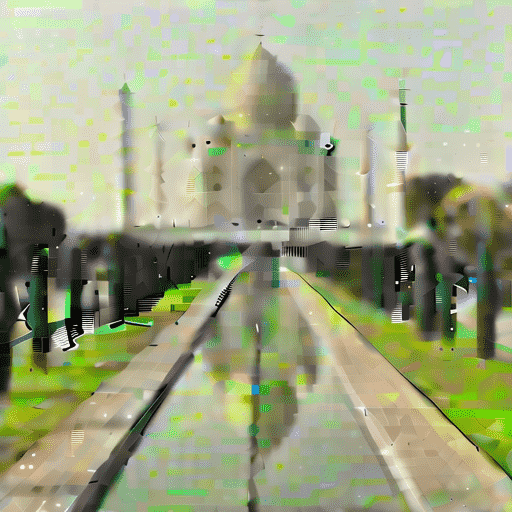}};
            \spy on (0.05,1.35) in node [left] at (1.6,0.2);
        \end{tikzpicture}

        \\[-5pt]

        \rotatebox{90}{\phantom{AAA.}\scriptsize{(4) Vanilla-VAE Flux }} &

        \begin{tikzpicture}[spy using outlines={}]
            \node {\includegraphics[width=\ww,frame]{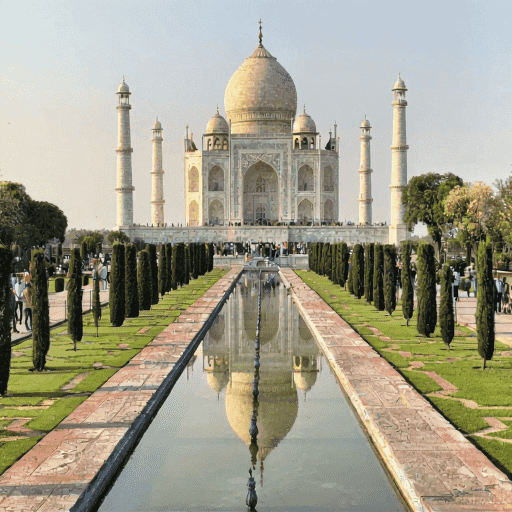}};
        \end{tikzpicture} &

        \begin{tikzpicture}[spy using outlines={}]
            \node {\includegraphics[width=\ww,frame]{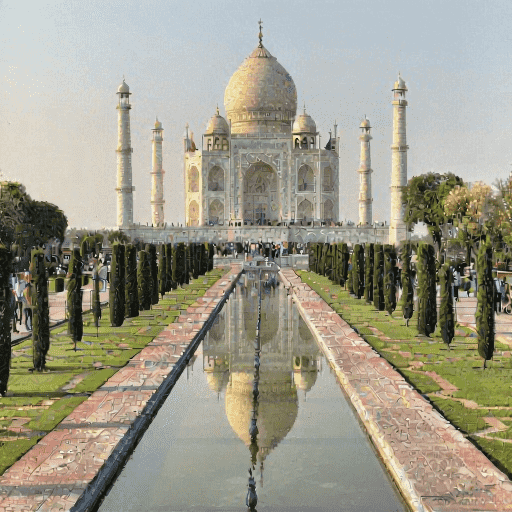}};
        \end{tikzpicture} &

        \begin{tikzpicture}[spy using outlines={}]
            \node {\includegraphics[width=\ww,frame]{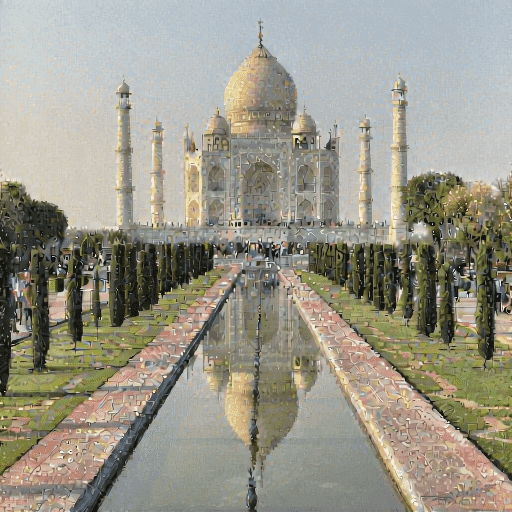}};
        \end{tikzpicture} &

        \begin{tikzpicture}[spy using outlines=]
            \node {\includegraphics[width=\ww,frame]{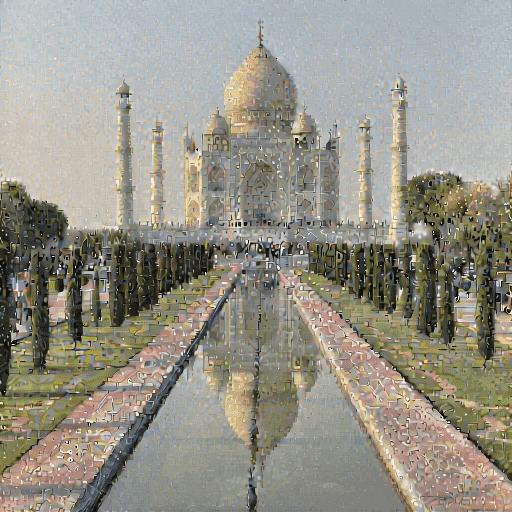}};
        \end{tikzpicture} &

        \begin{tikzpicture}[spy using outlines={circle, thick, red, magnification=2.5,size=0.9cm, connect spies}]
            \node {\includegraphics[width=\ww,frame]{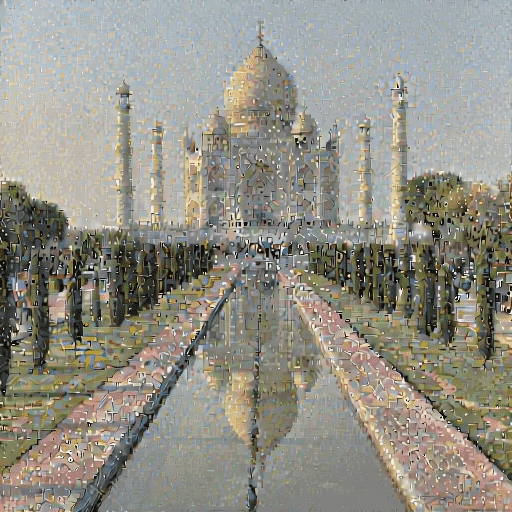}};
            \spy on (0.05,1.35) in node [left] at (1.6,0.2);
        \end{tikzpicture}
        
        \\[-5pt]

        \rotatebox{90}{\phantom{AAA.}\scriptsize{\textbf{(5) REED-VAE SD2 }}} &

        \begin{tikzpicture}[spy using outlines={}]
            \node {\includegraphics[width=\ww,frame]{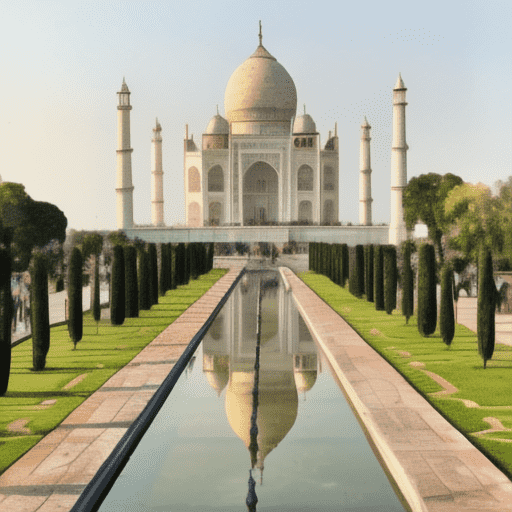}};
        \end{tikzpicture} &

        \begin{tikzpicture}[spy using outlines={}]
            \node {\includegraphics[width=\ww,frame]{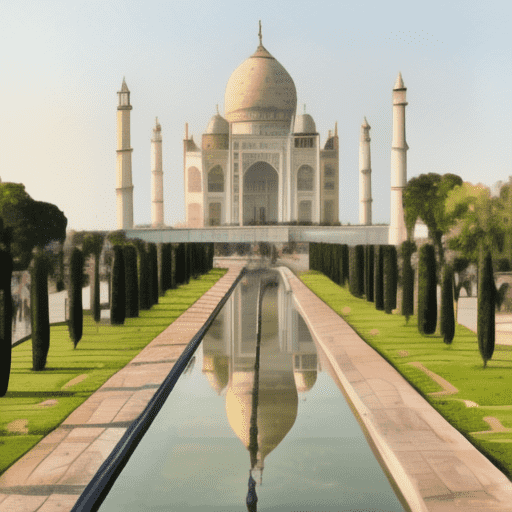}};
        \end{tikzpicture} &

        \begin{tikzpicture}[spy using outlines={}]
            \node {\includegraphics[width=\ww,frame]{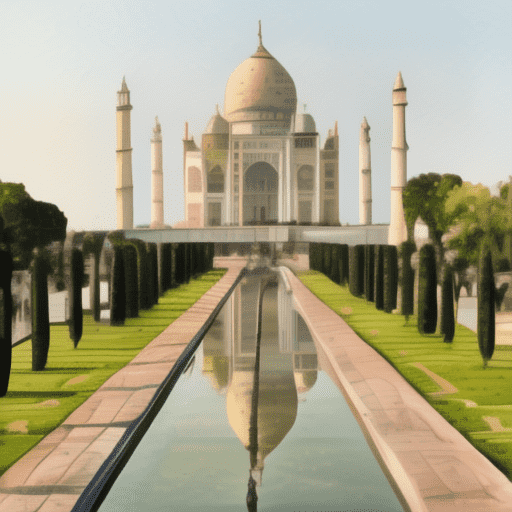}};
        \end{tikzpicture} &

        \begin{tikzpicture}[spy using outlines={}]
            \node {\includegraphics[width=\ww,frame]{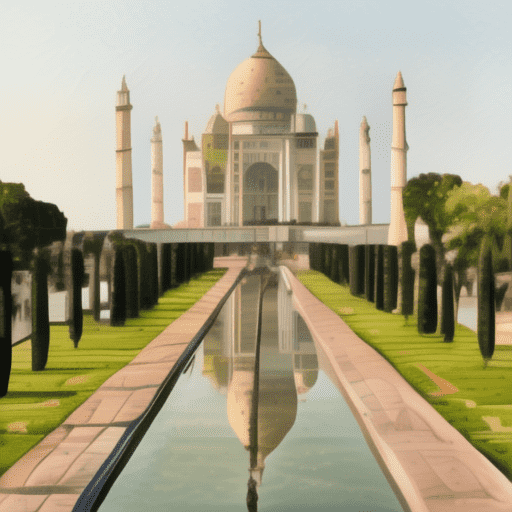}};
        \end{tikzpicture} &

        \begin{tikzpicture}[spy using outlines={circle, thick, red, magnification=2.5,size=0.9cm, connect spies}]
            \node {\includegraphics[width=\ww,frame]{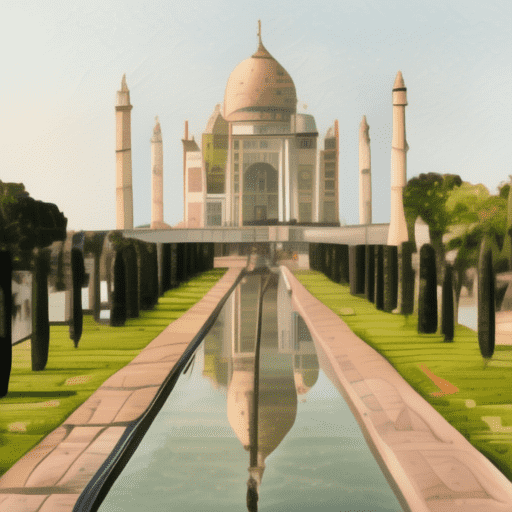}};
            \spy on (0.05,1.35) in node [left] at (1.6,0.2);
        \end{tikzpicture}
        
        \\
        
        &
        \scriptsize{5 } &
        \scriptsize{10 } &
        \scriptsize{15 } &
        \scriptsize{20 } &
        \scriptsize{25 } 

        \\
        &
        \multicolumn{5}{l}{\phantom{.}
        \begin{tikzpicture}
            \draw[->](0,0)--(16.4,0);
        \end{tikzpicture}}

        \\
        &
        \multicolumn{5}{c}{
        \scriptsize{Num. Encode/Decode Iterations}
        }
        
    \end{tabular}
    \caption{Additional comparison on iterative encode/decode task with more recent latent diffusion models, reported at 5,10,15,20,25 iterations.}

    \label{fig:new_models_tajmahal}
\end{figure*}

At the time of writing, Stable Diffusion 2.1 (SD 2.1) \cite{rombach2022high} was one of the most advanced and widely-used diffusion models for image generation and editing.
While SD2 remains an important and widely-used model, more recent LDM variants such as SDXL \cite{podell2023sdxl}, SD3 \cite{esser2024scaling}, and Flux \cite{flux2023}, have since been released.
Specifically, SD3 and Flux use more advanced latent spaces with improved, 16-channel VAEs that may behave differently from the 4-channel VAE used in SD2.
We conduct additional experiments to confirm that a similar iterative degradation problem does occur in these newer models as well.
We perform an iterative encode/decode task on the images in the ImagenHub dataset (179 images) and compute metrics with the original image.
The results (\Cref{table:new_ldms}, \Cref{fig:new_models_eiffel}, \Cref{fig:new_models_tajmahal}) show that in all the newer models, noise and artifacts still accumulate after $5+$ iterations.
Flux seems to be the most resistant, yet still accumulates a fair amount of noise.
When SD2 is paired with our REED-VAE, it consistently outperforms even these newer LDM variants in most metrics.
Flux is the only model that surpasses SD2 $+$ REED-VAE in certain cases.
Specifically, the updated Flux model demonstrates improved performance in terms of LPIPS and FID at the 5 iteration mark, though this advantage is not sustained at later iterations. 
Therefore, despite their more advanced latent spaces, at higher iterations these newer models still exhibit the problem of iterative degradation, and the REED training algorithm trained on their respective VAEs will likely improve performance in them as well.
Even when trained on the simpler SD2, the pipeline with REED-VAE is able to outperform all models at higher iterations.

\begin{figure*}[htpb]
    \centering
    \setlength{\tabcolsep}{-2pt}
    \renewcommand{\arraystretch}{0.5}
    \setlength{\ww}{0.185\textwidth}

    \begin{tabular}{p{0.4cm} c c c p{0.4cm} c c}
        & \begin{tikzpicture}[spy using outlines=]
            \node {\includegraphics[width=\ww,frame]{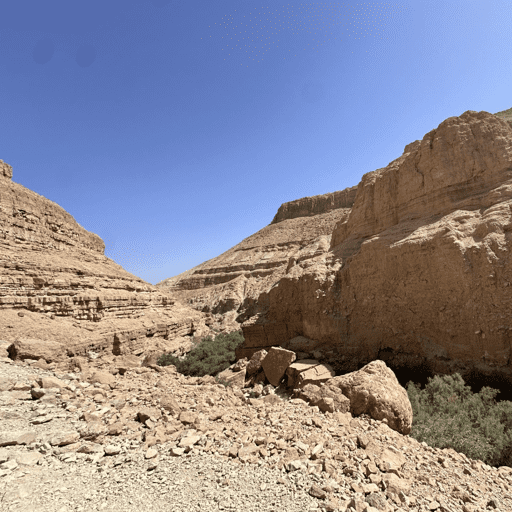}};
        \end{tikzpicture} &&&&&
        \\[-2pt]
        &\scriptsize{Input} &&&& \\
        \rotatebox{90}{\phantom{AAA.}\scriptsize{(1) Vanilla-VAE NTI }} &

        \begin{tikzpicture}[spy using outlines={}]
            \node {\includegraphics[width=\ww,frame]{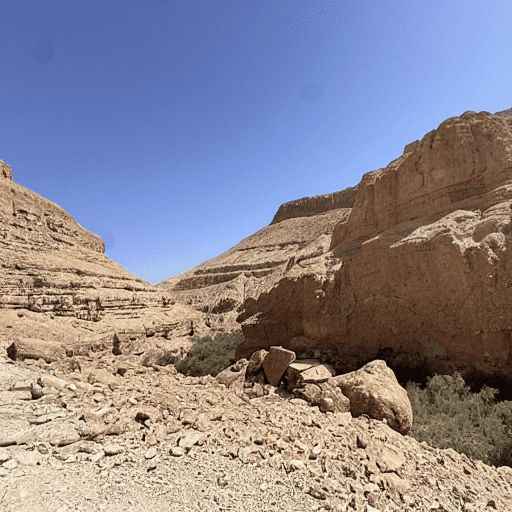}};
        \end{tikzpicture} &

        \begin{tikzpicture}[spy using outlines={}]
            \node {\includegraphics[width=\ww,frame]{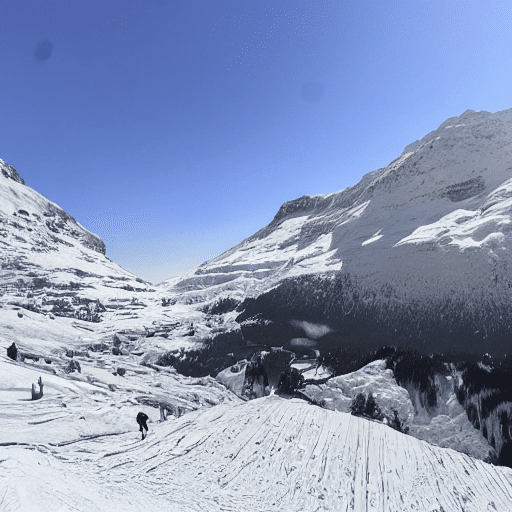}};
        \end{tikzpicture} &

        \begin{tikzpicture}[spy using outlines={}]
            \node {\includegraphics[width=\ww,frame]{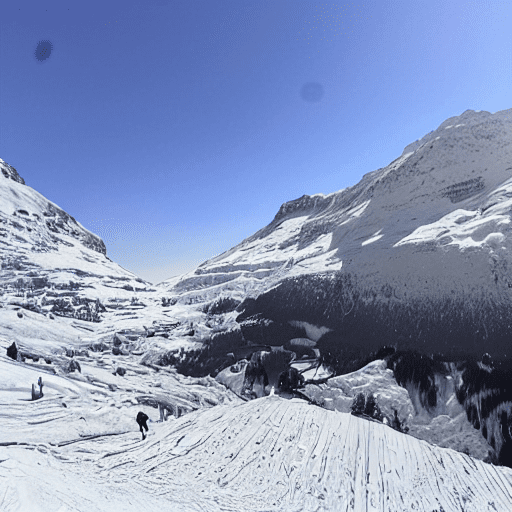}};
        \end{tikzpicture} &

        &

        \begin{tikzpicture}[spy using outlines=]
            \node {\includegraphics[width=\ww,frame]{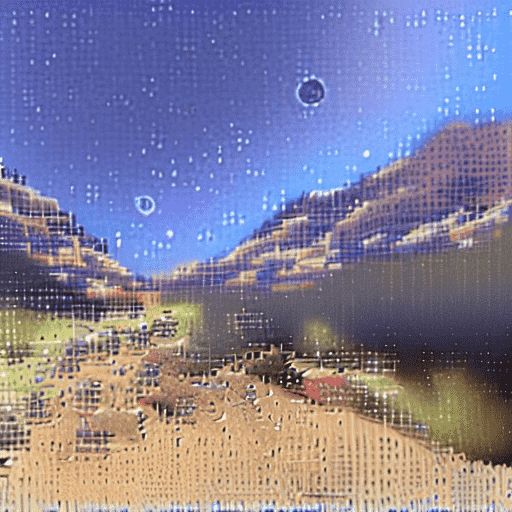}};
        \end{tikzpicture} &

        \begin{tikzpicture}[spy using outlines=]
            \node {\includegraphics[width=\ww,frame]{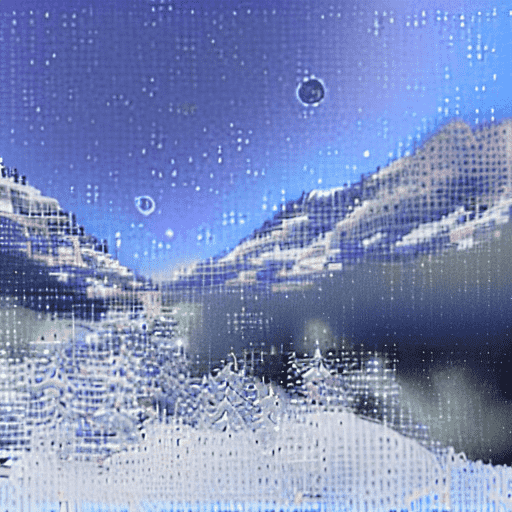}};
        \end{tikzpicture}
        
        \\[-5pt]

        \rotatebox{90}{\phantom{AAA.}\scriptsize{(2) REED-VAE NTI }} &

        \begin{tikzpicture}[spy using outlines={}]
            \node {\includegraphics[width=\ww,frame]{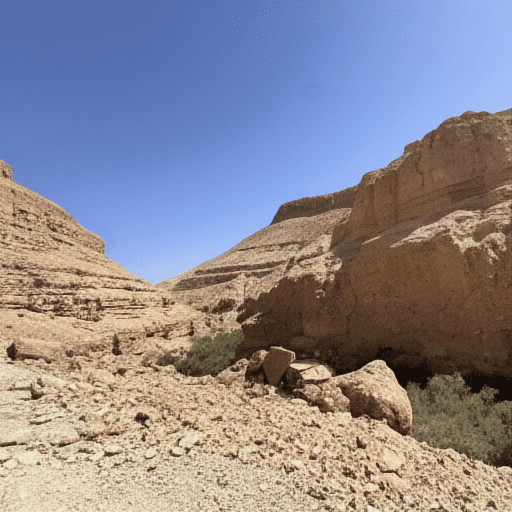}};
        \end{tikzpicture} &

        \begin{tikzpicture}[spy using outlines={}]
            \node {\includegraphics[width=\ww,frame]{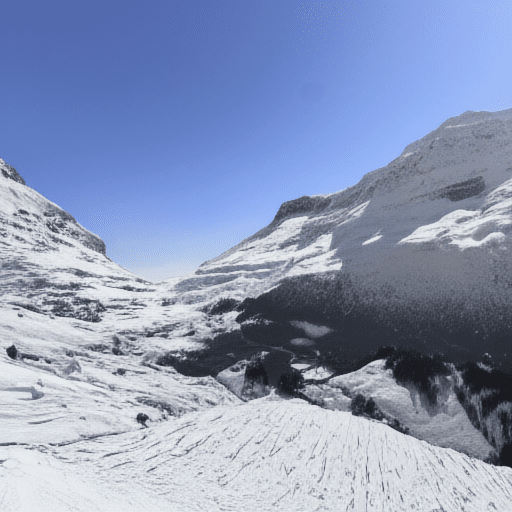}};
        \end{tikzpicture} &

        \begin{tikzpicture}[spy using outlines={}]
            \node {\includegraphics[width=\ww,frame]{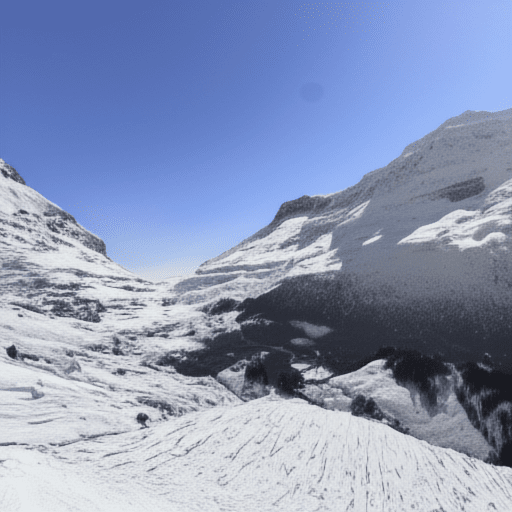}};
        \end{tikzpicture} &
        &
        \begin{tikzpicture}[spy using outlines=]
            \node {\includegraphics[width=\ww,frame]{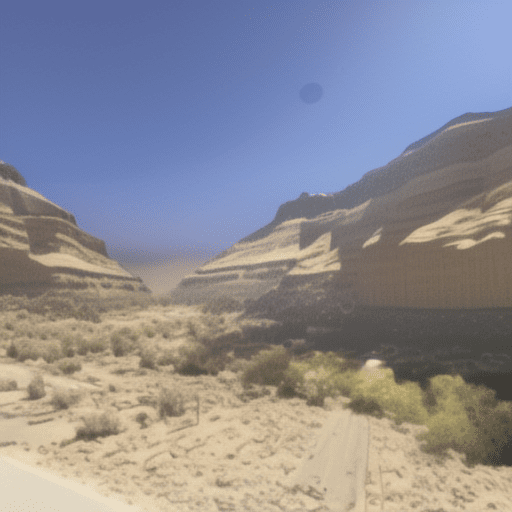}};
        \end{tikzpicture} &

        \begin{tikzpicture}[spy using outlines=]
            \node {\includegraphics[width=\ww,frame]{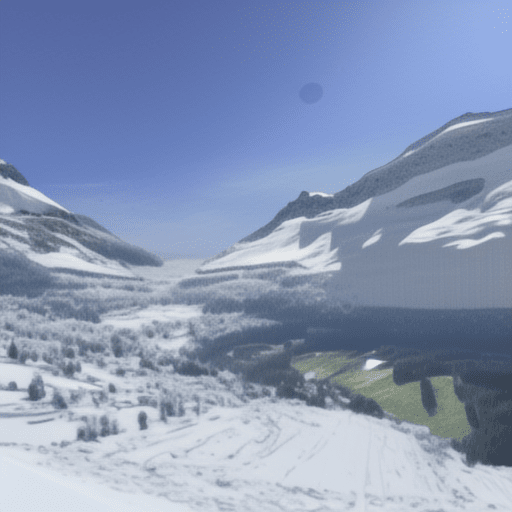}};
        \end{tikzpicture}
        
        \\

        &
        \scriptsize{Original Inversion } &
        \scriptsize{Edit 1 } &
        \scriptsize{Inversion 1 } &
        \scriptsize{  (...) } &
        \scriptsize{Inversion 6 } &
        \scriptsize{Edit 6 } \\

    \end{tabular}
    \caption{Iterative edits using Null-Text Inversion. Prompts: ``a landscape with \textbf{desert} mountains'' $\to$ ``a landscape with \textbf{snowy} mountains''. Despite regenerating latents through the inversion process, visual artifacts accumulate, particularly in later iterations (e.g. noise patterns and loss of fidelity to the original image). This illustrates that DDIM inversion-based methods do not mitigate the degradation that occurs in iterative editing tasks, underscoring the need for REED-VAE.}

    \label{fig:iterative_nulltext_edit}
\end{figure*}

\subsection{More ablation results}
\begin{figure*}[htpb]
    \centering
    \setlength{\tabcolsep}{-2pt}
    \renewcommand{\arraystretch}{0.5}
    \setlength{\ww}{0.175\textwidth}

    \begin{tabular}{p{0.4cm} c c c c}
        \rotatebox{90}{\phantom{AWWWW.}\scriptsize{ Input }} &
        \begin{tikzpicture}[spy using outlines={}]
            \node {\includegraphics[width=\ww,frame]{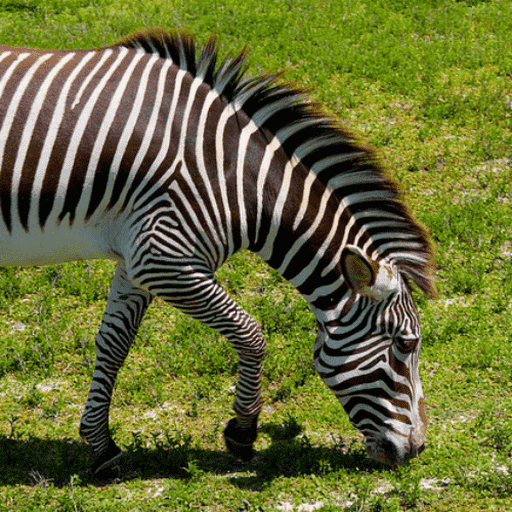}};
        \end{tikzpicture} &&&
        \\[-2pt]
        \rotatebox{90}{\phantom{AAAA.}\scriptsize{(1) Vanilla-VAE }} &

        \begin{tikzpicture}[spy using outlines={}]
            \node {\includegraphics[width=\ww,frame]{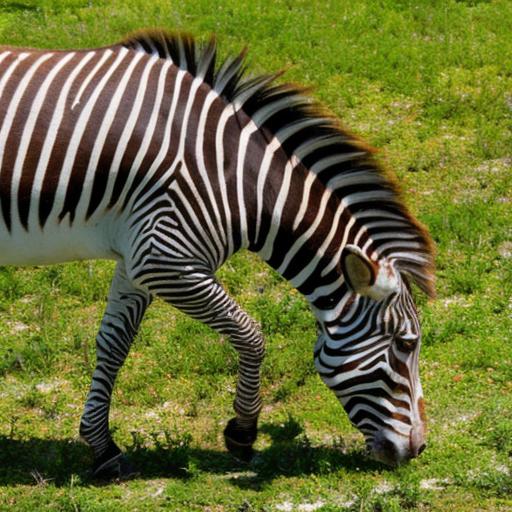}};
        \end{tikzpicture} &

        \begin{tikzpicture}[spy using outlines={}]
            \node {\includegraphics[width=\ww,frame]{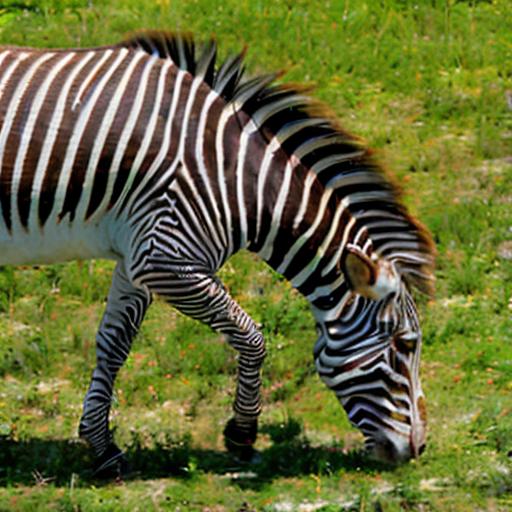}};
        \end{tikzpicture} &

        \begin{tikzpicture}[spy using outlines={}]
            \node {\includegraphics[width=\ww,frame]{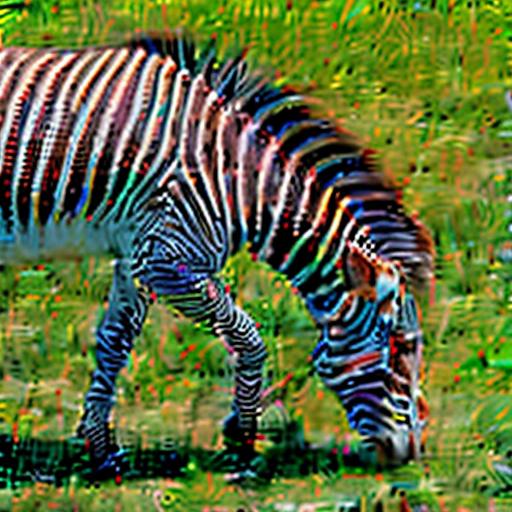}};
        \end{tikzpicture} &

        \begin{tikzpicture}[spy using outlines=]
            \node {\includegraphics[width=\ww,frame]{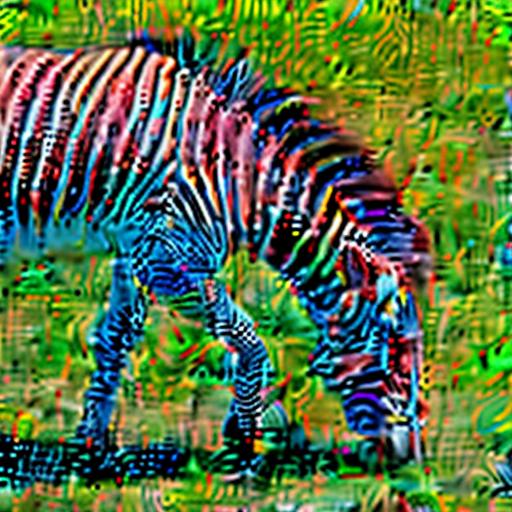}};
        \end{tikzpicture}
        
        \\[-5pt]

        \rotatebox{90}{\phantom{AAAAA.}\scriptsize{(2) IT (k=2) }} &

        \begin{tikzpicture}[spy using outlines={}]
            \node {\includegraphics[width=\ww,frame]{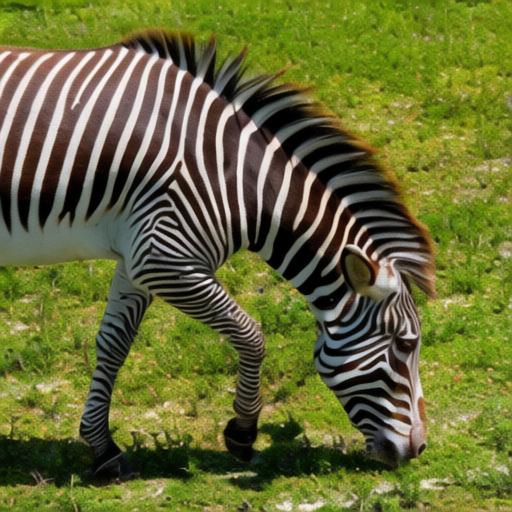}};
        \end{tikzpicture} &

        \begin{tikzpicture}[spy using outlines={}]
            \node {\includegraphics[width=\ww,frame]{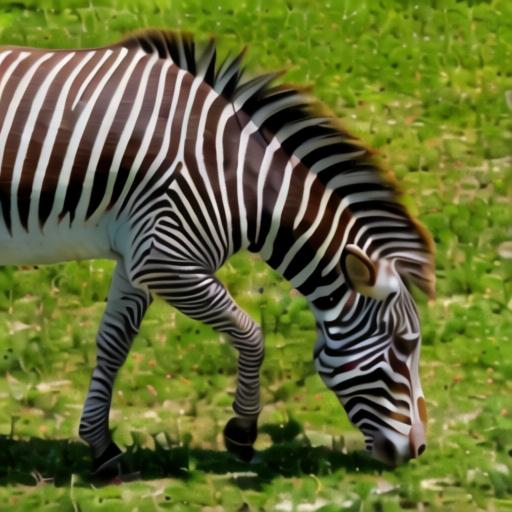}};
        \end{tikzpicture} &

        \begin{tikzpicture}[spy using outlines={}]
            \node {\includegraphics[width=\ww,frame]{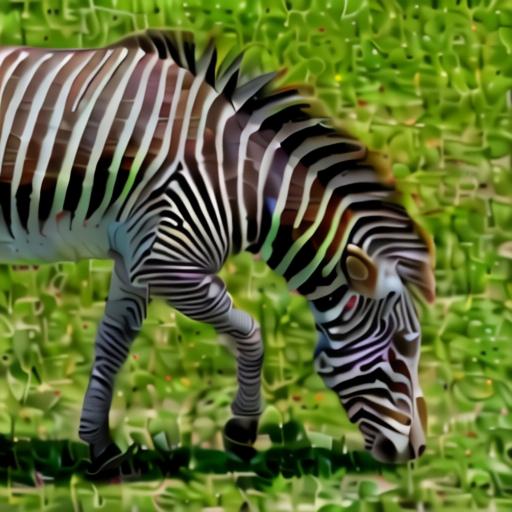}};
        \end{tikzpicture} &

        \begin{tikzpicture}[spy using outlines=]
            \node {\includegraphics[width=\ww,frame]{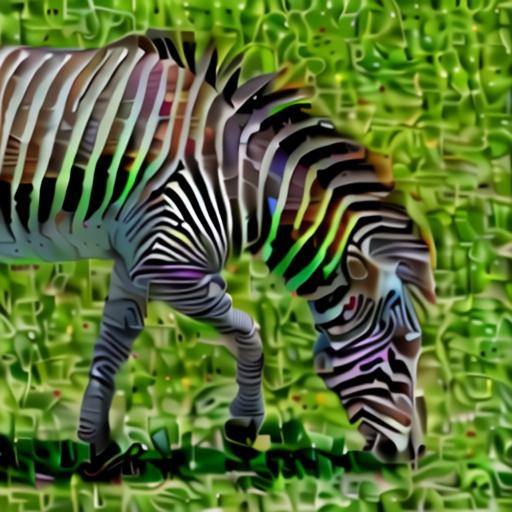}};
        \end{tikzpicture}
        
        \\[-5pt]

        \rotatebox{90}{\phantom{AAAAA.}\scriptsize{(3) IT (k=5) }} &

        \begin{tikzpicture}[spy using outlines={}]
            \node {\includegraphics[width=\ww,frame]{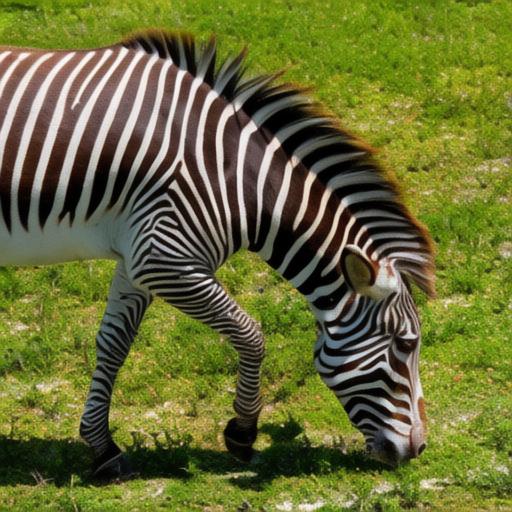}};
        \end{tikzpicture} &

        \begin{tikzpicture}[spy using outlines={}]
            \node {\includegraphics[width=\ww,frame]{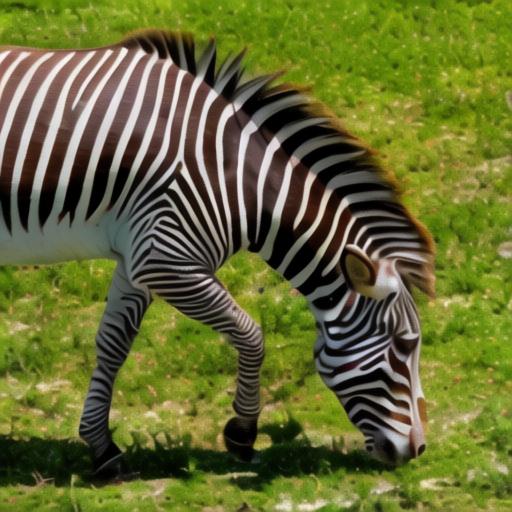}};
        \end{tikzpicture} &

        \begin{tikzpicture}[spy using outlines={}]
            \node {\includegraphics[width=\ww,frame]{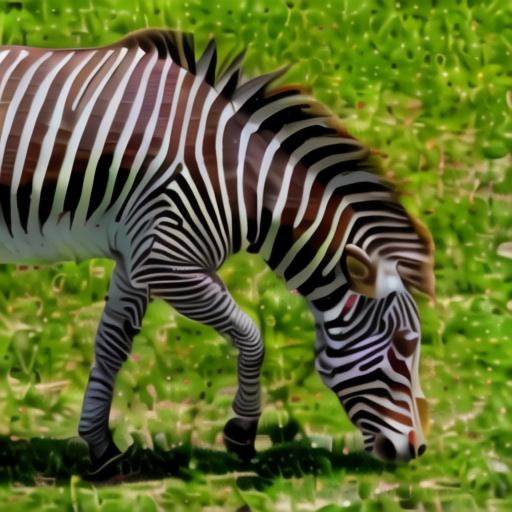}};
        \end{tikzpicture} &

        \begin{tikzpicture}[spy using outlines=]
            \node {\includegraphics[width=\ww,frame]{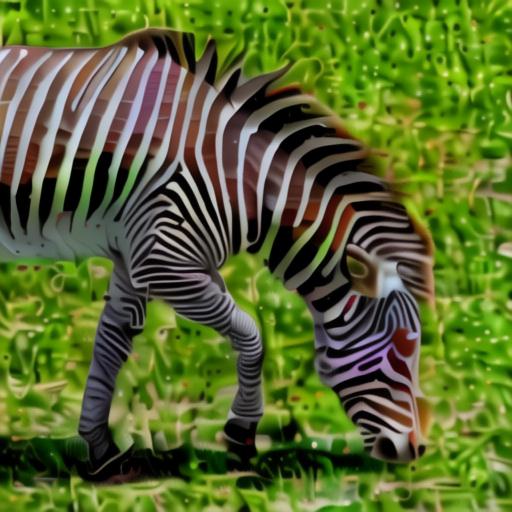}};
        \end{tikzpicture}
        
        \\[-5pt]

        \rotatebox{90}{\phantom{AAAA.}\scriptsize{(4) IT+FSL (k=5) }} &

        \begin{tikzpicture}[spy using outlines={}]
            \node {\includegraphics[width=\ww,frame]{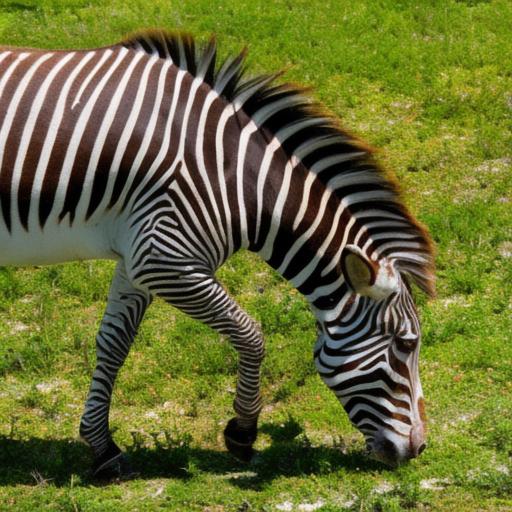}};
        \end{tikzpicture} &

        \begin{tikzpicture}[spy using outlines={}]
            \node {\includegraphics[width=\ww,frame]{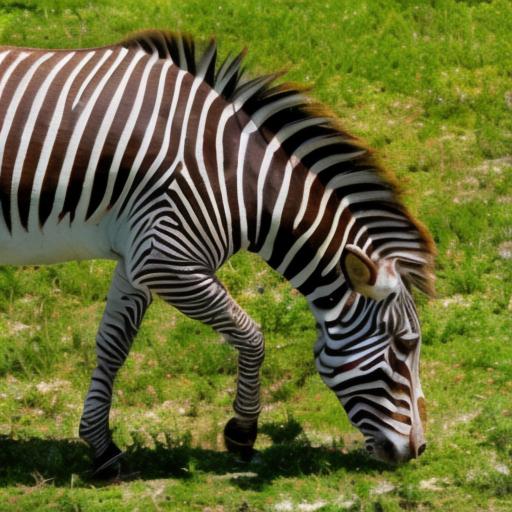}};
        \end{tikzpicture} &

        \begin{tikzpicture}[spy using outlines={}]
            \node {\includegraphics[width=\ww,frame]{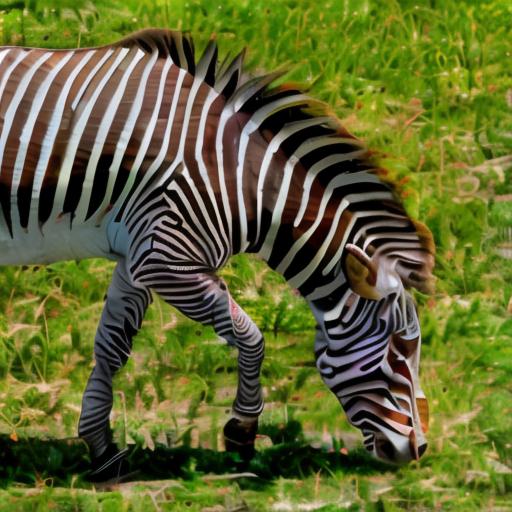}};
        \end{tikzpicture} &

        \begin{tikzpicture}[spy using outlines=]
            \node {\includegraphics[width=\ww,frame]{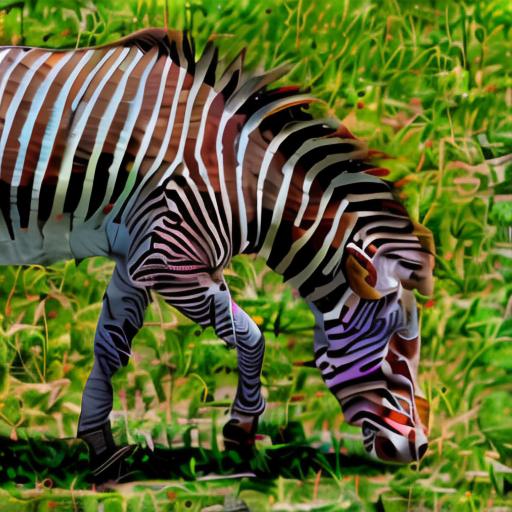}};
        \end{tikzpicture}
        
        \\[-5pt]

        \rotatebox{90}{\phantom{AAA.}\scriptsize{\textbf{(5) IT+FSL+DI (k=5) }}} &

        \begin{tikzpicture}[spy using outlines={}]
            \node {\includegraphics[width=\ww,frame]{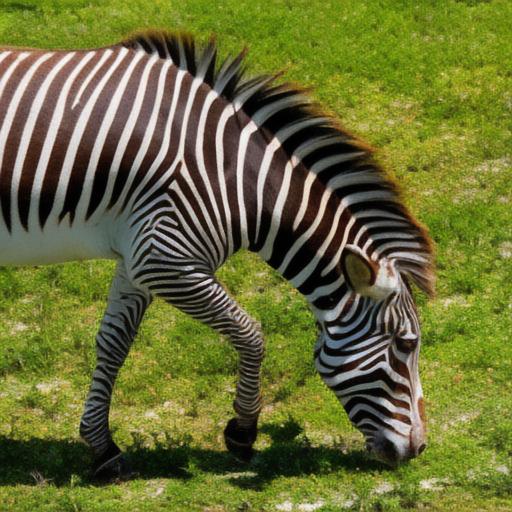}};
        \end{tikzpicture} &

        \begin{tikzpicture}[spy using outlines={}]
            \node {\includegraphics[width=\ww,frame]{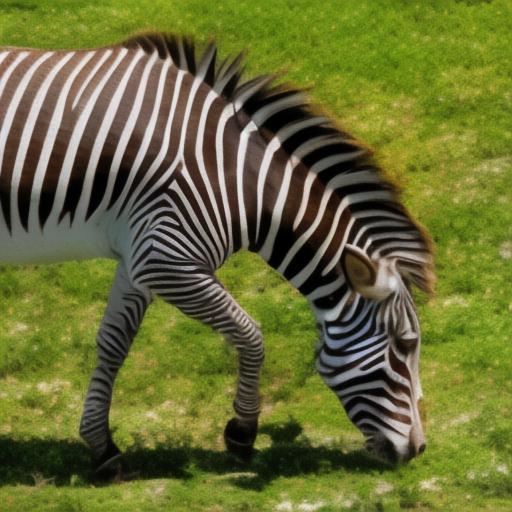}};
        \end{tikzpicture} &

        \begin{tikzpicture}[spy using outlines={}]
            \node {\includegraphics[width=\ww,frame]{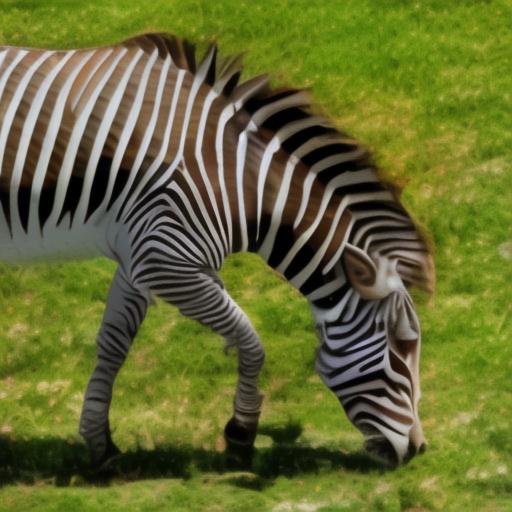}};
        \end{tikzpicture} &

        \begin{tikzpicture}[spy using outlines={}]
            \node {\includegraphics[width=\ww,frame]{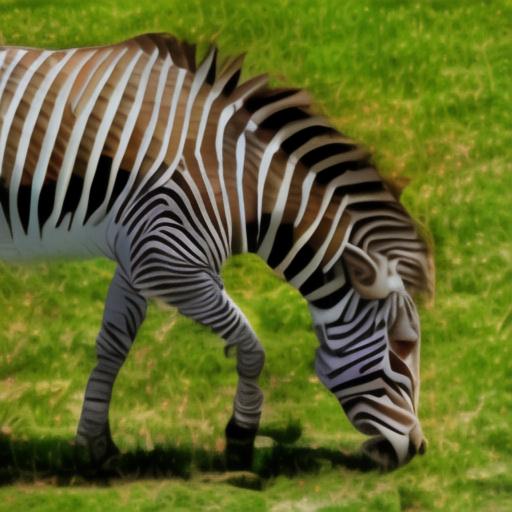}};
        \end{tikzpicture}
        
        \\
        
        &
        \scriptsize{1 } &
        \scriptsize{5 } &
        \scriptsize{15 } &
        \scriptsize{25 } 

        \\
        &
        \multicolumn{4}{l}{\phantom{.}
        \begin{tikzpicture}
            \draw[->](0,0)--(12.7,0);
        \end{tikzpicture}}

        \\
        &
        \multicolumn{4}{c}{
        \scriptsize{Num. Encode/Decode Iterations}
        }
        
    \end{tabular}
    \caption{Ablation on individual components of REED-VAE on a sample image from our evaluation set. We compare the Vanilla-VAE (1), REED with static Iterative Training (IT) at $k=2$ (2) and $k=3$ (3), REED with IT at $k=5$ and the First-Step Loss (FSL) (4), and the full REED-VAE model with IT at $k=5$, FSL, and Dynamic Incrementation (DI). It can be seen that the full REED-VAE model (IT+DI+FSL ($k=5$)) is best able to maintain image features and colors, even at 25 iterations.}
    \label{fig:ablation_zebra}
\end{figure*}
We provide a visual example of the improvement provided by each component in our final REED-VAE in Figure.\ref{fig:ablation_zebra}.

\subsection{Comparison with Inversion-Based Methods}
In addition to the iterative inversion experiment discussed in the main paper, we also provide results for iterative NTI \cite{mokady2023null} combined with P2P editing \cite{hertz2022prompt}.
As shown in \Cref{fig:iterative_nulltext_edit}, NTI introduces significant artifacts over iterations that are noticably reduced when NTI is paired with REED-VAE.
The experiment follows a similar iterative setup: NTI is first used to invert the image back to its latent representation, after which P2P editing is applied based on the provided text prompt.
These iterative steps are repeated for 6 iterations in the example figure.
The results demonstrate that NTI-based methods struggle to maintain fidelity across iterations, accumulating noise and distortions.
These findings highlight the limitation of iterative editing that exists in DDIM-Inversion-based methods as well as non-inversion diffusion-based editing methods, demonstrating the importance and relevance of REED-VAE even with such newer models.

\begin{figure*}[htpb]
    \centering
    \setlength{\tabcolsep}{-2pt}
    \renewcommand{\arraystretch}{0.5}
    \setlength{\ww}{0.175\textwidth}

    \begin{tabular}{c c c c c}
        \begin{tikzpicture}[spy using outlines=]
            \node {\includegraphics[width=\ww,frame]{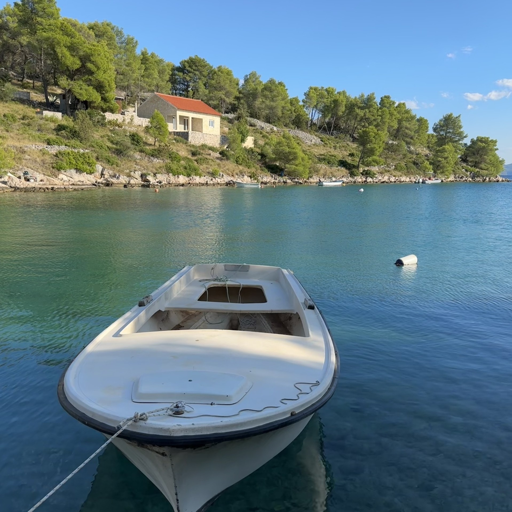}};
        \end{tikzpicture} &&&&
        \\[-3pt]
        \scriptsize{Input Image } &&&&
        \\[-3pt]

        \begin{tikzpicture}[spy using outlines={}]
            \node {\includegraphics[width=\ww,frame]{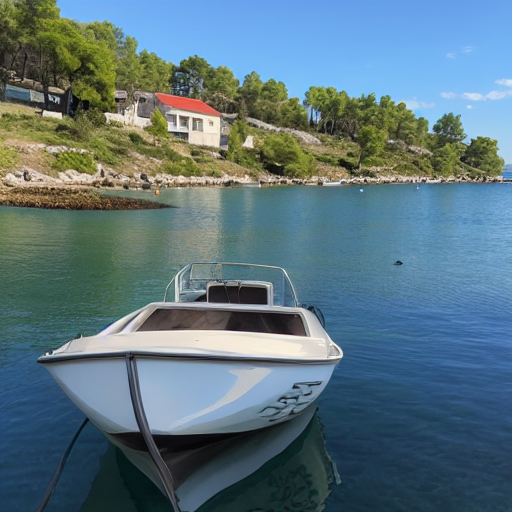}};
        \end{tikzpicture} &

        \begin{tikzpicture}[spy using outlines={}]
            \node {\includegraphics[width=\ww,frame]{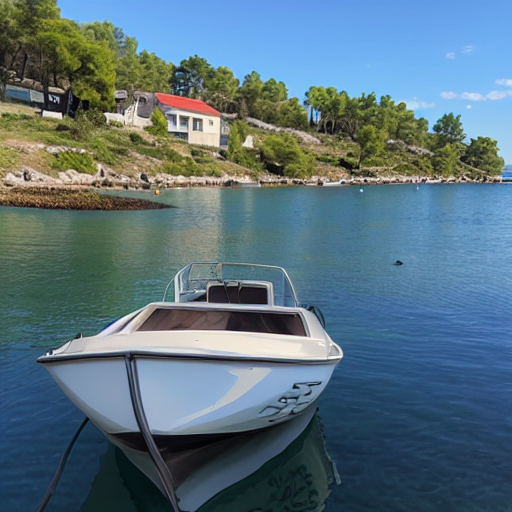}};
        \end{tikzpicture} &

        \begin{tikzpicture}[spy using outlines={}]
            \node {\includegraphics[width=\ww,frame]{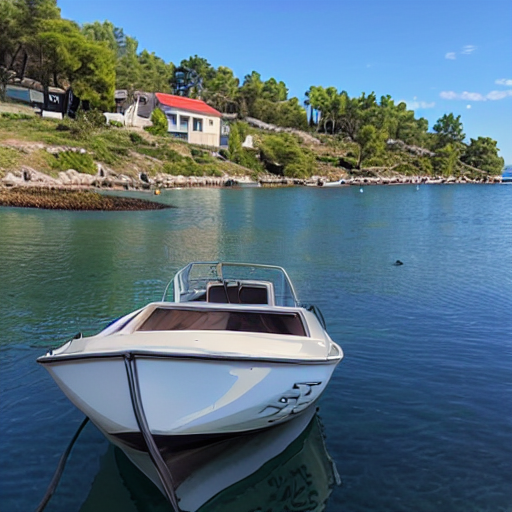}};
        \end{tikzpicture} &

        \begin{tikzpicture}[spy using outlines=]
            \node {\includegraphics[width=\ww,frame]{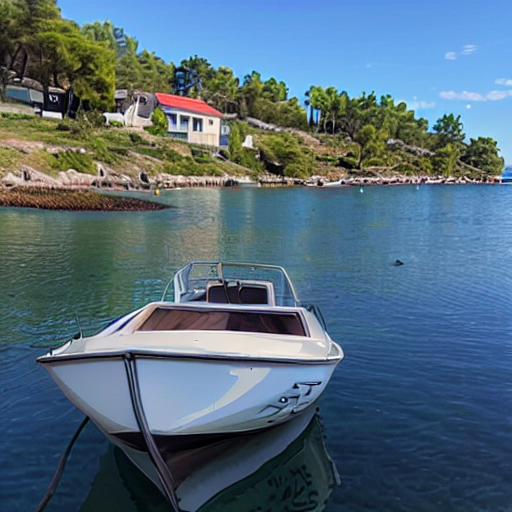}};
        \end{tikzpicture} &

        \begin{tikzpicture}[spy using outlines=]
            \node {\includegraphics[width=\ww,frame]{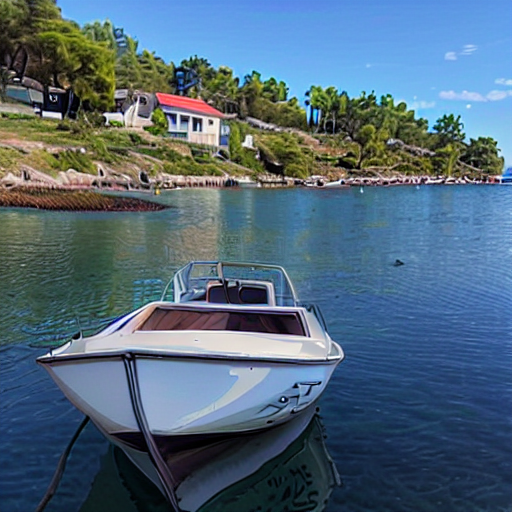}};
        \end{tikzpicture} 

        \\[-3pt]        
        \scriptsize{1 } &
        \scriptsize{2 } &
        \scriptsize{3 } &
        \scriptsize{4 } &
        \scriptsize{5 } 
        \\[-3pt]
        
        \begin{tikzpicture}[spy using outlines=]
            \node {\includegraphics[width=\ww,frame]{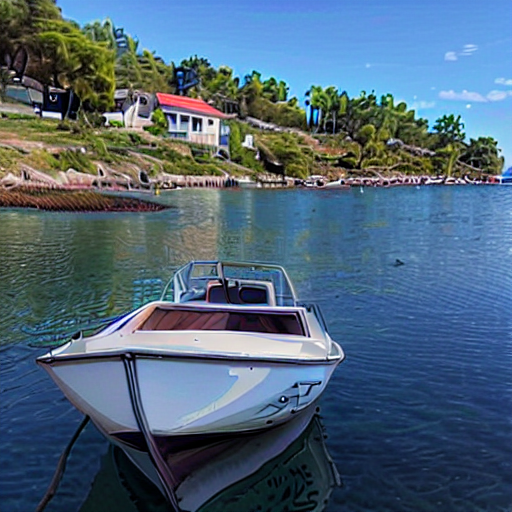}};
        \end{tikzpicture} &

        \begin{tikzpicture}[spy using outlines=]
            \node {\includegraphics[width=\ww,frame]{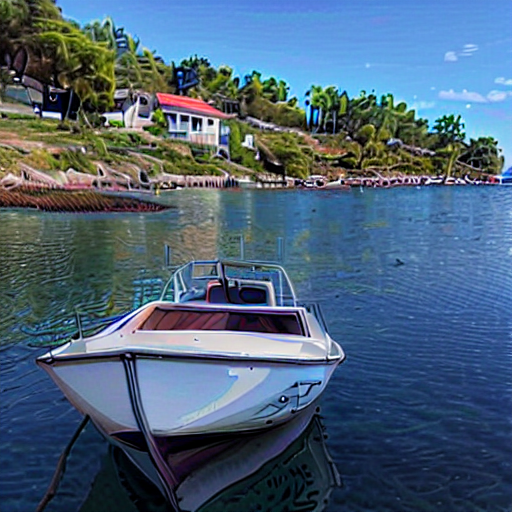}};
        \end{tikzpicture} &

        \begin{tikzpicture}[spy using outlines=]
            \node {\includegraphics[width=\ww,frame]{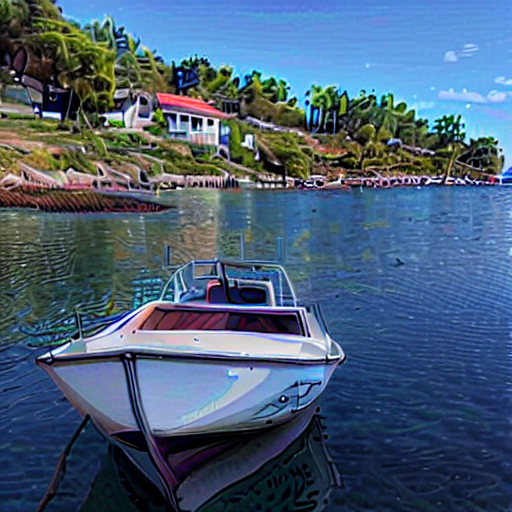}};
        \end{tikzpicture} &

        \begin{tikzpicture}[spy using outlines=]
            \node {\includegraphics[width=\ww,frame]{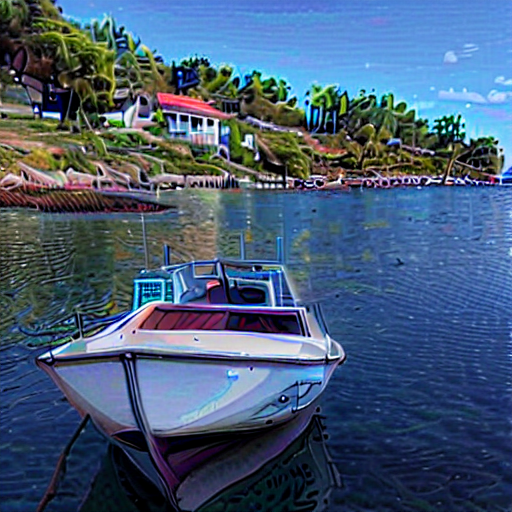}};
        \end{tikzpicture} &

        \begin{tikzpicture}[spy using outlines=]
            \node {\includegraphics[width=\ww,frame]{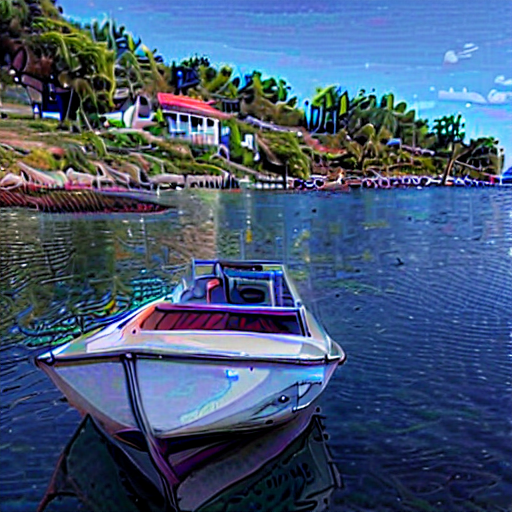}};
        \end{tikzpicture}
        
        \\[-3pt]
        \scriptsize{6 } &
        \scriptsize{7 } &
        \scriptsize{8 } &
        \scriptsize{9 } &
        \scriptsize{10 } 
        \\[-3pt]

        \begin{tikzpicture}[spy using outlines=]
            \node {\includegraphics[width=\ww,frame]{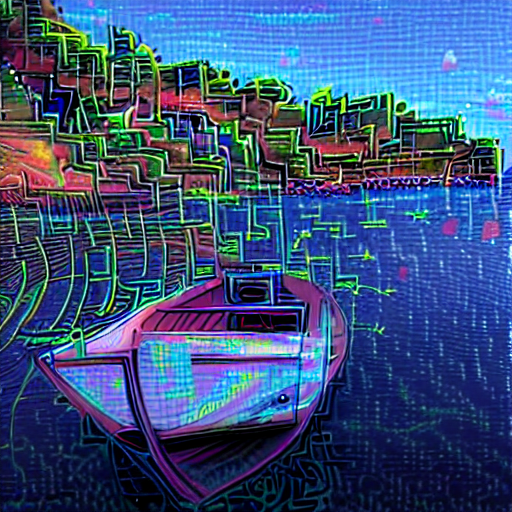}};
        \end{tikzpicture} &

        \begin{tikzpicture}[spy using outlines=]
            \node {\includegraphics[width=\ww,frame]{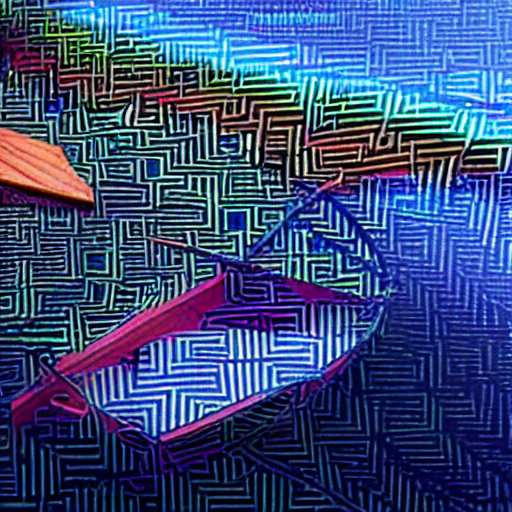}};
        \end{tikzpicture} &

        \begin{tikzpicture}[spy using outlines=]
            \node {\includegraphics[width=\ww,frame]{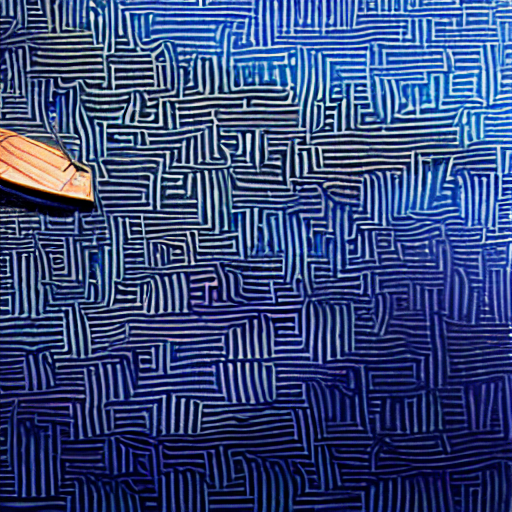}};
        \end{tikzpicture} &

        \begin{tikzpicture}[spy using outlines=]
            \node {\includegraphics[width=\ww,frame]{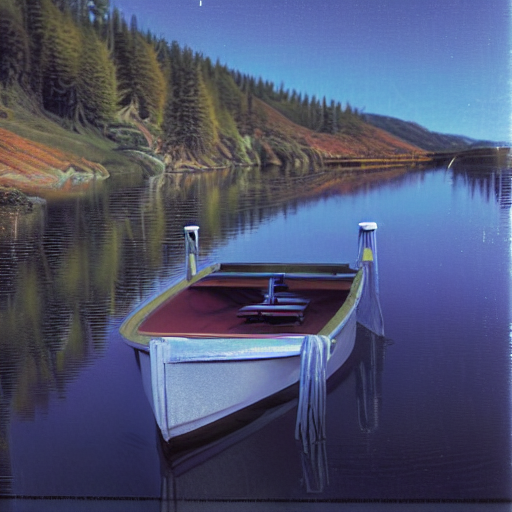}};
        \end{tikzpicture} &

        \begin{tikzpicture}[spy using outlines=]
            \node {\includegraphics[width=\ww,frame]{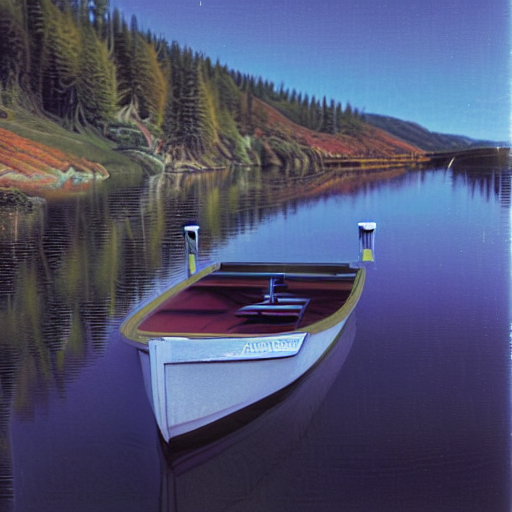}};
        \end{tikzpicture}
        
        \\[-3pt]
        \scriptsize{11 } &
        \scriptsize{12 } &
        \scriptsize{13 } &
        \scriptsize{14 } &
        \scriptsize{15 } 
        \\[-3pt]
        
        \begin{tikzpicture}[spy using outlines=]
            \node {\includegraphics[width=\ww,frame]{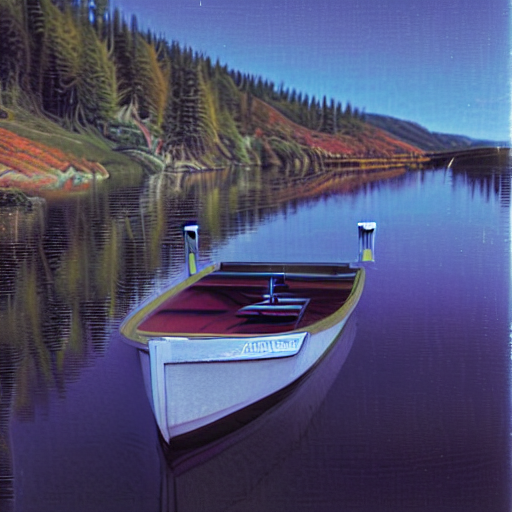}};
        \end{tikzpicture} &

        \begin{tikzpicture}[spy using outlines=]
            \node {\includegraphics[width=\ww,frame]{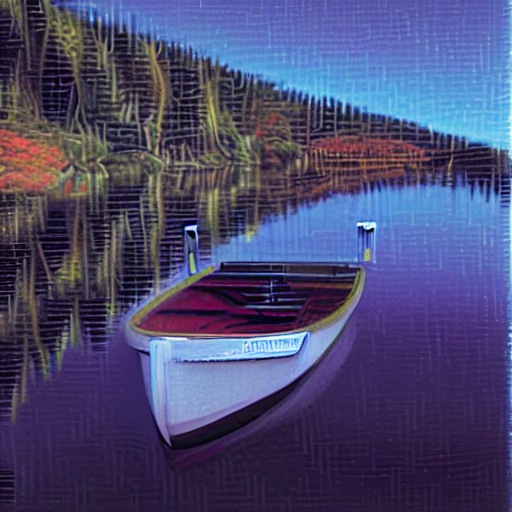}};
        \end{tikzpicture} &

        \begin{tikzpicture}[spy using outlines=]
            \node {\includegraphics[width=\ww,frame]{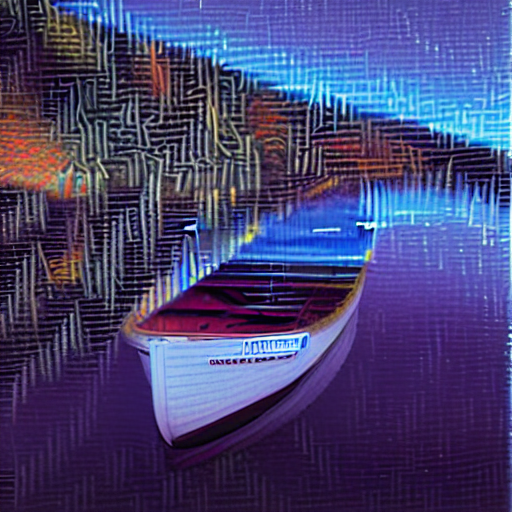}};
        \end{tikzpicture} &

        \begin{tikzpicture}[spy using outlines=]
            \node {\includegraphics[width=\ww,frame]{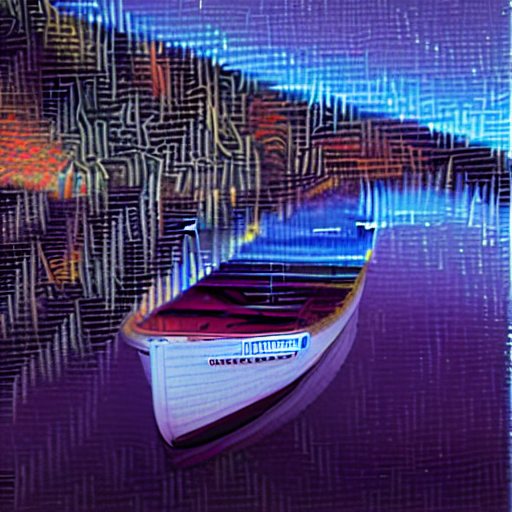}};
        \end{tikzpicture} &

        \begin{tikzpicture}[spy using outlines=]
            \node {\includegraphics[width=\ww,frame]{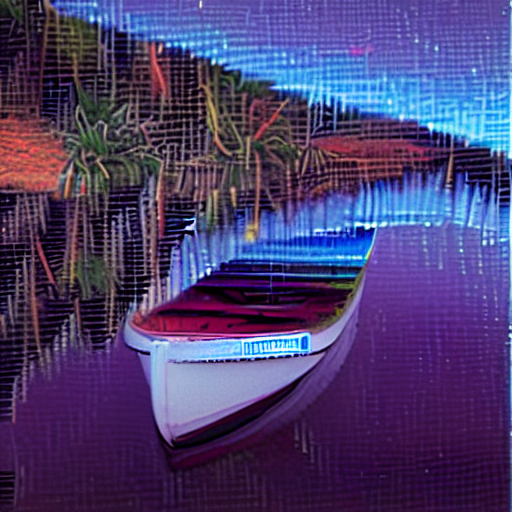}};
        \end{tikzpicture}
        
        \\[-3pt]
        \scriptsize{16 } &
        \scriptsize{17 } &
        \scriptsize{18 } &
        \scriptsize{19 } &
        \scriptsize{20 }
        \\[-3pt]

        \begin{tikzpicture}[spy using outlines=]
            \node {\includegraphics[width=\ww,frame]{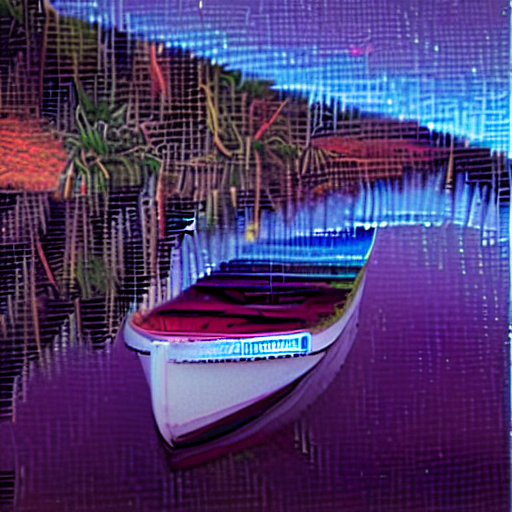}};
        \end{tikzpicture} &

        \begin{tikzpicture}[spy using outlines=]
            \node {\includegraphics[width=\ww,frame]{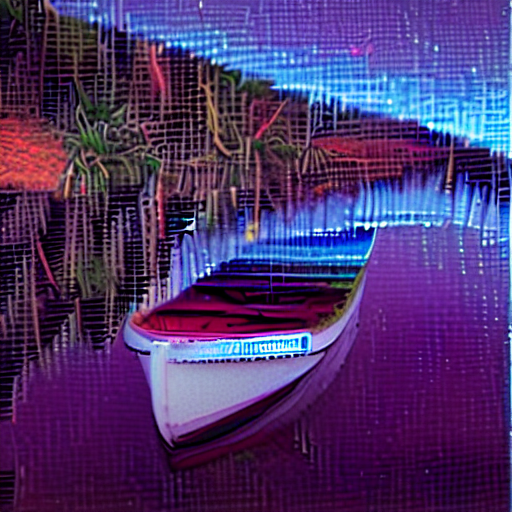}};
        \end{tikzpicture} &

        \begin{tikzpicture}[spy using outlines=]
            \node {\includegraphics[width=\ww,frame]{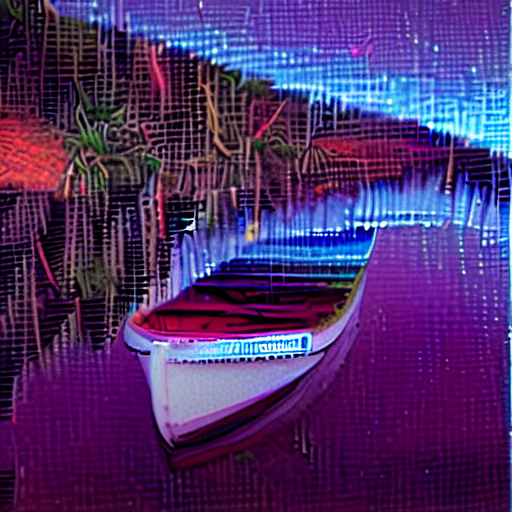}};
        \end{tikzpicture} &

        \begin{tikzpicture}[spy using outlines=]
            \node {\includegraphics[width=\ww,frame]{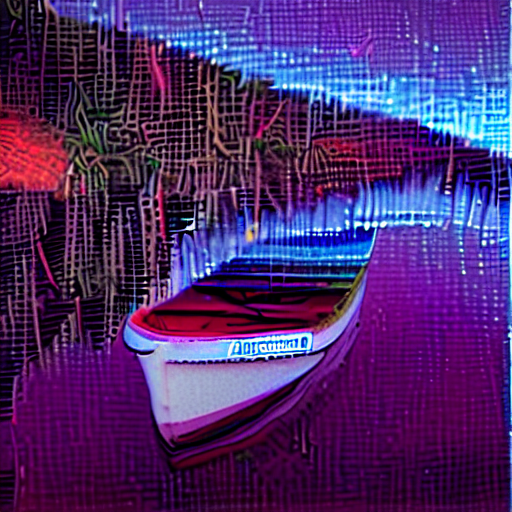}};
        \end{tikzpicture} &

        \begin{tikzpicture}[spy using outlines=]
            \node {\includegraphics[width=\ww,frame]{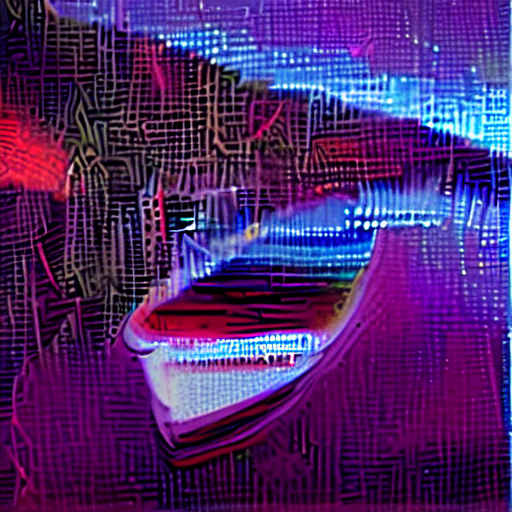}};
        \end{tikzpicture}
        
        \\[-3pt]
        \scriptsize{21 } &
        \scriptsize{22 } &
        \scriptsize{23 } &
        \scriptsize{24 } &
        \scriptsize{25 }
        \\[-3pt]
        
    \end{tabular}
    \caption{Full sequence of iterative inversion reconstructions using Vanilla-VAE with Null-Text Inversion (NTI) \cite{mokady2023null}. The input image undergoes NTI-based inversion followed by reconstruction with the same source prompt for 25 iterations. Early iterations (1-10) retain reasonable fidelity, but progressive iterations introduce artifacts, noise patterns, and distortions, culminating in severe degradation by iteration 25.}

    \label{fig:nti_editing_vanilla_full}
\end{figure*}

\begin{figure*}[htpb]
    \centering
    \setlength{\tabcolsep}{-2pt}
    \renewcommand{\arraystretch}{0.5}
    \setlength{\ww}{0.175\textwidth}

    \begin{tabular}{c c c c c}
        \begin{tikzpicture}[spy using outlines=]
            \node {\includegraphics[width=\ww,frame]{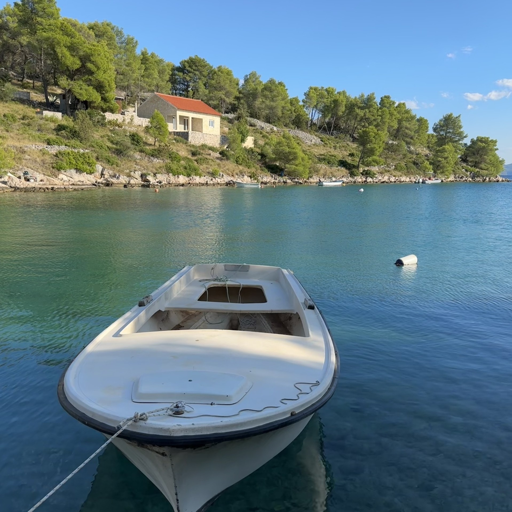}};
        \end{tikzpicture} &&&&
        \\[-3pt]
        \scriptsize{Input Image } &&&&
        \\[-3pt]

        \begin{tikzpicture}[spy using outlines={}]
            \node {\includegraphics[width=\ww,frame]{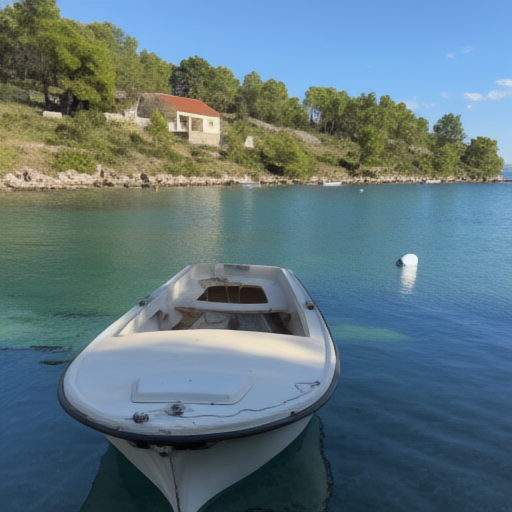}};
        \end{tikzpicture} &

        \begin{tikzpicture}[spy using outlines={}]
            \node {\includegraphics[width=\ww,frame]{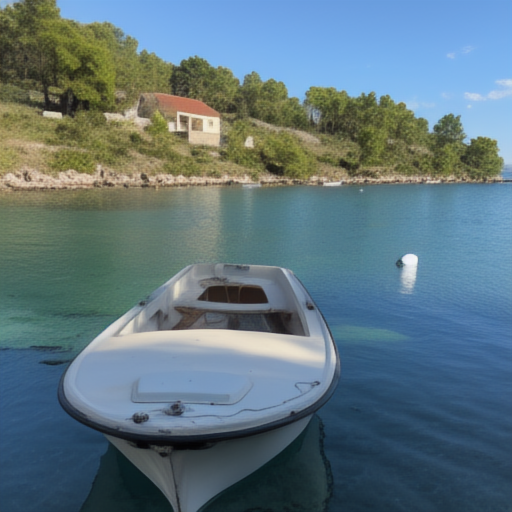}};
        \end{tikzpicture} &

        \begin{tikzpicture}[spy using outlines={}]
            \node {\includegraphics[width=\ww,frame]{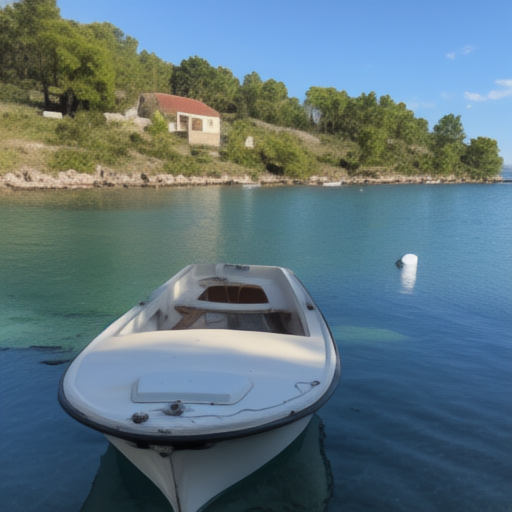}};
        \end{tikzpicture} &

        \begin{tikzpicture}[spy using outlines=]
            \node {\includegraphics[width=\ww,frame]{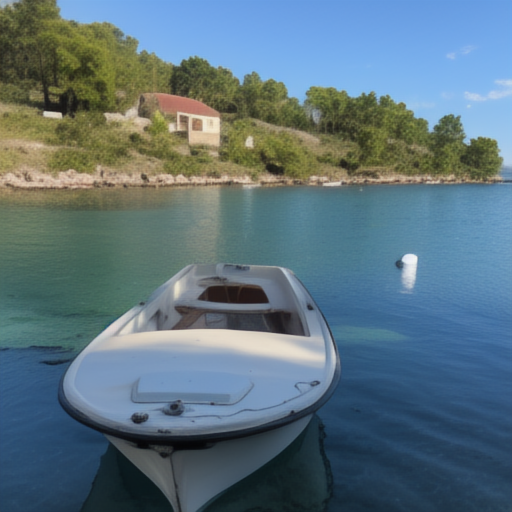}};
        \end{tikzpicture} &

        \begin{tikzpicture}[spy using outlines=]
            \node {\includegraphics[width=\ww,frame]{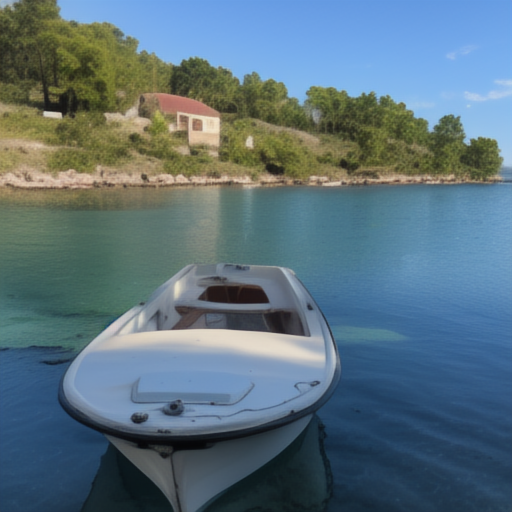}};
        \end{tikzpicture} 

        \\[-3pt]
        \scriptsize{1 } &
        \scriptsize{2 } &
        \scriptsize{3 } &
        \scriptsize{4 } &
        \scriptsize{5 } 
        \\[-3pt]

        \begin{tikzpicture}[spy using outlines=]
            \node {\includegraphics[width=\ww,frame]{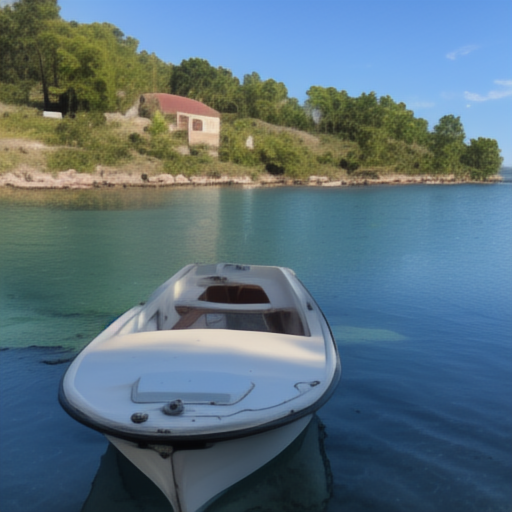}};
        \end{tikzpicture} &

        \begin{tikzpicture}[spy using outlines=]
            \node {\includegraphics[width=\ww,frame]{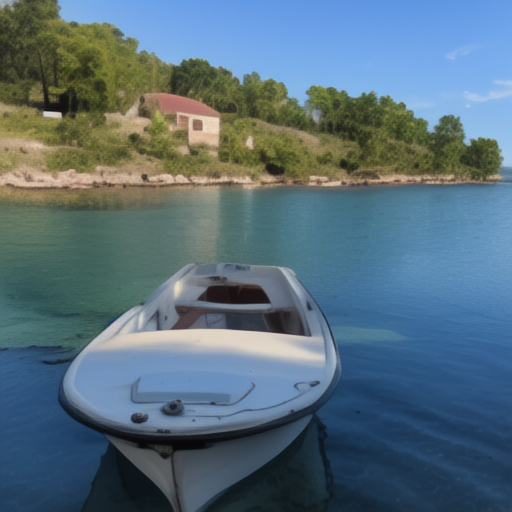}};
        \end{tikzpicture} &

        \begin{tikzpicture}[spy using outlines=]
            \node {\includegraphics[width=\ww,frame]{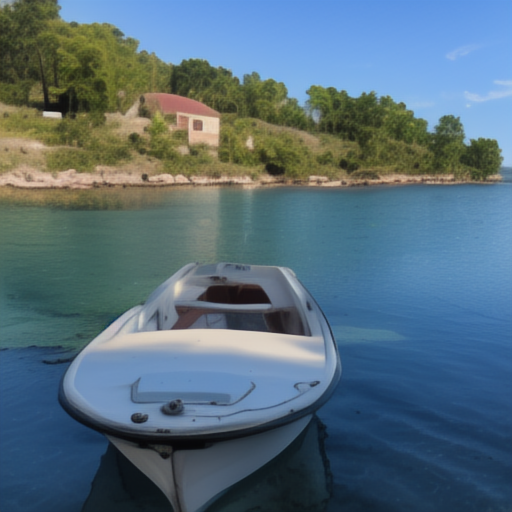}};
        \end{tikzpicture} &

        \begin{tikzpicture}[spy using outlines=]
            \node {\includegraphics[width=\ww,frame]{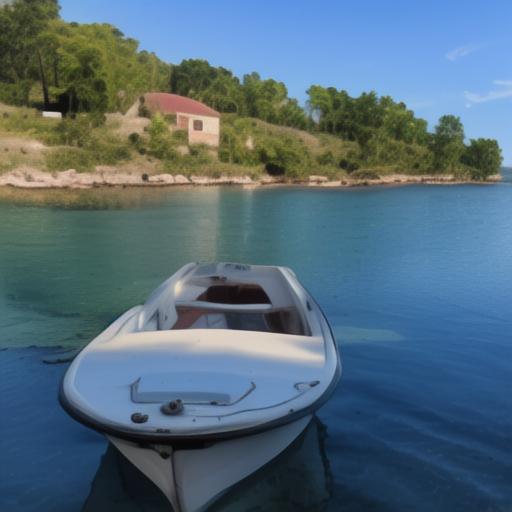}};
        \end{tikzpicture} &

        \begin{tikzpicture}[spy using outlines=]
            \node {\includegraphics[width=\ww,frame]{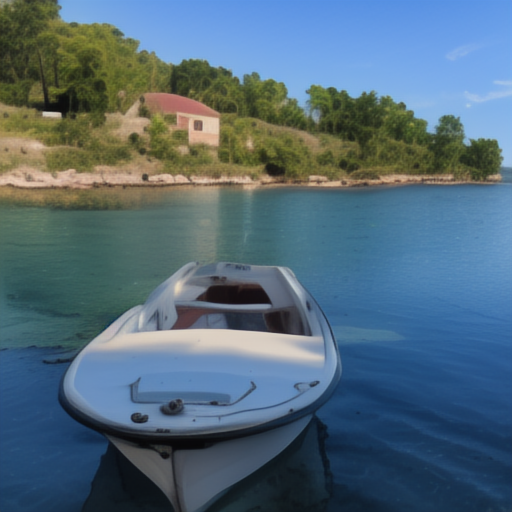}};
        \end{tikzpicture}
        
        \\[-3pt]
        \scriptsize{6 } &
        \scriptsize{7 } &
        \scriptsize{8 } &
        \scriptsize{9 } &
        \scriptsize{10 } 
        \\[-3pt]

        \begin{tikzpicture}[spy using outlines=]
            \node {\includegraphics[width=\ww,frame]{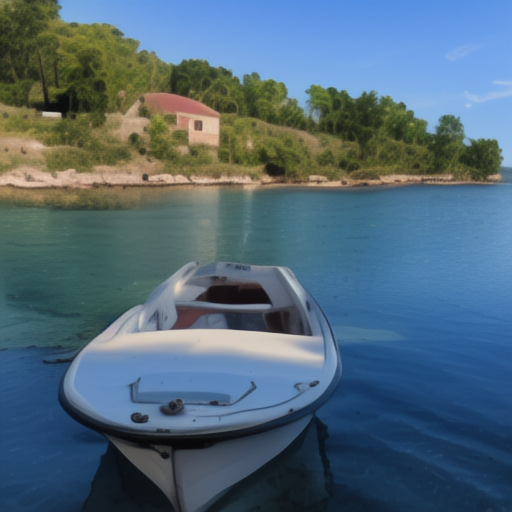}};
        \end{tikzpicture} &

        \begin{tikzpicture}[spy using outlines=]
            \node {\includegraphics[width=\ww,frame]{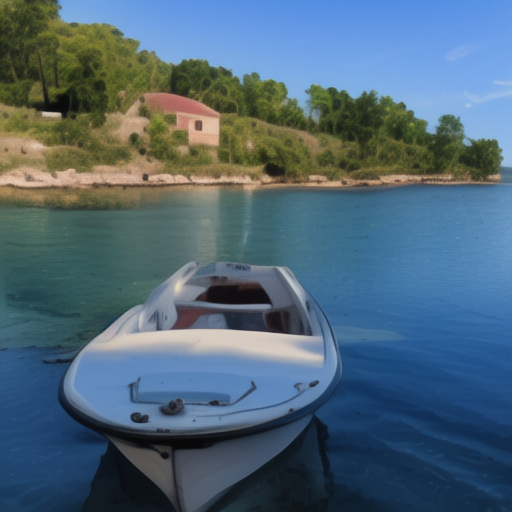}};
        \end{tikzpicture} &

        \begin{tikzpicture}[spy using outlines=]
            \node {\includegraphics[width=\ww,frame]{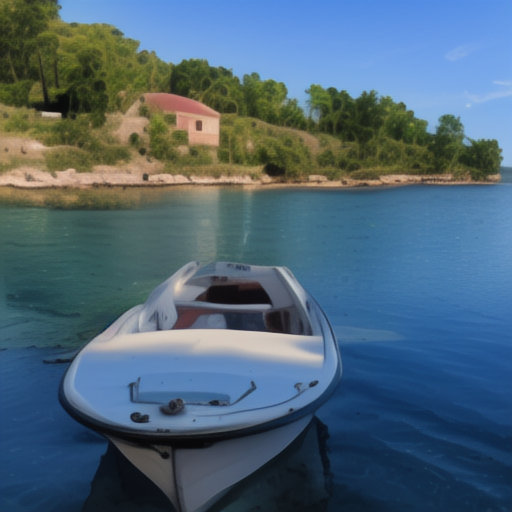}};
        \end{tikzpicture} &

        \begin{tikzpicture}[spy using outlines=]
            \node {\includegraphics[width=\ww,frame]{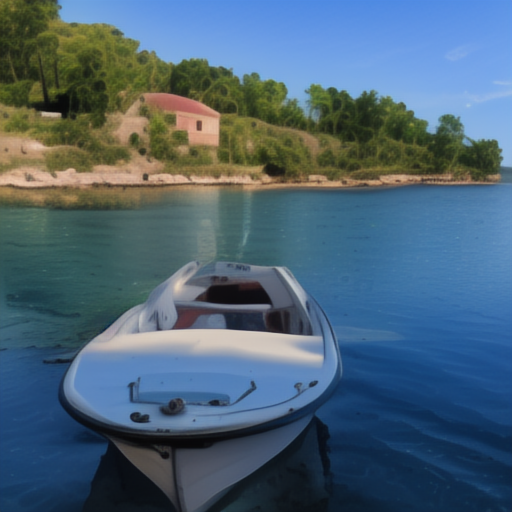}};
        \end{tikzpicture} &

        \begin{tikzpicture}[spy using outlines=]
            \node {\includegraphics[width=\ww,frame]{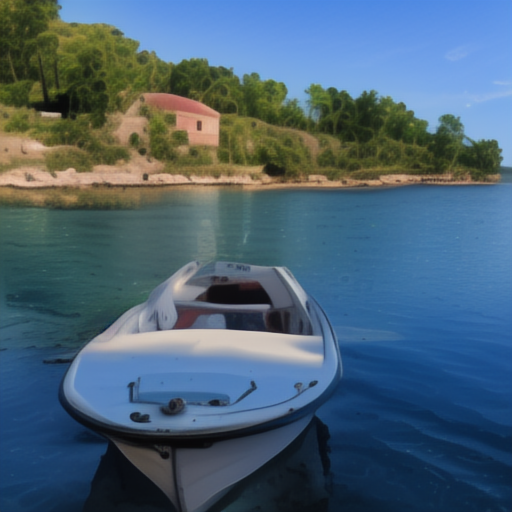}};
        \end{tikzpicture}
        
        \\[-3pt]
        \scriptsize{11 } &
        \scriptsize{12 } &
        \scriptsize{13 } &
        \scriptsize{14 } &
        \scriptsize{15 } 
        \\[-3pt]
        
        \begin{tikzpicture}[spy using outlines=]
            \node {\includegraphics[width=\ww,frame]{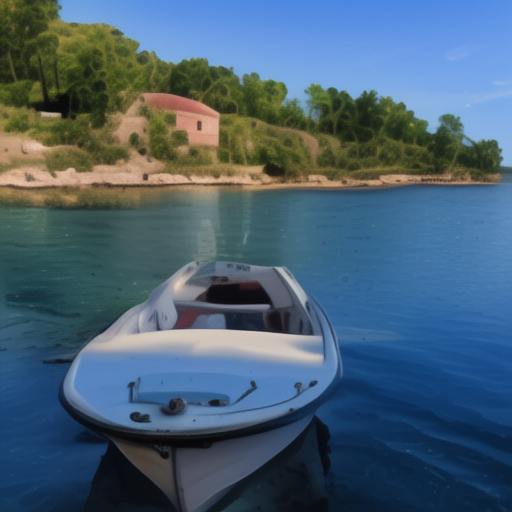}};
        \end{tikzpicture} &

        \begin{tikzpicture}[spy using outlines=]
            \node {\includegraphics[width=\ww,frame]{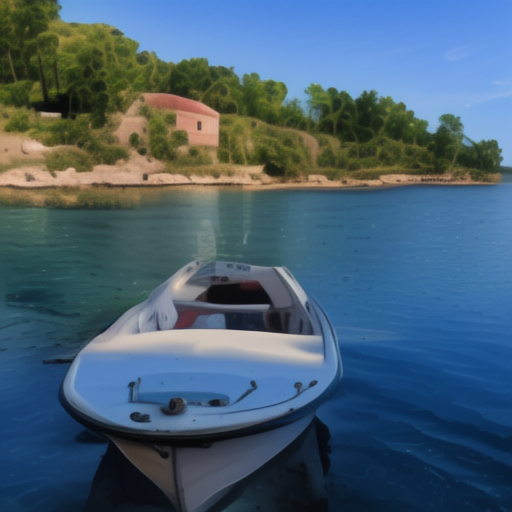}};
        \end{tikzpicture} &

        \begin{tikzpicture}[spy using outlines=]
            \node {\includegraphics[width=\ww,frame]{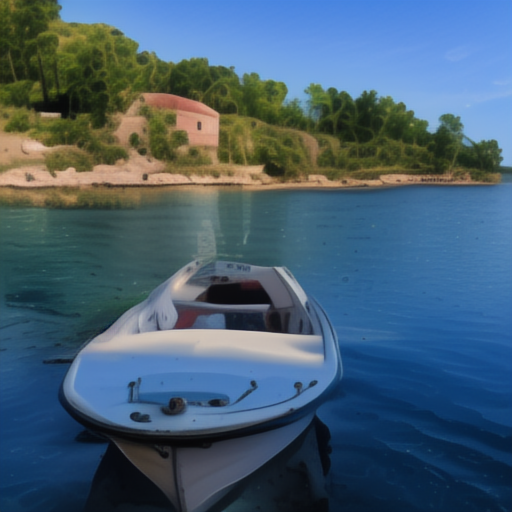}};
        \end{tikzpicture} &

        \begin{tikzpicture}[spy using outlines=]
            \node {\includegraphics[width=\ww,frame]{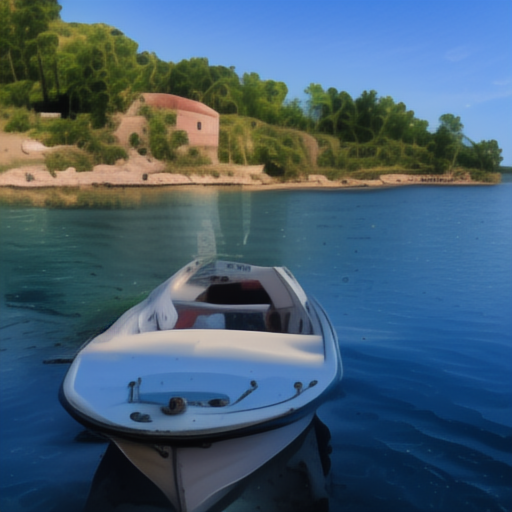}};
        \end{tikzpicture} &

        \begin{tikzpicture}[spy using outlines=]
            \node {\includegraphics[width=\ww,frame]{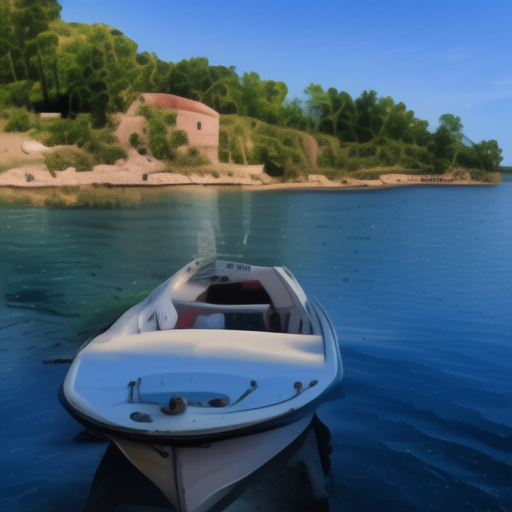}};
        \end{tikzpicture}
        
        \\[-3pt]
        \scriptsize{16 } &
        \scriptsize{17 } &
        \scriptsize{18 } &
        \scriptsize{19 } &
        \scriptsize{20 }
        \\[-3pt]

        \begin{tikzpicture}[spy using outlines=]
            \node {\includegraphics[width=\ww,frame]{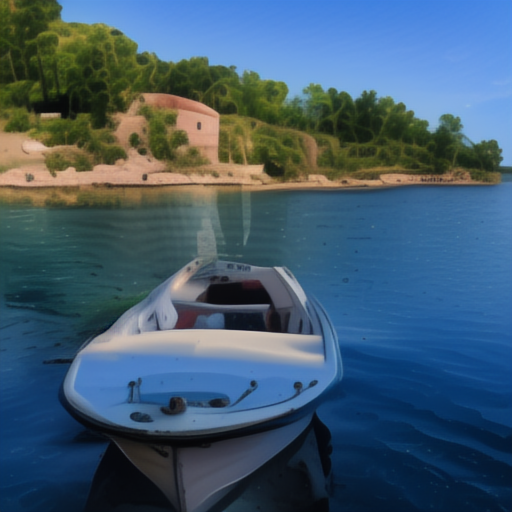}};
        \end{tikzpicture} &

        \begin{tikzpicture}[spy using outlines=]
            \node {\includegraphics[width=\ww,frame]{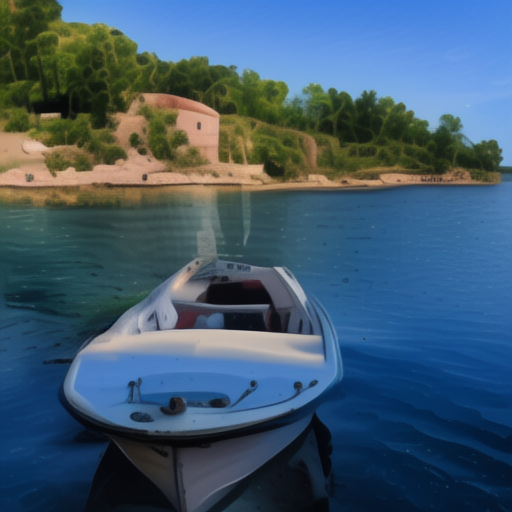}};
        \end{tikzpicture} &

        \begin{tikzpicture}[spy using outlines=]
            \node {\includegraphics[width=\ww,frame]{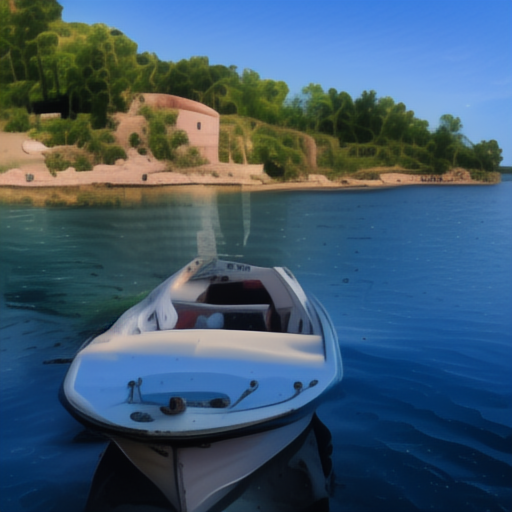}};
        \end{tikzpicture} &

        \begin{tikzpicture}[spy using outlines=]
            \node {\includegraphics[width=\ww,frame]{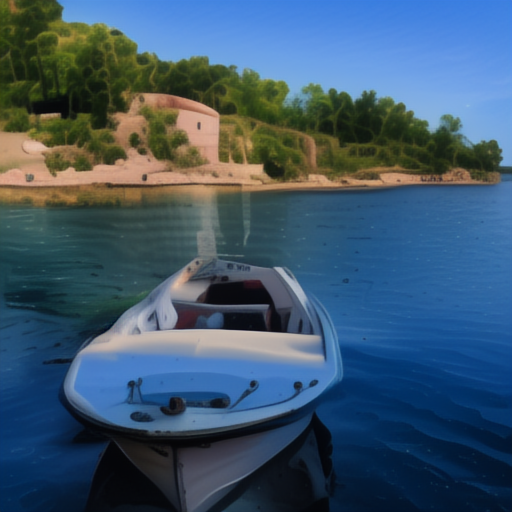}};
        \end{tikzpicture} &

        \begin{tikzpicture}[spy using outlines=]
            \node {\includegraphics[width=\ww,frame]{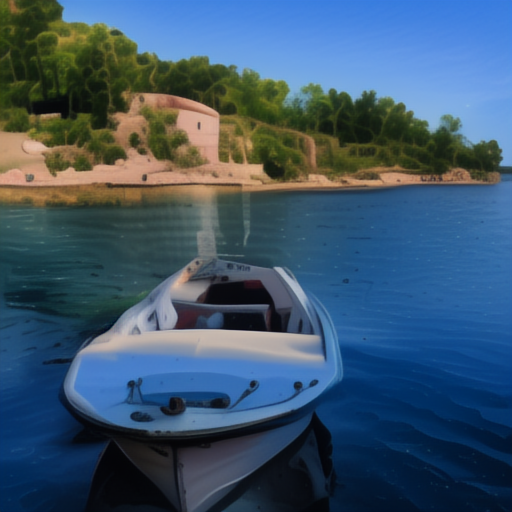}};
        \end{tikzpicture}
        
        \\[-3pt]
        \scriptsize{21 } &
        \scriptsize{22 } &
        \scriptsize{23 } &
        \scriptsize{24 } &
        \scriptsize{25 }
        \\[-3pt]
        
    \end{tabular}
    \caption{Full sequence of iterative inversion reconstructions using REED-VAE with Null-Text Inversion (NTI) \cite{mokady2023null}. The input image undergoes NTI-based inversion followed by reconstruction with the same source prompt for 25 iterations. REED-VAE maintains high fidelity to the input image across all iterations, with significantly reduced artifacts, noise, and distortions even at high iterations.}

    \label{fig:nti_editing_reed_full}
\end{figure*}

We also provide more extensive results from the iterative inversion experiment in \Cref{fig:nti_editing_vanilla_full} and \Cref{fig:nti_editing_reed_full}.
In the full sequence, it is evident that using NTI with the Vanilla-VAE (\Cref{fig:nti_editing_vanilla_full}) causes artifacts and noise patterns to progressively worsen in the first 10 iterations, until arriving at near complete noise at iteration 13.
Likely due to the involvement of the source prompt during inversion, the image is able to ``bounce back'', but this time severely diverged from the original imaging, displaying heavy distortions and color shifts.
Through iteration 25, the image continues to build noise in a more typical manner.
When NTI is combined with REED-VAE \Cref{fig:nti_editing_reed_full}, the fidelity to the input image is maintained throughout the entire 25 iterations.
Although there are some minor color distortions, the overall level of noise and artifacts is significantly reduced.
The improved fidelity to the input image with REED-VAE already visible in Inversion 1, as well as the lack of noise dominance around Inversion 13, suggest REED-VAE may contribute to a more efficient latent space organization that is more conducive to image editing and resilient to iterative operations.

\clearpage

\clearpage

\end{document}